\documentclass[epj]{svjour}
\usepackage{mdframed}
\usepackage{graphicx}
\usepackage{hyperref}
\usepackage{url}
\usepackage{natbib}

\usepackage{tablefootnote}
\usepackage{caption}
\usepackage{subcaption}
\usepackage{amsmath}
\usepackage{amssymb}
\usepackage[table]{xcolor}

\newcommand{\angstrom}{\textup{\AA}}

\begin{document}
\title{The external photoevaporation of planet-forming discs}
\author{Andrew J. Winter\inst{1,}\thanks{andrew.winter@uni-heidelberg.de} \and Thomas J. Haworth\inst{2,}
\thanks{t.haworth@qmul.ac.uk}%
}                     
%
%
\institute{Zentrum für Astronomie, Heidelberg University, Albert Ueberle Str. 2, 69120 Heidelberg, Germany \and Astronomy Unit, School of Physics and Astronomy, Queen Mary University of London, London E1 4NS, UK}
\date{Received: date / Revised version: date}
%
\abstract{Planet-forming disc evolution is not independent of the star formation and feedback process in giant molecular clouds. In particular, OB stars emit UV radiation that heats and disperses discs in a process called `external photoevaporation'. This process is understood to be the dominant environmental influence acting on planet-forming discs in typical star forming regions. Our best studied discs are nearby, in sparse stellar groups where external photoevaporation is less effective. However the majority of discs are expected to reside in much stronger UV environments. Understanding external photoevaporation is therefore key to understanding how most discs evolve, and hence how most planets form. Here we review our theoretical and observational understanding of external photoevaporation. We also lay out key developments for the future to address existing unknowns and establish the full role of external photoevaporation in the disc evolution and planet formation process. 
\PACS{
      {PACS-key}{discribing text of that key}   \and
      {PACS-key}{discribing text of that key}
     } 
} 
\maketitle

\tableofcontents

\section{Introduction}
\label{sec:intro}

Planets form in the discs of dust and gas known as `planet-forming' or `protoplanetary' discs. These discs are found around most young stellar objects (YSOs) in star forming regions up to ages of $\sim 3{-}10$~Myr \citep[e.g.][]{Haisch01, Ribas15}. Since 2015, \textit{Atacama Large Millimetre/sub-Millimetre Array} (ALMA) has instigated a revolution in our understanding of these objects. In particular, the high resolution sub-mm continuum images, tracing the dust content of protoplanetary discs, not only offer an insight into the dust masses and radii \citep[e.g.][]{Ansdell16, Ansdell17, Eisner18, vanTerwisga20, Ansdell20}, but also a wealth of internal sub-structures which serve as a window into planet formation processes \citep[e.g.][]{HLTau15, 2018ApJ...869L..41A, 2021ApJ...916L...2B}. The nearby (distance $D\lesssim 150$~pc) discs in the low-mass star forming regions such as Taurus and Lupus can be sufficiently resolved to expose rings, gaps and spirals in dust, as well as kinematic signatures in gas \citep{2018ApJ...860L..12T, 2019NatAs...3.1109P, 2019Natur.574..378T, 2022arXiv220309528P, 2022arXiv220603236C} which may either be signatures of planets  or the processes that govern their formation.

Prior to ALMA, some of the first images of discs were obtained around stars in the Orion Nebula cluster (ONC) at a distance of $D\sim 400$~pc away, much further than the famous HL Tau \citep{Sargent87, HLTau15}. These discs exhibit non-thermal emission in the radio \citep{Churchwell87, Felli93, Felli93b, Forbrich21}, with resolved ionisation fronts and disc silhouettes seen at optical wavelengths \citep{O'dell93, O'Dell94, Prosser94, McCaughrean96}. Identified as `solar-system sized condensations' by \citet{Churchwell87}, they were estimated to be undergoing mass loss at a rate of $\dot{M} \sim 10^{-6}\,M_\odot$~yr$^{-1}$. \citet{O'dell93} confirmed them to be irradiated protoplanetary discs, dubbing them with the contraction `proplyds'. While this term has now been dropped for the broader class of protoplanetary disc, it is still used in relation to discs with a bright, resolved ionisation front owing to the external UV radiation field. These objects are of great importance in unravelling what now appears to be an important process in understanding planet formation: \textit{external photoevaporation}. 

The process of external photoevaporation is distinguished from the internal photoevaporation (or more commonly just `photoevaporation') of protoplanetary discs via the source that is responsible for driving the outflows. Both processes involve heating of the gas by photons, leading to thermal winds that deplete the discs. However, internal photoevaporation is driven by some combination of FUV (far-ultraviolet), EUV (extreme-ultraviolet) and X-ray photons originating from the central host star \citep[e.g.][]{2008ApJ...688..398E, 2019MNRAS.487..691P, 2022arXiv220409704S}, depleting the disc from inside-out \citep[e.g.][]{Clarke01, Owen10, 2012MNRAS.422.1880O, Jennings18, Coleman22}. \textit{External} photoevaporation, by contrast, is driven by an OB star external to the host star-disc system. The high FUV and EUV luminosities of OB stars, combined with the heating of outer disc material that experiences weaker gravity, can result in extremely vigorous outside-in depletion \citep{Johnstone98}. Indeed, as inferred by \citet{Churchwell87}, the brightest proplyds in the ONC exhibit wind-driven mass loss rates of up to $\dot{M}_{\rm{ext}} \sim 10^{-6}\,M_\odot$~yr$^{-1}$ \citep[see also e.g.][]{Henney98, Henney99}. 

At present, external photoevaporation remains an under-studied component of the planet formation puzzle. It has long been thought that the majority of star formation occurs in clusters or associations which include OB stars \citep{Miller78} and that mass loss due to external photoevaporation can be more rapid than the $\dot{M}_\mathrm{acc/int} \sim 10^{-8}\,M_\odot$~yr$^{-1}$ typical of stellar accretion \citep[e.g.][]{Manara12} and internal photoevaporation \citep[e.g.][]{Owen10}. Thus it is reasonable to ask why external photoevaporation has, until now, not featured heavily in models for planet formation. While there is no one answer to this question, two important historical considerations are compelling in this context:
\begin{itemize}
    \item The first is a selection bias. Low mass star forming regions (SFRs) are more numerous, so they are also the closest to us \citep[although nearby star formation is probably also affected by the expansion of a local supernova bubble --][]{Zucker22}. Surveys with, for example, ALMA, have rightly prioritised bright and nearby protoplanetary discs in these low mass SFRs. However, when accounting for the relative number of stars, these regions are not typical birth environments for stars and their planetary systems (see Section~\ref{sec:demographics_SFRs}). This has motivated recent studies of more typical birthplaces that experience strong external irradiation from neighbouring massive stars \citep[e.g.][]{Ansdell17,Eisner18,vanTerwisga19}.
    
    \item The second is the well-known `proplyd lifetime problem' in the ONC \citep[e.g.][]{Storzer99b}. The presence or absence of a near infrared (NIR) excess, indicating inner material, was one of the earliest available probes for studying disc physics. The high fraction ($\sim 80$~percent) of stars with a NIR excess in the ONC \citep{Hillenbrand98} was an apparent paradox that appeared to undermine the efficacy of external photoevaporation in influencing disc populations. We discuss this problem in Section~\ref{sec:stellar_dynamics}. 
\end{itemize}

Given the apparently high fraction of planet forming discs that undergo at least some degree of external photoevaporation, this process has particular relevance for modern efforts to connect observed disc populations to exoplanets \citep[e.g.][]{2018haex.bookE.143M}. Now is therefore an opportune time to take stock of the findings from the past three decades. 


In this review, we first summarise the theory of external photoevaporation in Section~\ref{sec:theory}, including both analytic estimates that provide an intuition for the problem as well as state-of-the-art simulations and microphysics. We address the observational signatures of disc winds in Section~\ref{sec:obs}, including inferences of individual mass loss rates. We consider the role of external photoevaporation for disc evolution and planet formation in Section~\ref{sec:planet_formation}, including evidence from disc surveys. In Section~\ref{sec:SFRs} we contextualise these studies in terms of the physics and demographics of star forming environments. Finally, we summarise the current understandings and open questions in Section~\ref{sec:concluding_summary}. 


\section{Theory of externally driven disc winds}
\label{sec:theory}

\subsection{The most basic picture}
\begin{figure}
    \centering
    \vspace{-0.5cm}
    \includegraphics[width=\columnwidth]{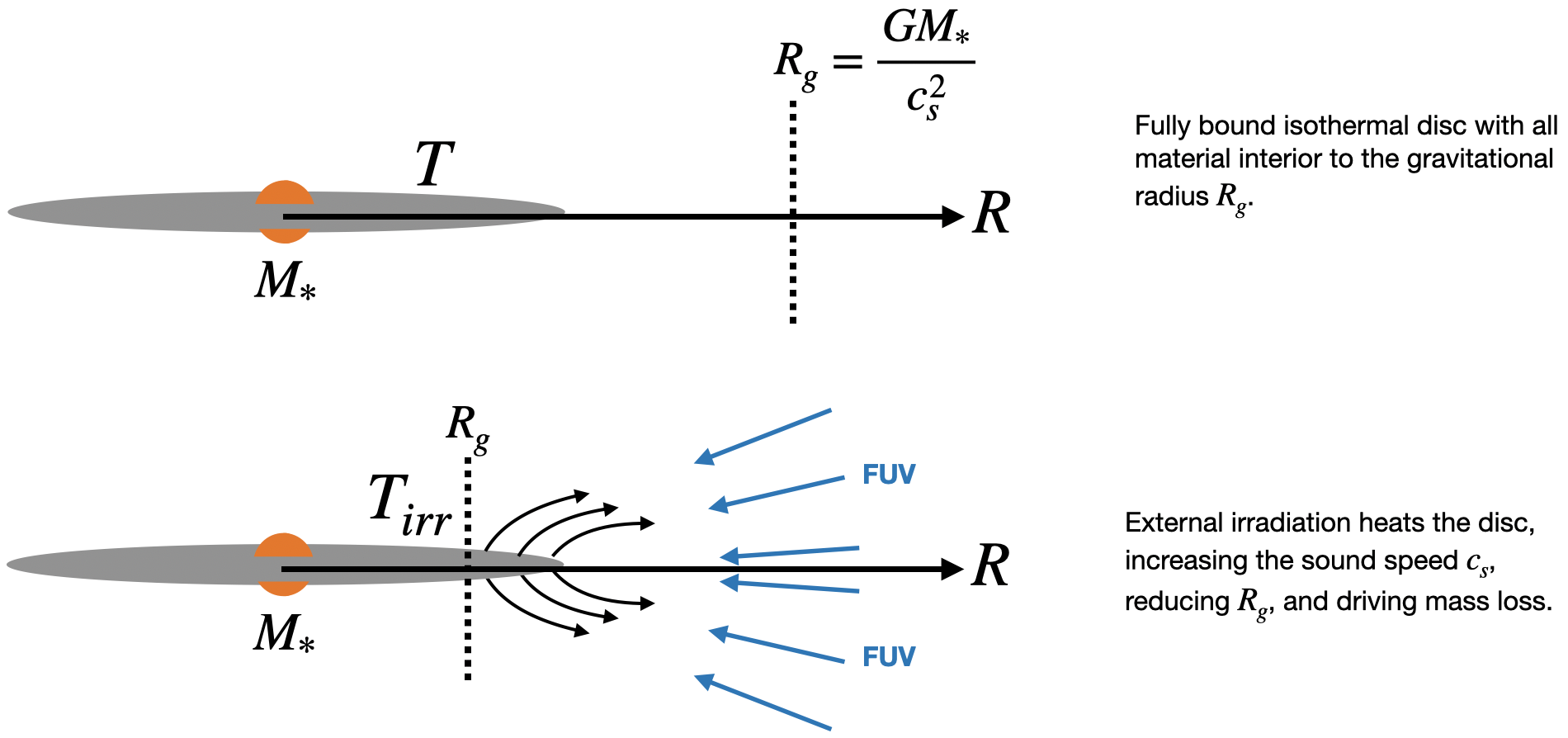}
    \caption{A schematic of the basic picture of external photoevaporation in terms of the gravitational radius. The gravitational radius is that beyond which mean thermal motions (propagating at the sound speed) exceed the escape velocity and are unbound. In this basic picture, a disc smaller than the gravitational radius will hence not lose mass.  External UV irradiation heats the disc, leading to faster mean thermal motions (a higher sound speed), driving the gravitational radius to smaller radii and unbinding material in the disc.  }
    \label{fig:basicRgrav}
\end{figure}

We begin by reviewing our theoretical understanding of external photoevaporation. At the very basic level one can determine whether material will be unbound from a potential by comparing the mean thermal velocity of particles (i.e. the sound speed) in the gas with the local escape velocity. In an isothermal system, equating the sound speed and escape velocity yields the gravitational radius, beyond which material is unbound
\begin{equation}
    R_{\rm{g}} = \frac{GM_*}{c_{\rm{s}}^2}
\end{equation}
where $M_*$ is the point source potential mass, and $c_{\rm{s}}$ is the sound speed. Therefore if an isothermal disc were to extend beyond $R_{\rm{g}}$, then material would be lost in a wind. Consider now an isothermal disc that is entirely interior to the gravitational radius and is hence fully bound (upper panel of Figure \ref{fig:basicRgrav}). If this disc is then externally irradiated, the temperature and hence sound speed increases, driving down the gravitational radius to smaller values, potentially moving interior to the edge of the disc and unbinding its outer parts (lower panel of Figure \ref{fig:basicRgrav}). 
The details of external photoevaporation do get subtantially more complicated than the above picture, for example with pressure gradients helping to launch winds interior to $R_{\rm{g}}$. However, this picture provides a neat basic insight into how external photoevaporation can instigate mass loss in otherwise bound circumstellar planet-forming discs.

\subsection{EUV and FUV driven flows}

\label{sec:EUVvFUV}


\begin{figure}
     \centering
     \begin{subfigure}[b]{0.49\textwidth}
         \centering
         \includegraphics[width=\textwidth]{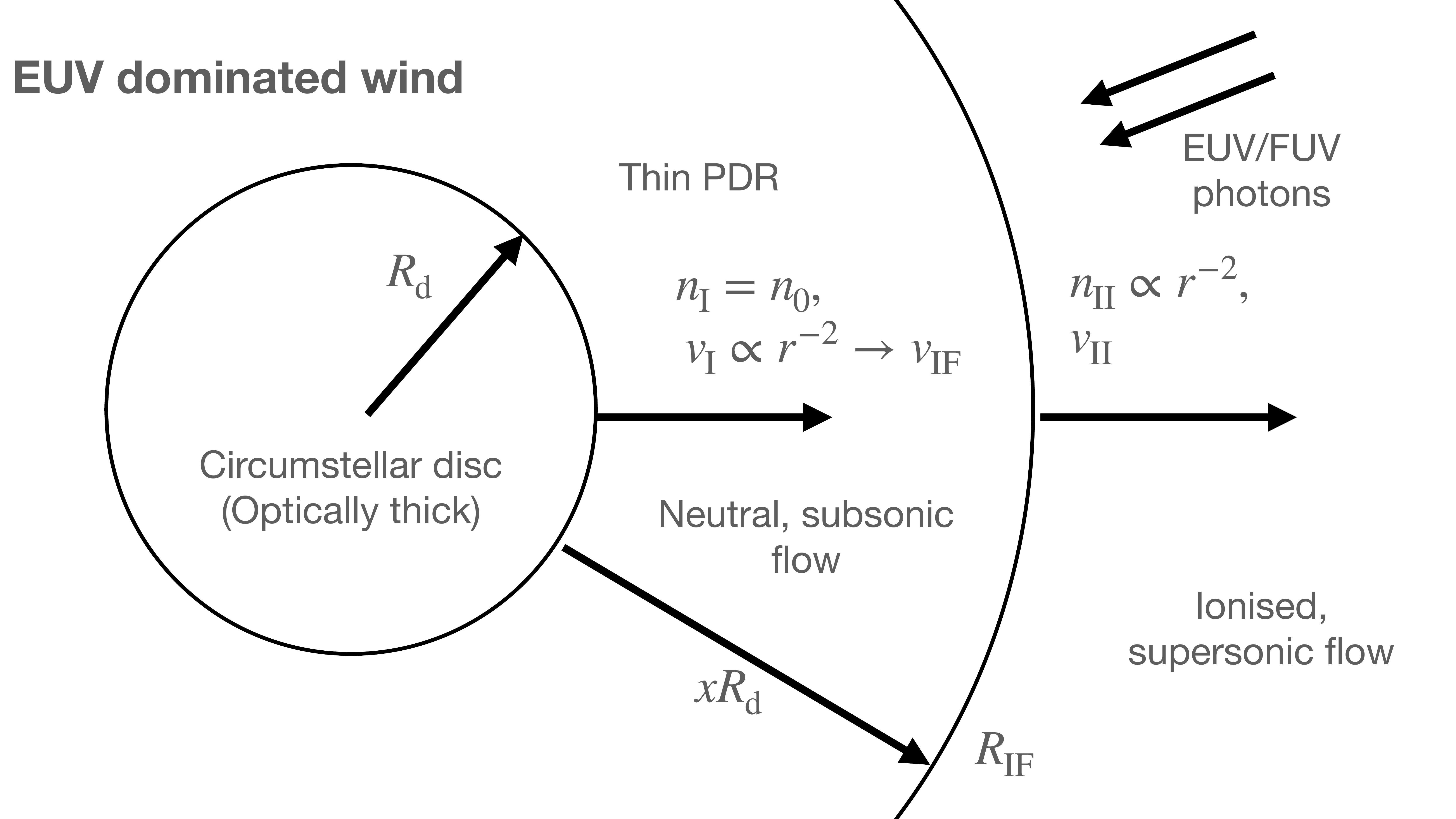}
         \caption{}
         \label{fig:EUVschem}
     \end{subfigure}
     \hfill
     \begin{subfigure}[b]{0.49\textwidth}
         \centering
         \includegraphics[width=\textwidth]{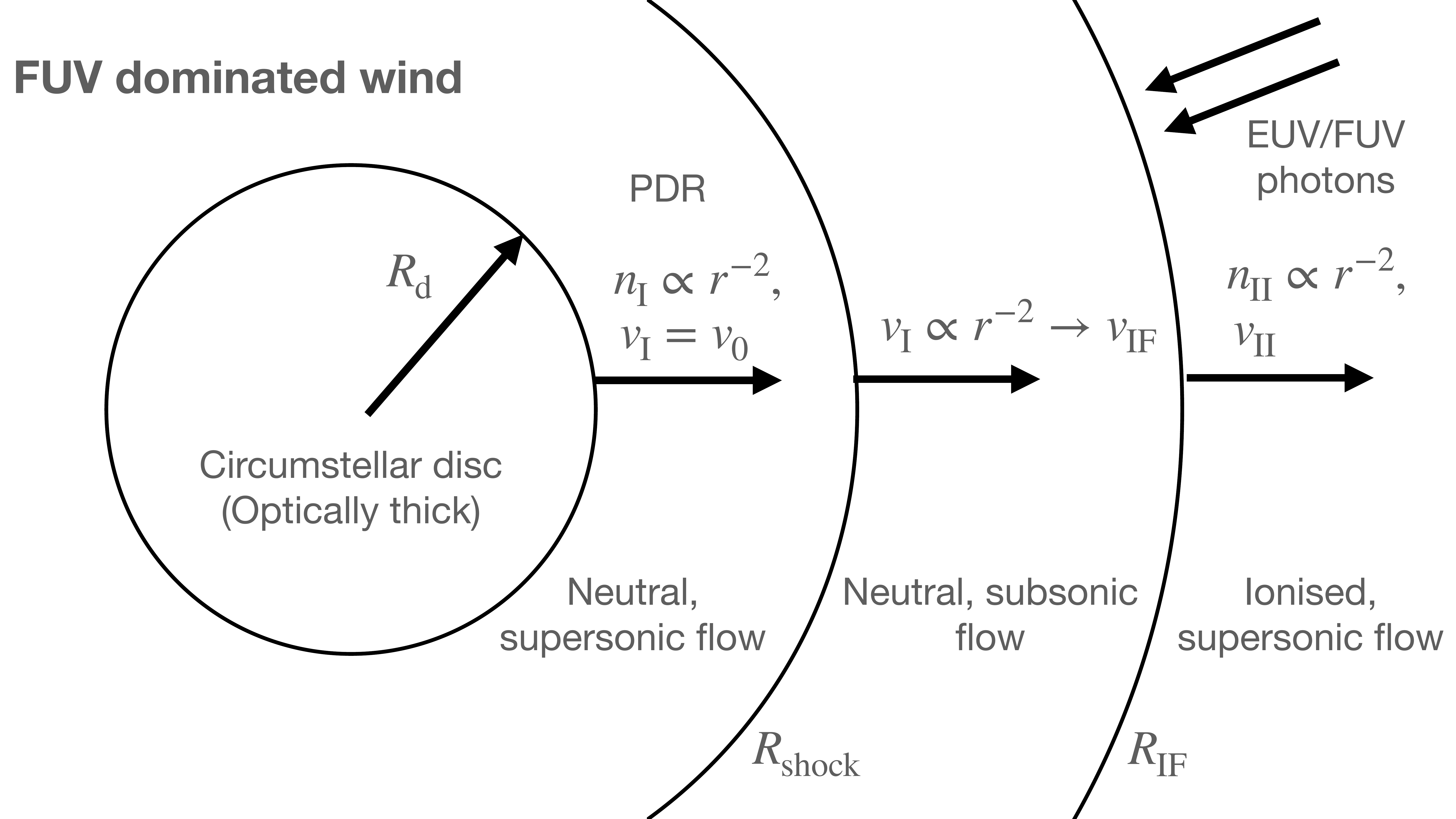}
         \caption{
    }
         \label{fig:FUVschem}
     \end{subfigure}
     
       \label{fig:schem_wind}
     \caption{Schematic diagrams of the flow structure in EUV (Figure~\ref{fig:EUVschem}) and FUV (Figure~\ref{fig:FUVschem}) externally driven winds, adapted from \citet{Johnstone98}. In the EUV driven wind, the flow from the disc edge at radius $R_\mathrm{d}$ travels at a subsonic velocity through the thin photodissociation region (PDR) of thickness $xR_\mathrm{d}$ with $x\lesssim 1.5$ before reaching the ionisation front (IF). Mass loss is therefore determined by the thermal pressure at the IF. For $x\gtrsim 1.5$, the wind is launched from the disc edge at a supersonic velocity, producing a shock front that is reached before the IF. In this case, the mass loss rate is determined by the thermal conditions in the PDR. }
\end{figure}

\subsubsection{Flux units}
\label{sec:flux_units}
Before considering the physics of EUV and FUV driven winds, a note on the units canonically used to measure UV fluxes is necessary. While the ionising flux is usually measured in photon counts per square centimetre per second, FUV flux is normally expressed in terms of the Habing unit, written $G_0$ \citep{1968BAN....19..421H}. This is the flux integral over $912-2400\angstrom$, normalized to the value in the solar neighbourhood, i.e.
\begin{equation}
    \left(\frac{F_\mathrm{FUV}}{1\,G_0}\right) = \int_{912\angstrom}^{2400\angstrom} \frac{F_\lambda d\lambda}{1.6\times10^{-3}\,\textrm{erg\,s}^{-1}\,\textrm{cm}^{-2}}.
\end{equation}
Another measure of the FUV field strength is the Draine unit \citep{1978ApJS...36..595D}, which is a factor 1.71 larger than the Habing unit. Hence $10^3 \, G_0 \approx 585$~Draines. We highlight both because two similar units that vary by a factor of order unity can and does lead to confusion. For reference, the UV environments that discs are exposed to in star forming regions ranges from $\ll 1$ (i.e.  embedded discs) to $\sim10^{7}$G$_0$ (discussed further in Section~\ref{sec:SFRs}). {For the sake of clarity, we will hereafter consider low FUV environments to be those with $F_\mathrm{FUV}\lesssim 100\, G_0$, intermediate environments greater than this up to $F_\mathrm{FUV} \lesssim 5000 \, G_0$ and high FUV environments for any FUV fluxes $F_\mathrm{FUV}\gtrsim 5000\, G_0$.}

\subsubsection{Flow geometry}

The basic physical picture of an externally photoevaporating protoplanetary disc was laid out by \citet{Johnstone98}, and has remained largely the same since. We summarise that picture here because it is useful for what follows, but refer interested readers to the more detailed discussion in that original work as well as that of \citet{Storzer99b}. 

The heating mechanism that launches the thermal wind may be driven by ionising EUV photons (energies  $h\nu >13.6$~eV), heating gas to temperatures $T\sim 10^4$~K, or photodissociation FUV photons ($6\,\rm{eV}< h\nu < 13.6$~eV), yielding temperatures of $T\sim 100{-}1000$~K \citep{Tielens85, Hollenbach97}. The EUV photons penetrate down to the ionisation front, at radius $R_\mathrm{IF} = (1+x)R_\mathrm{d}$ from the disc-hosting star, with disc outer radius $R_\mathrm{d}$. The atomic gas outside of $R_\mathrm{IF}$ is optically thin for the FUV photons, which penetrate down to $R_\mathrm{d}$ producing a neutral PDR of thickness $x R_\mathrm{d}$. Whichever photons drive the photoevaporative wind, the flow geometry far from the disc surface is approximately spherically symmetric, since it is accelerated by the radial pressure gradient (cf. the Parker wind). We will start with this approximation of spherical geometry, which guides the following analytic intuition. 

\subsubsection{EUV driven winds}

For an EUV driven wind, the gravitational radius is $R_\mathrm{g}\sim 10$~au {for a solar mass star (and scales linearly with stellar mass)}, such that we are generally in a regime in the outer disc where $R_\mathrm{d}\gg R_\mathrm{g}$, and a wind can be launched. If the EUV flux is sufficiently strong, the ionisation front (IF) sits close to the disc surface making the PDR thin ($x \lesssim 1.5$; see Section~\ref{sec:FUVwind}). In this case, the basic geometry of the system is shown in Figure~\ref{fig:EUVschem}. The thermal pressure at the disk surface is determined by the ionisation rate, with the flow proceeding subsonically through the PDR. If we assume isothermal conditions in the PDR, then the density $n_\mathrm{I}$ is constant: $n_\mathrm{I} = n_0 = N_\mathrm{D} /xR_\mathrm{d}$, where $N_\mathrm{D}$ is the column density of the PDR, which is the column density required to produce an optical depth $\tau_\mathrm{FUV} \sim 1$. This column density is dominated by the base of the wind $N_\mathrm{D}\approx n_0 R_\mathrm{d}$, but is dependent on the microphysics and dust content (see Section~\ref{sec:microphysics}). For our purposes, we will simply adopt  $N_\mathrm{D} \sim 10^{21}$~cm$^{-2}$ \citep[although see the direct calculations of][for example]{Storzer99b}.

Since density in the PDR is constant, the velocity in the flow $v_\mathrm{I} \propto r^{-2}$ to maintain a constant mass flux. In order to conserve mass and momentum flux, the velocity at the ionisation front must be $v_\mathrm{IF} = c_\mathrm{s,I}^2/2c_\mathrm{s,II} \sim 0.5$~km~s$^{-1}$, where $c_\mathrm{s,I} \approx 3$~km~s$^{-1}$ is the sound speed in the PDR and $c_\mathrm{s,II} \approx 10$~km~s$^{-1}$ beyond the IF. We can write the mass loss rate:
\begin{equation}
    \dot{M}_\mathrm{EUV} = 4\pi \mathcal{F} (1+x)^2 R_\mathrm{d}^2 n_\mathrm{I} m_\mathrm{I} \frac{c_\mathrm{s,I}^2}{2c_\mathrm{s,II}} ,
\end{equation}
where $\mathcal{F}$ is a geometric factor and $m_\mathrm{I}$ is the mean molecular mass in the PDR, and as before $x$ is the relative thickness of the PDR with respect to the disc radius $R_\mathrm{d}$. Since $n_\mathrm{I} \propto 1/x$ this mass loss rate appears to diverge in the limit of a thin PDR. However, $x$ must satisfy the recombination condition such that:
\begin{equation}
\label{eq:recomb}
    \frac{f_\mathrm{r} \Phi }{4\pi d^2} = \int_{R_\mathrm{IF}}^\infty \alpha_\mathrm{B} n_{\mathrm{II}}^2 \,\mathrm{d}r = R_\mathrm{IF}^4\left(\frac{m_\mathrm{I} v_{\rm{IF}}}{m_\mathrm{II} v_\mathrm{II}}\right)^2 \int_{R_\mathrm{IF}}^\infty \alpha_\mathrm{B} \frac{n_{\rm{I}}^2}{r^4} \,\mathrm{d}r 
\end{equation}where $\Phi$ is the EUV counts of the ionising source at distance $d$, $f_\mathrm{r}$ is the fraction of photons unattenuated by the interstellar medium (ISM), and $\alpha_\mathrm{B} = 2.6 \times 10^{-13}$~cm$^{3}$~s$^{-1}$ is the recombination coefficient for hydrogen with temperature $10^4$~K \citep[e.g.][]{Osterbrock89}. Substituting $R_\mathrm{IF} = (1+x)R_\mathrm{d}$ into equation~\ref{eq:recomb} and adopting typical values, we can write a defining equation for $x$ that can be solved numerically. However, in the limit of $x\ll 1$  we can also simply estimate:
\begin{equation}
\label{eq:Mdot_EUV}
     \dot{M}_\mathrm{EUV} = 5.8 \times 10^{-8}  \epsilon_\mathrm{EUV}  \mathcal{F}  \left( \frac{d}{0.1\, \mathrm{pc}}\right)^{-1} \left(\frac{ f_\mathrm{r} \Phi}{10^{49} \, \mathrm{s}^{-1}} \right)^{1/2}  \left( \frac{ R_\mathrm{d}}{100 \, \mathrm{au}}\right)^{3/2} \, M_\odot \, \rm{yr}^{-1}, 
\end{equation}which is the solution for an ionised globule with no PDR physics \citep{Bertoldi90}. Here, $\epsilon_\mathrm{EUV} \sim 1$ is a correction factor that absorbs uncertainties in the PDR physics. We notice that the EUV driven wind is super-linearly dependent on $R_\mathrm{d}$, but only scales with the square root of the EUV photon count.

\subsubsection{FUV driven winds}
\label{sec:FUVwind}

We will here proceed under the assumption that the outer radius and FUV flux is sufficient to produce a supersonic, FUV heated wind. This means that $R_\mathrm{d}\gg R_\mathrm{g}$, where $R_\mathrm{g}$ is the gravitational radius in the PDR.  In this case, where the wind mass loss is determined by FUV heating, the neutral wind must launch at (constant) supersonic velocity $v_\mathrm{I} = v_0\gtrsim c_\mathrm{s,I}$, {where $v_0$ is the launching velocity} from the disc surface {with number density $n_0$, while $v_\mathrm{I}$ is the wind velocity in the PDR as before.} To conserve mass, the density in the PDR drops as $n_\mathrm{I} \propto r^{-2}$. The wind travels faster than at the IF ($v_\mathrm{IF} = c_\mathrm{s,I}^2/2c_\mathrm{s,II}$, as before) and therefore must eventually meet a shock front at radius $R_\mathrm{shock}$. Assuming this shock is isothermal, the density $n_\mathrm{I}$ increases by a factor $\sim M^2 $, where $M = v_0/c_{\mathrm{s,I}}$ is the Mach number. If the region between the shock and ionisation fronts is isothermal, then $n_\mathrm{I}$ is then constant and $v_\mathrm{I} \propto r^{-2}$ to conserve mass. This geometry is shown in Figure~\ref{fig:FUVschem}.

Solving mass and momentum conservation requirements for the flow, we have:
\begin{equation}
R_\mathrm{shock} = \sqrt{ \frac{v_0}{2c_\mathrm{s, II}}} R_\mathrm{IF},
\end{equation}which immediately puts a minimum distance below which EUV mass loss dominates over FUV. If $R_\mathrm{shock}< R_\mathrm{d}$ then the shock front is inside the disc, the flow is subsonic at the base, and we are back in the EUV driven wind regime. Given that $\sqrt{{v_0}/{2c_\mathrm{s, II}}} \sim 0.4{-}0.6$ (with $v_0 \sim 3{-}6$~km~s$^{-1}$), then we require $R_\mathrm{IF}\gtrsim 2.5 R_\mathrm{d}$ for an FUV driven wind to be launched. Coupled with equation~\ref{eq:recomb} this gives the minimum distance required for the launching of FUV driven winds. Conversely, a maximum distance exists from the requirement that the gas can be sufficiently heated to escape the host star, meaning that FUV driven winds only occur at intermediate separations from a UV source.

Under the assumption that the FUV flux is sufficient to launch a wind and $R_\mathrm{IF}\gtrsim 2.5 R_\mathrm{d}$, then the overall mass loss in the flow does not care about what is going outside of the base of the FUV-launched wind. The mass loss rate in the FUV dominated case is simply:
\begin{equation}
\label{eq:Mdot_FUV}
    \dot{M}_\mathrm{FUV} = 4\pi \mathcal{F}  R_\mathrm{d}^2 n_0  v_0 m_\mathrm{I} \approx  1.9 \times  10^{-7}  \epsilon_\mathrm{FUV}  \mathcal{F} \left( \frac{R_\mathrm{d}}{100\,\rm{au}} \right) \, M_\odot \, \rm{yr}^{-1}, 
\end{equation}{where $\mathcal{F}$ is the geometric correction and $m_\mathrm{I}$ is the mean molecular mass in the PDR as before.} All of the difficult physics is contained within a convenient correction factor $\epsilon_\mathrm{FUV}$. In reality, this expression is only helpful in the limit of a very extended disc, and computing the mass loss rate in general requires a more detailed treatment that we will discuss in Section~\ref{sec:microphysics}. 

Nonetheless, we gain some insight from the estimate from equation~\ref{eq:Mdot_FUV}. First, we see that the mass loss rate is not dependent on the FUV flux $F_\mathrm{FUV}$. In reality, there is some dependence due to the increased temperature in the PDR with increasing $F_\mathrm{FUV}$, but this dependence is weak for $F_\mathrm{FUV}\gtrsim 10^4\, G_0$ \citep{Tielens85}. From equation~\ref{eq:Mdot_EUV}, we also see that the mass loss rate $ \dot{M}_\mathrm{FUV}$ scales less steeply with $R_\mathrm{d}$ than $ \dot{M}_\mathrm{EUV}$ does. This means that once a disc has been sufficiently truncated by these external winds, the FUV dictates the mass loss rate. Since the time-scale for this depletion can be short (see Section~\ref{sec:gas_evolution}), we expect FUV flux to dominate mass loss over the disc lifetime for reasonable EUV fluxes. 

While this picture is useful to gain some insight into the physics of external photoevaporation, accurately computing the mass loss rate in the wind requires more detailed numerical modelling of the PDR physics. We consider efforts in this direction to date as follows.

\subsection{Microphysics of external photoevaporation}
\label{sec:microphysics}
One of the biggest challenges surrounding external photoevaporation is that it depends upon a wide array of complicated microphysics. The wind is launched primarily by the FUV radiation field and determining the temperature in this launching region, which is critical, requires solving photodissociation region (PDR) microphysics which in itself can consist of many hundreds of species and reactions and complicated thermal processes \citep[e.g.][]{Tielens85, Hollenbach97}. In particular, as we will discuss below, the line cooling is difficult to estimate in 3D, meaning most PDR codes are limited to 1D \citep[e.g.][]{2007A&A...467..187R}. The dust and PAH abundance in externally driven wind also play key roles in determining the mass loss rate, but may differ substantially from the abundances in the ISM \citep[e.g.][]{2013ApJ...765L..38V, Facchini16}. In addition to the complicated PDR-dynamics, EUV photons establish an ionisation front downstream in the wind which affects the observational characteristics. Here we introduce some of the key aspects of the microphysics of external photoevaporation in more detail.

\subsubsection{The theory of dust grain entrainment in external photoevaporative winds}
We begin by considering the dust microphysics of external photoevaporation. First it is necessary to provide some context by discussing briefly how dust evolves in the disc itself. Canonically, in the ISM the dust-to-gas mass ratio is $10^{-2}$ and grains typically follow a size distribution of the form 
\begin{equation}
    \frac{\mathrm{d} n(a_\mathrm{s})}{\mathrm{d} a_\mathrm{s}} \propto a_\mathrm{s}^{-q}
\end{equation}
\citep{1977ApJ...217..425M} with grain sizes spanning $a_\mathrm{s}\sim10^{-3}-1$\,$\mu$m \citep{2001ApJ...548..296W} and $q\approx3.5$. In protoplanetary discs the dust grains grow to larger sizes which eventually (when the Stokes number is of order unity) dynamically decouples them from the gas, leading to radial drift inwards to the inner disc. This growth proceeds more quickly in the inner disc \citep[e.g.][]{Birnstiel12} and so there is a growth/drift front that proceeds from the inner disc outwards. It is not yet clear how satisfactory our basic models of this process are, particularly in terms of the timescale on which it operates, since if left unhindered by pressure bumps in the disc it quickly results in most of the larger drifting dust being deposited onto the central star \citep[see e.g.][]{Birnstiel12, 2015PASP..127..961A, 2020MNRAS.498.2845S}. However for our purposes the key point is that the abundance of smaller ($\sim \mu$m) grains in the disc ends up depleted relative to the ISM due to grain growth. 

The nature of dust in the external photoevaporative wind is important for three key reasons
\begin{enumerate}
    \item The dust in the wind sets the extinction in the wind and hence has a significant impact on the mass loss rate
    \item The extraction of dust in the wind could have implications for the mass reservoir in solids for terrestrial planet formation and/or the cores of gas giants. 
    \item The entrainment of dust in winds could provide observational diagnostics of the external photoevaporation process (we will discuss this further in section \ref{sec:obs}). 
\end{enumerate}
and so the key questions are what size, and how much, dust is entrained in an external photoevaporative wind. This problem was addressed in the semi-analytic work of  \cite{Facchini16}. They solved the flow structure semi-analytically (we discuss semi-analytic modelling of external photoevaporative winds further in section \ref{sec:semianalytic}) and calculated the maximum entrained grain size. The efficiency of dust entrainment in the wind is dependent on the balance of the \citet{Epstein24} drag exerted on the dust grain with density $\rho_\mathrm{s}$ by the the outflowing gas of velocity $v_\mathrm{th}$ versus the gravity from the host star. Thus the condition for a dust grain to be lost to external photoevaporation is \citep{Facchini16}:
\begin{equation}
    a_\mathrm{s} < a_\mathrm{ent} =\frac{1}{4\pi \mathcal{F} } \frac{v_\mathrm{th}\dot{M}_\mathrm{ext}}{G M_* \rho_\mathrm{s}} ,
    \label{equn:entrainedSize}
\end{equation} 
where $4\pi \mathcal{F}$ is the solid angle subtended by the wind \citep[see][]{Adams04}.

The main outcome of the above is that only small grains are entrained in the wind and the mean cross section is reduced. Therefore, when grain growth proceeds to the disc outer edge, the dust-to-gas mass ratio, mean cross section, and hence extinction in the wind drops substantially. This makes external photoevaporation more effective than previously considered when the dust in the wind was treated as ISM-like. This lower cross section in the wind is now accounted for in numerical models of external photoevaporation, assuming some constant low value \citep[e.g. the FRIED grid of mass loss rates][discussed more in \ref{sec:RHDmodels}]{Haworth18b}. However, what is still missed in models is that the cross section in the wind is actually a function of the mass loss rate \citep{Facchini16} and so needs to be solved iteratively with the dynamics.

\cite{2021MNRAS.508.2493O} also recently introduced dynamically decoupled dust into 1D isothermal models of externally irradiated discs (discussed more in section \ref{sec:boundaryconditions}) finding that it does indeed lead to a radial decrease in the maximum grain size which could be searched for observationally.  This gradient in grain sizes has been observed by \citet{2001Sci...292.1686T} and \citet{Miotello12}, who studied the large proplyd 114-426 on the near side of the ONC, calculating the attenuation of the background radiation through translucent parts of the disc. They found a maximum grain size decreasing from 0.7\,$\mu$m to 0.2\,$\mu$m, moving away from the disc outer edge over a distance of around 250\,au, consistent with \cite{2021MNRAS.508.2493O}.

\subsubsection{Photodissociation region physics for external photoevaporation}
\label{sec:PDRmicrophysics}
The FUV excited photodissociation region (PDR) microphysics determines the composition, temperature and therefore the dynamics of the inner parts of external photoevaporative winds. As discussed above, this FUV/PDR part of the inner wind can determine the mass loss rate from the disc. This is not a review on PDR physics \citep[for further information see e.g. ][]{Tielens85, Hollenbach97, 2008ARA&A..46..289T} but given its importance for setting the temperature, and therefore the dynamics, we provide a brief overview of some relevant processes. 

\begin{figure}
    \centering
    \vspace{-0.5cm}
    \includegraphics[width=0.48\columnwidth]{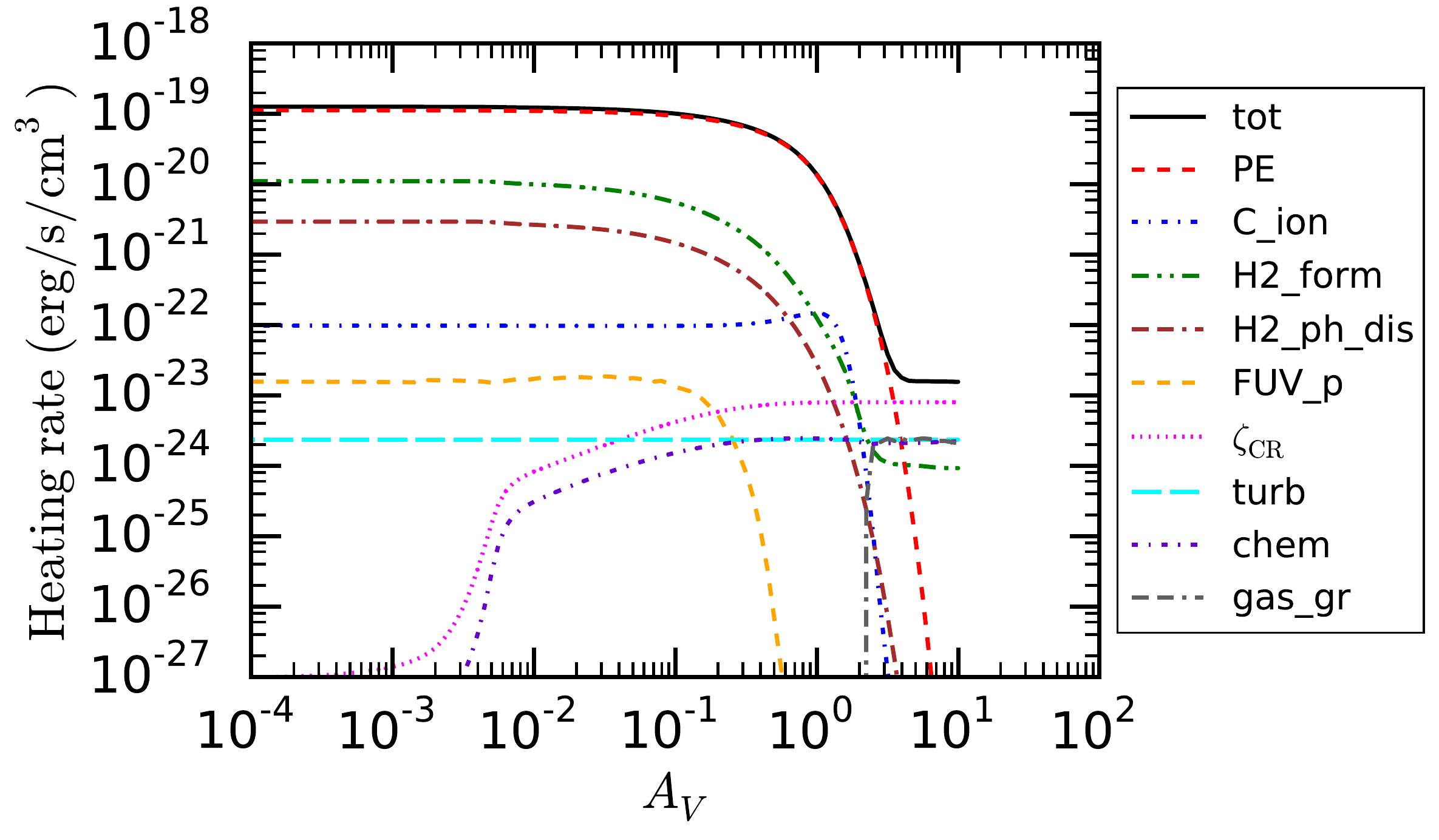}    
    \includegraphics[width=0.48\columnwidth]{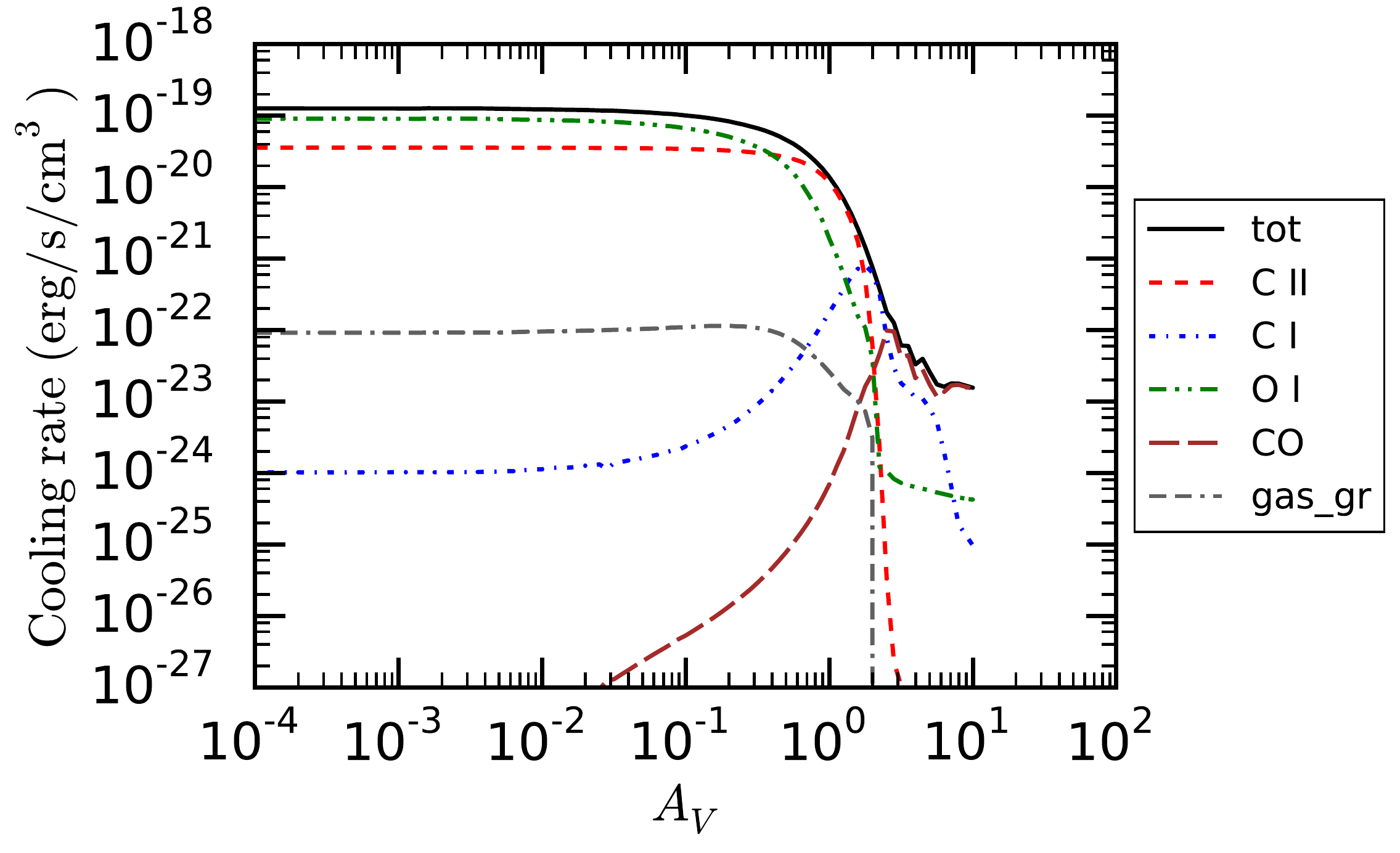}    
    \vspace{-0.2cm}
    \caption{A summary of the key heating and cooling mechanisms in a medium irradiated by a $300\,G_0$ FUV radiation field. PAH driven photoelectric heating dominates until high $A_V$, where cosmic rays take over. The key cooling mechanism in the wind is the escape of line photons. From \protect\cite{Facchini16}.}
    \label{fig:PdrHeatingCooling}
\end{figure}

We focus primarily on the main heating and cooling contributions. These are summarised as a function of extinction for an external FUV field of 300\,G$_0$ in Figure \ref{fig:PdrHeatingCooling}, which is taken from \cite{Facchini16}. Note that, as we will discuss below, the exact form of these plots depends on the FUV field strength and the assumed composition, e.g. the metallicity, dust grain properties and polycyclic aromatic hydrocarbon (PAH) abundance. 

\begin{figure}
    \centering
    \vspace{-0.5cm}
    \includegraphics[width=0.6\columnwidth]{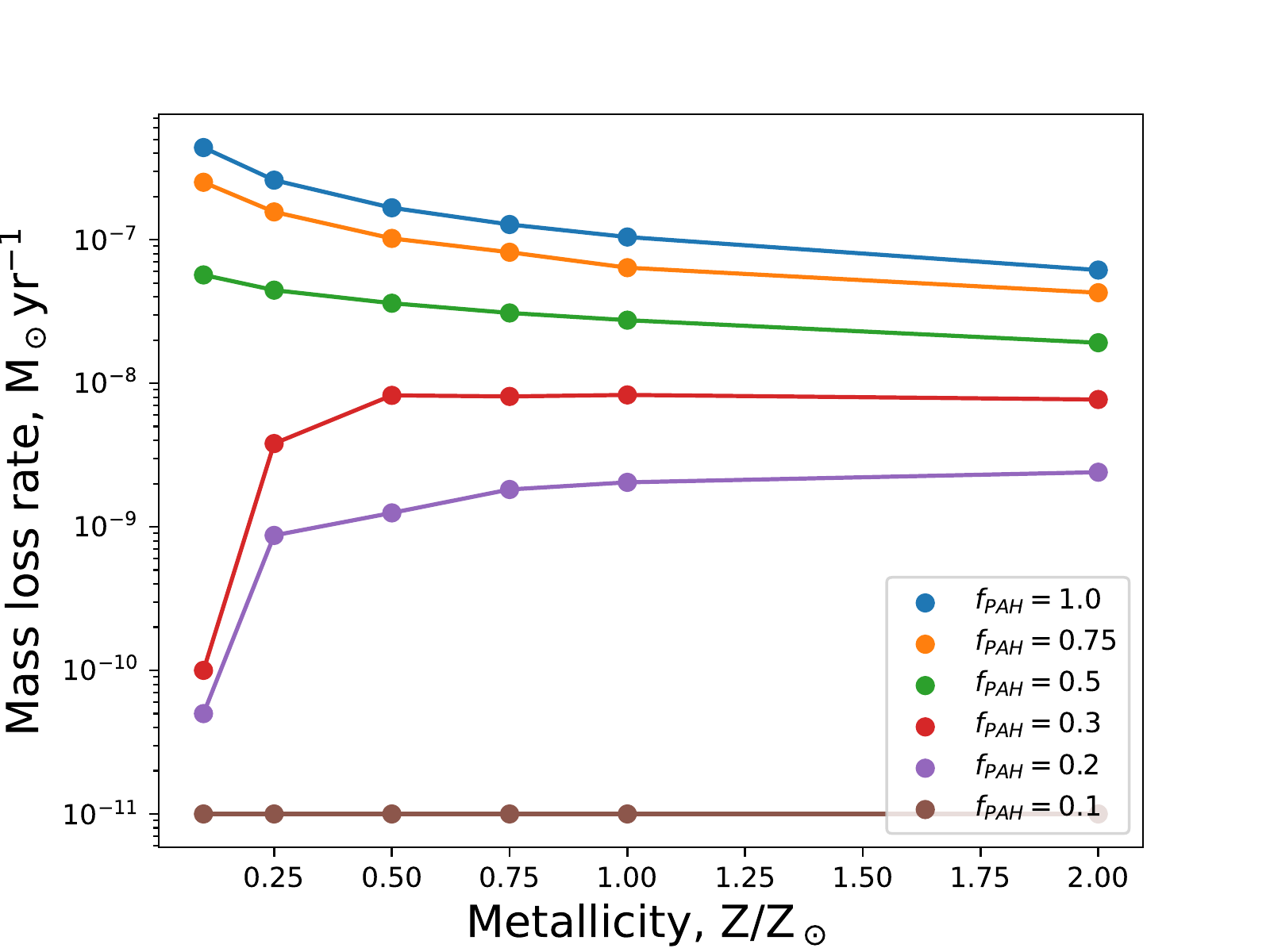}
    \caption{External photoevaporative mass loss rate as a function of metallicity ($Z/Z_\odot$) for a 100\,au disc around a 1\,M$_\odot$ star irradiated by a $1000 \, G_0$ radiation field. Each coloured line represents a different value of the base PAH-to-dust mass ratio scaling $f_\mathrm{PAH}$. These are extensions of the FRIED PDR-dynamical models of \protect\cite{Haworth18b}. When the overall metallicity is scaled, there are changes to both the heating and cooling contributions that broadly cancel out. Conversely, varying the PAH-to-dust ratio (which is very uncertain) can lead to large changes in the mass loss rate.  Note that these calculations have a floor value of $10^{-11}$\,M$_\odot$\,yr$^{-1}$.  }
    \label{fig:Metallicity_PAH}
\end{figure}

The heating mechanism that is anticipated to be most important for external photoevaporation is photoelectric heating (see the left hand panel of Figure \ref{fig:PdrHeatingCooling}) that occurs when PAHs lose electrons following photon absorption, increasing the gas kinetic energy \citep{2008ARA&A..46..289T}. The impact this can have on the mass loss rate is illustrated in Figure \ref{fig:Metallicity_PAH}, which shows the results of numerical models of an externally photoevaporating 100\,au disc around a 1\,M$_\odot$ star in a $1000 \, G_0$ FUV environment as a function of metallicity.  Each coloured set of points connected by a line represents a different PAH-to-dust ratio. Reducing the PAH-to-dust ratio has a much larger impact on the mass loss rate than changing the overall metallicity. These models are previously unpublished extensions of the 1D PDR-dynamical calculations of \cite{Haworth18b}, which are discussed further in \ref{sec:RHDmodels}. When the metallicity is reduced the PAH abundance and heating is also lowered, but so is the line cooling. Changes in metallicity therefore only lead to relatively small changes to the mass loss rate as the heating and cooling changes compensate. Conversely, changing only the PAH-to-dust ratio can lead to dramatic changes in the mass loss rate. 

A key issue for the study of external photoevaporation is that the PAH abundance in the outer parts of discs and in winds is very poorly constrained.  For the proplyd  HST 10 in the ONC, \cite{2013ApJ...765L..38V} inferred a PAH abundance relative to gas around a factor 50 lower than the ISM and a factor 90 lower than NGC 7023 \citep{2012PNAS..109..401B}. Note that the models in Figure \ref{fig:Metallicity_PAH} use a dust-to-gas mass ratio of $3\times10^{-4}$ so an $f_PAH$ of 0.1 in Figure \ref{fig:Metallicity_PAH}  corresponds to a PAH-to-dust ratio of $1/330$. PAH detections around T Tauri stars are generally relatively rare \citep{2006A&A...459..545G, 2007A&A...476..279G}, which leads us to expect that the PAH abundance is depleted in discs irrespective of external photoevaporation. This lower PAH abudance  would mean less heating due to external photoevaporation, resulting in lower external photoevaporative mass loss rates.  Conversely, \cite{2021A&A...653A..21L} demonstrated that PAH \textit{emission} from the inner disc could be suppressed when PAHs aggregate into clumps, which also crucially would not suppress the heating contribution from PAHs (Lange, priv. comm.). However, it is unclear if that same model for PAH clustering applies at larger radii in the disc, let alone in the wind itself and so this is to be addressed in future work (Lange, priv. comm.). 

Given its potential role as the dominant heating mechanism, determining the PAH abundance in the outer regions of discs is vital for understanding the magnitude of mass loss rates and so should be considered a top priority in the study of external photoevaporation. The \textit{James Webb Space Telescope} (JWST) should be able to constrain abundances by searching for features such as the 15-20\,$\mu$m emission lines \citep[e.g.][]{Boulanger98, 1999ESASP.427..579T, 2000A&A...354L..17M,  2010A&A...511A..32B, 2015PASP..127..584R,  2022A&A...661A..80J}.  \cite{2022arXiv220208252E} also demonstrated with synthetic observations that the upcoming \textit{Twinkle} \citep{2019ExA....47...29E} and \textit{Ariel} \citep{2018ExA....46..135T, 2021arXiv210404824T} missions should be able to detect the PAH 3.3\,$\mu$m feature towards discs, at least out to 140\,pc. Even if detections with \textit{Twinkle}/\textit{Ariel} would not succeed in high UV environments because of the larger distances to those targets, constaining PAH abundances of the outer disc regions in lower UV nearby regions would also provide valuable constraints. These will be an important step for calibrating models, refining the mass loss rate estimates and hence our understanding of external photoevaporation.

Often the dominant cooling in PDRs is the escape of line photons from species such as CO, C, O and C$^+$. Evaluating this is the most challenging component of PDR calculations, since to estimate the degree of line cooling, line radiative transfer in principle needs to sample all directions ($4\pi$ steradians) from every single point in the calculation. For this reason, most PDR studies to date (even without dynamics) have been 1D, where it is assumed that exciting UV radiation and cooling radiation can only propagate along a single path, with all other trajectories infinitely optically thick \citep[e.g.][]{1999ApJ...527..795K, 2006ApJS..164..506L, 2006MNRAS.371.1865B,  2007A&A...467..187R}. Most dynamical models of external photoevaporation with detailed PDR microphysics have also therefore been 1D. For example \cite{Adams04} used the \cite{1999ApJ...527..795K} 1D PDR code to pre-tabulate PDR temperatures as inputs for 1D semi-analytic dynamical models (we will discuss these in more detail in section \ref{sec:semianalytic}). Note that 2D models of other features of discs have circumvented this issue by assuming a dominant cooling direction, for example vertically through the disc \citep[e.g.][]{2016A&A...586A.103W}, or radially in the case of internal photoevaporation calculations \citep{2017ApJ...847...11W}. This approach is not applicable in multidimensional simulations of an externally driven wind, where there is no  obvious or universally applicable dominant cooling direction. 

The \textsc{3d-pdr} code developed by \cite{2012MNRAS.427.2100B} and based on the \textsc{ucl-pdr} code \citep{2006MNRAS.371.1865B}  was the first code (and to our knowledge remains the only code) able to treat PDR models in 3D. It utilises a \textsc{healpix} scheme \citep{2005ApJ...622..759G} to estimate the line cooling in 3D without assuming preferred escape directions. \textsc{healpix} breaks the sky into samples of equal solid angle at various levels of refinement. For applications to external photoevaporation, \textsc{3d-pdr} was coupled with the Monte Carlo radiative transfer and hydrodynamics code \textsc{torus} \citep{2019A&C....27...63H} in the \textsc{torus-3dpdr} code \citep{2015MNRAS.454.2828B}, making 2D and 3D calculations possible in principle, which we will discuss more in section \ref{sec:RHDmodels}. However, the computational expense of doing 3D ray tracing from every cell in a simulation iteratively with a hydrodynamics calculation is prohibitely expensive. Finding ways to emulate the correct temperature without solving the full PDR chemistry may offer a way to alleviate this problem \citep[e.g.][]{2021A&A...653A..76H}.

\subsection{1D Semi-analytic models of the external photoevaporative wind flow structure}
\label{sec:semianalytic}
In Section \ref{sec:microphysics} we discussed the importance of PDR microphysics for determining the temperature structure and hence the flow structure of externally irradiated discs. We also noted that PDR calculations are computationally expensive and are usually limited to 1D geometries. Until recently, calculations of the mass loss rate that utilise full PDR physics have also been confined to 1D and solved semi-analytically. Here we briefly review those approaches.

\begin{figure}
    \centering
    \vspace{-0.1cm}    
    \includegraphics[width=0.5\columnwidth]{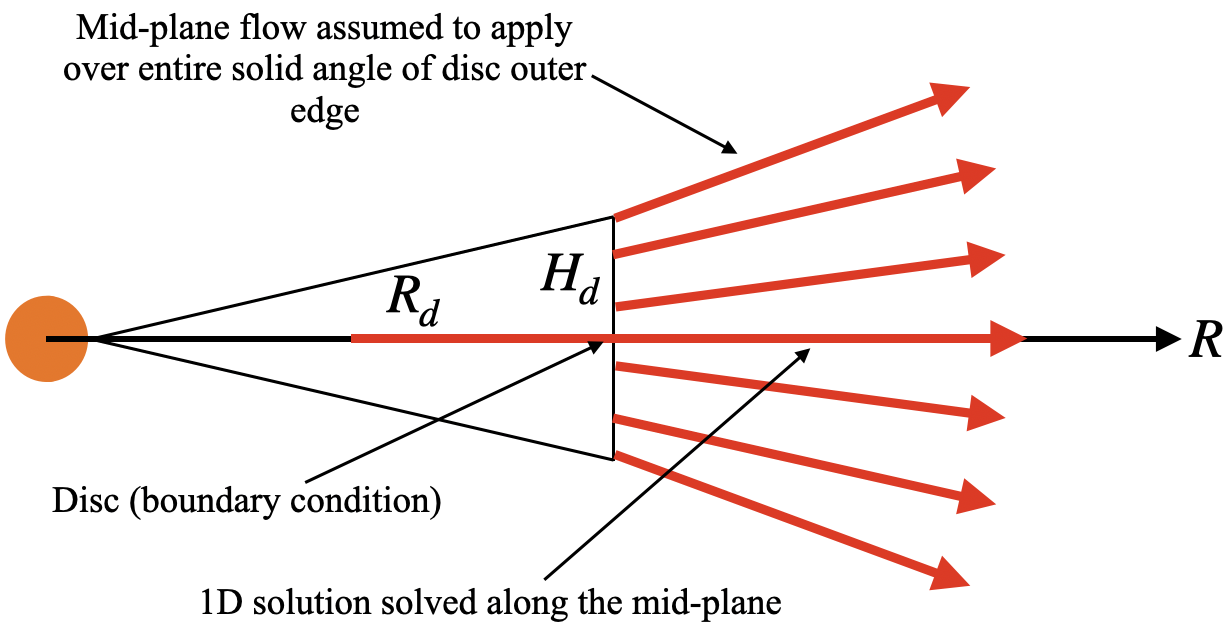}
    \vspace{-0.1cm}
    \caption{A schematic of the 1D semi-analytic model structure and how it is used to estimate total mass loss rate estimates. The flow solution is solved along the disc mid-plane with appropriate boundary conditions, e.g. at the disc outer edge and at some critical point in the wind. This mid-plane flow is then assumed to apply over the entire solid angle subtended by the disc outer edge. }
    \label{fig:1Dgeometry}
\end{figure}

First we describe the 1D approach to models of external photoevaporation and the justification for such a geometry. 1D models essentially follow the structure radially outwards along the disc mid-plane into the wind, as illustrated in Figure \ref{fig:1Dgeometry}. The grid is spherical, buts the assumptionis that the flow only applies over the solid angle subtended by the disc outer edge at $R_{\rm{d}}$. That fraction of $4\pi$ steradians is
\begin{equation}
    \mathcal{F} = \frac{H_{\rm{d}}}{\sqrt{H_{\rm{d}}^2+R_{\rm{d}}^2}}
\end{equation}
for a disc scale height $H_\mathrm{d}$ (again see Figure \ref{fig:1Dgeometry}). The mass loss rate at a point in the flow at $R$ with velocity $\dot{R}$ and density $\rho$ is then
\begin{equation}
    \dot{M} = 4\pi R^2\mathcal{F} \rho \dot{R}.  
\end{equation}

The 1D geometry is justified based on the expectation that it is there that material is predominantly lost from the disc outer edge. That expectation arises because:
\begin{enumerate}
    \item Material towards the disc outer edge is the least gravitationally bound. 
    \item The vertical scale height is much smaller than the radial, which results in a higher density at the radial sonic point than the vertical one \citep{Adams04}.  
\end{enumerate}
This is demonstrated analytically in the case of compact discs in the appendix of \cite{Adams04}, who show that the ratio of mass lost from the disc surface to disc edge is
\begin{equation}
    \frac{\dot{M}_\mathrm{surface}}{\dot{M}_\mathrm{edge}} \approx \left(\frac{R_{\rm{d}}}{R_{\rm{g}}}\right)^{1/2}. 
\end{equation}
where $\dot{M}_\mathrm{surface}$ and $\dot{M}_\mathrm{edge}$ are the mass loss rates from the disc upper layers and mass loss rates from the outer edge respectively. As before, $R_{\rm{d}}$ and $R_{\rm{g}}$ are the disc outer radius and gravitational radius respectively. That is for larger, more strongly heated discs there is a more significant contribution from the disc surface. This has also been tested and validated in 2D radiation hydrodynamic simulations by \cite{2019MNRAS.485.3895H} (that we will discuss fully in Section~\ref{sec:RHDmodels}) who showed that, at least in a $10^{3}\, G_0$ environment, the majority of the mass loss comes from the disc outer edge and the rest from the outer 20\,percent of the disc surface. Mass loss rates in 2D and analogous 1D models were also determined to be similar to within a factor two, with the 1D mass loss rates being the lower values. Mass loss rates computed in one dimension are therefore expected to be somewhat conservative but reasonable approximations.

\cite{Adams04} took a semi-analytic approach to solving for the flow structure by using pre-tabulated PDR temperatures from the code of \cite{1999ApJ...527..795K} and using those in the flow equations. They found that the flow structure is analogous to a \cite{1965SSRv....4..666P} wind, but non-isothermal and with centrifugal effects. At each point in the flow, the pre-tabulated PDR temperatures are interpolated as a function of local density incident FUV and extinction. The boundary conditions used were the conditions at the disc outer edge and the sonic point in the flow. They demonstrated both that FUV driven winds are dominant for setting the mass loss rate and that winds could be driven interior to the gravitational radius, down to  $\sim 0.1-0.2\,R_{\mathrm{g}}$ \citep[see also e.g.][]{Woods96, Liffman03}.

\cite{Facchini16} took a similar approach to 1D semi-analytic models with pre-tabulated PDR temperatures from \cite{2012MNRAS.427.2100B}. As already discussed above, their main focus was on dust entrainment and the impact of grain growth in the disc on the dust properties in the wind. They found that the entrainment of only small grains, coupled with grain growth in the disc, reduces the extinction in the wind and can enhance the mass loss rate. In addition, they used a different approach to the outer boundary condition, finding a critical point in the modified Parker wind solution and taking into account deviations from isothermality at that point. They then integrated from that critical point, inwards to the disk. Thanks to this different approach \cite{Facchini16} were able to compute solutions over a wider parameter space than before, particularly down to low FUV field strengths. 

Semi-analytic models have offered a powerful and efficent tool for estimating mass loss rates in different regimes. However, there are still regions of parameter space where solutions are not possible and semi-analytic models are still only limited to 1D. To alleviate those issues we require radiation hydrodynamic simulations.

\subsection{Radiation hydrodynamic models of external photoevaporation}
\label{sec:RHDmodels}

Above we discussed semi-analytic calculations of the external wind flow structure. Those have the advantage that they are quick to calculate. However they are limited by being restricted to 1D solutions and by solutions not always being calculable. This leaves a demand for radiation hydrodynamic calculations capable of solving the necessary radiative transfer and PDR temperature structure in conjunction with hydrodynamics. Such calculations can solve for the flow structure in 2D/3D and in any scenario. 

The radiation hydrodynamics of external photoevaporation is one of the more challenging problems in numerical astrophysics because the key heating responsible for launching the wind is described by PDR physics. That is, we are required to iteratively solve a chemical network that is sensitive to the temperature with a thermal balance that is sensitive to the chemistry. To make matters worse, the cooling in PDRs is non-local, being dependent on the escape probability of line photons into $4\,\pi$ steradians from any point in the simulation (see Section \ref{sec:PDRmicrophysics}). In other scenarios this does not cause significant issues if there is a clear dominant escape direction. For example, within a protoplanetary disc the main cooling can quite reasonably be assumed to occur vertically through the disc, since other trajectories have a longer path length (and column) through the disc \citep[``1+1D'' models, e.g.][]{2008ApJ...683..287G, 2016A&A...586A.103W}. Similarly, for internal winds there are models with radiation hydrodynamics and PDR chemistry, where the line cooling is evaluated along single paths radially on a spherical grid \citep[][]{2017ApJ...847...11W, 2018ApJ...857...57N,2018ApJ...865...75N, 2019ApJ...874...90W}.  In the complicated structure of an external photoevaporative wind, however, this sort of geometric argument cannot be applied and so multiple samplings of the sky ($4\pi$ steradians) are required from every point in a simulation. 

\begin{figure}
    \centering
    \includegraphics[width=\columnwidth]{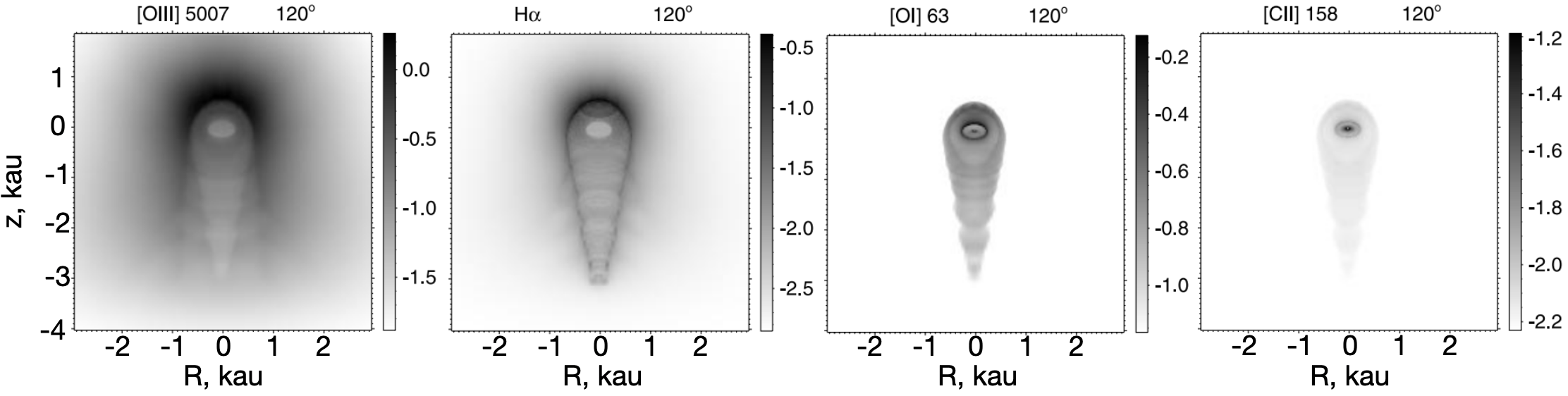}
    \caption{A gallery of intensity maps ($\log_{10}$erg\,s$^{-1}$\,cm$^{-2}$\,strad$^{-1}$) resulting from the 2D-axysmmetric radiation hydrodynamic simulations of proplyds by \protect\cite{2000ApJ...539..258R}. The simulations utilised a simplified microphysics approach which enabled them to model proplyds in 2D-axisymmetry, with UV radiation incident from the top. These multi-wavelength synthetic line emission maps were compared and found to be consistent with the properties of ONC proplyds.  }
    \label{fig:RichlingYorkeIntensities}
\end{figure}

Although 3D cooling is ideally required, simulations have been performed using approximations to the cooling in high UV radiation fields in particular, where the PDR is small. For example \cite{2000ApJ...539..258R} ran 2D axisymmetric simulations of discs irradiated face-on. In their calculations the optical depth and cooling is estimated using a single ray from the cell to the irradiating UV source (this same path is used for calculating the exciting UV and cooling radiation). They also employed a more simple PDR microphysics model compared to work at the time such as \cite{Johnstone98}, which enabled the move from 1D to 2D-axisymmetry. \cite{2000ApJ...539..258R} studied the mass loss of proplyds as well as the observational characteristics using intensity maps derived from their dynamical models, some of which are illustrated in Figure \ref{fig:RichlingYorkeIntensities}. They found, in the first geometrically realistic EUV+FUV irradiation models, that rapid disc dispersal gives morphologies in various lines similar to those observed in the ONC.  

The \textsc{torus-3dpdr} code \citep{2015MNRAS.454.2828B} is a key recent development in the direct radiation hydrodynamic modelling of external photoevaporation. It was constructed by merging components of the first fully 3D photodissociation region code \textsc{3d-pdr} \citep{2012MNRAS.427.2100B} with the \textsc{torus} Monte Carlo radiative transfer and hydrodynamics code \citep{2019A&C....27...63H}. \textsc{3d-pdr} (and hence \textsc{torus-3dpdr}) address the 3D line cooling issue using a \textsc{healpix} scheme, which breaks the sky into regions of equal solid angle. \textsc{torus-3dpdr} has been used to run a range of 1D studies of external photoevaporation. It has been shown to be consistent with semi-analytic calculations \citep{2016MNRAS.463.3616H}. It was used to study external photoevaporation in the case of very low mass stars, with a focus on Trappist-1 \citep{2018MNRAS.475.5460H}. The approach was to run a grid of models to provide the mass loss rate as a function of the UV field strength, disc mass/radius for a 0.08\,M$_\odot$ star and interpolate over that grid in conjunction with a disc viscous evolutionary code based on that of \citet{Clarke07} to evolve the disc. The usefulness of such a grid led to the FRIED (\textbf{F}UV \textbf{R}adiation \textbf{I}nduced \textbf{E}vaporation of \textbf{D}iscs) grid of mass loss rates that has since been employed in a wide range of disc evolutionary calculations by various groups \citep[e.g.][others]{Winter19a, Concha-Ramirez19,Sellek20}. Note, given the discussion on the importance of PAHs above that the FRIED models use a dust-to-gas ratio a factor 33 lower than the canonical $10^{-2}$ for the ISM and a PAH-to-dust mass ratio  a further factor of 10 lower than ISM-like. This is conservative (i.e. a PAH-to-gas ratio of 1/330 that in the ISM) compared to the factor 50 or so PAH depletion measured by \cite{2013ApJ...765L..38V}. So models predict that the PDR heating is still capable of driving significant mass loss, even when the PAH abundance is heavily depleted compared to the ISM.

These applications were all still 1D though and so there is a growing theoretical framework that is based on that geometric simplification. \cite{2019MNRAS.485.3895H} ran 2D-axisymmetric external photoevaporation calculations (with 3D line cooling, utilising the symmetry of the problem) and found that 1D calculations are, if anything, conservative since the 2D mass loss rates were slightly higher. True 3D calculations with 3D line cooling and the disc not irradiated isotropically or face-on, are yet to be performed. Though in principle \textsc{torus-3dpdr} is capable of this, in practice the 3D ray tracing of the \textsc{healpix} scheme makes such calculations computationally expensive. 

\subsubsection{The disc-wind transition}
\label{sec:boundaryconditions}
Most of the models discussed so far impose a disc as a boundary condition from which a wind solution is derived. In reality the wind is not launched from some arbitrarily imposed point in an irradiated disc. \cite{2021MNRAS.508.2493O} recently implemented a model without an imposed inner boundary and a smooth transition from disc to wind using a slim disc approach \citep{1988ApJ...332..646A}. In this first approach they assume an isothermal disc, but demonstrate that while the transition from disc to wind is narrow, it is not negligibly thin. They also introduced dust of different sizes into their model, which predicts a radial gradient in grain size in the outer disc/inner wind. Although the fixed inner boundary models are valid for computing steady state mass loss rates for disc evolutionary models, a worthwhile future development will be to include detailed microphysics in a slim-disc approach like that of \cite{2021MNRAS.508.2493O}.

\begin{mdframed}
\vspace{-0.8cm}
\subsection{Summary and open questions for the theory of externally driven disc winds}
\begin{enumerate}
    \item Numerical models of external photoevaporation require some of the most challenging radiation hydrodynamics models in astrophysics. This is primarily because it is necessary to include 3D PDR microphysics, including line cooling with no obvious dominant cooling direction. 
    \item Limited by the above, 1D models of external photoevaporation are now well established and are used to estimate disc mass loss rates. But 2D and 3D simulations are still limited.\\
\end{enumerate}    
    Some of the many open questions and necessary improvements to current models are
\begin{enumerate}
    \item \textit{What is the PAH abundance in external photoevaporative winds and the outer regions of discs? This is key to setting the wind temperature and mass loss rate.} 
    \item \textit{Including mass loss rate dependent dust-to-gas mass ratios and maximum grain sizes (and hence extinction) in numerical models of external photoevaporation. At present a single representative cross section is assumed, irrespective of the mass loss rate.} 
    \item \textit{Can accurate temperatures from PDR microphysics be computed at vastly reduced computational expense (e.g. via emulators)? }
    \item \textit{What is the interplay between internal and external winds?}
    \item \textit{3D simulations of external photoevaporation with full PDR-dynamics}
    \item \textit{Non-isothermal slim-disc models of externally photoevaporating discs. }
\end{enumerate}
\end{mdframed}

\section{Observational properties of externally photoevaporating discs}
\label{sec:obs}

Here we discuss observations to date of individual externally photoevaporating discs. We discuss the diversity in their properties such as UV environment and age, and summarise key diagnostics. 

\subsection{Defining the term proplyd}
\begin{figure}
    \centering
    \includegraphics[width=0.46\columnwidth]{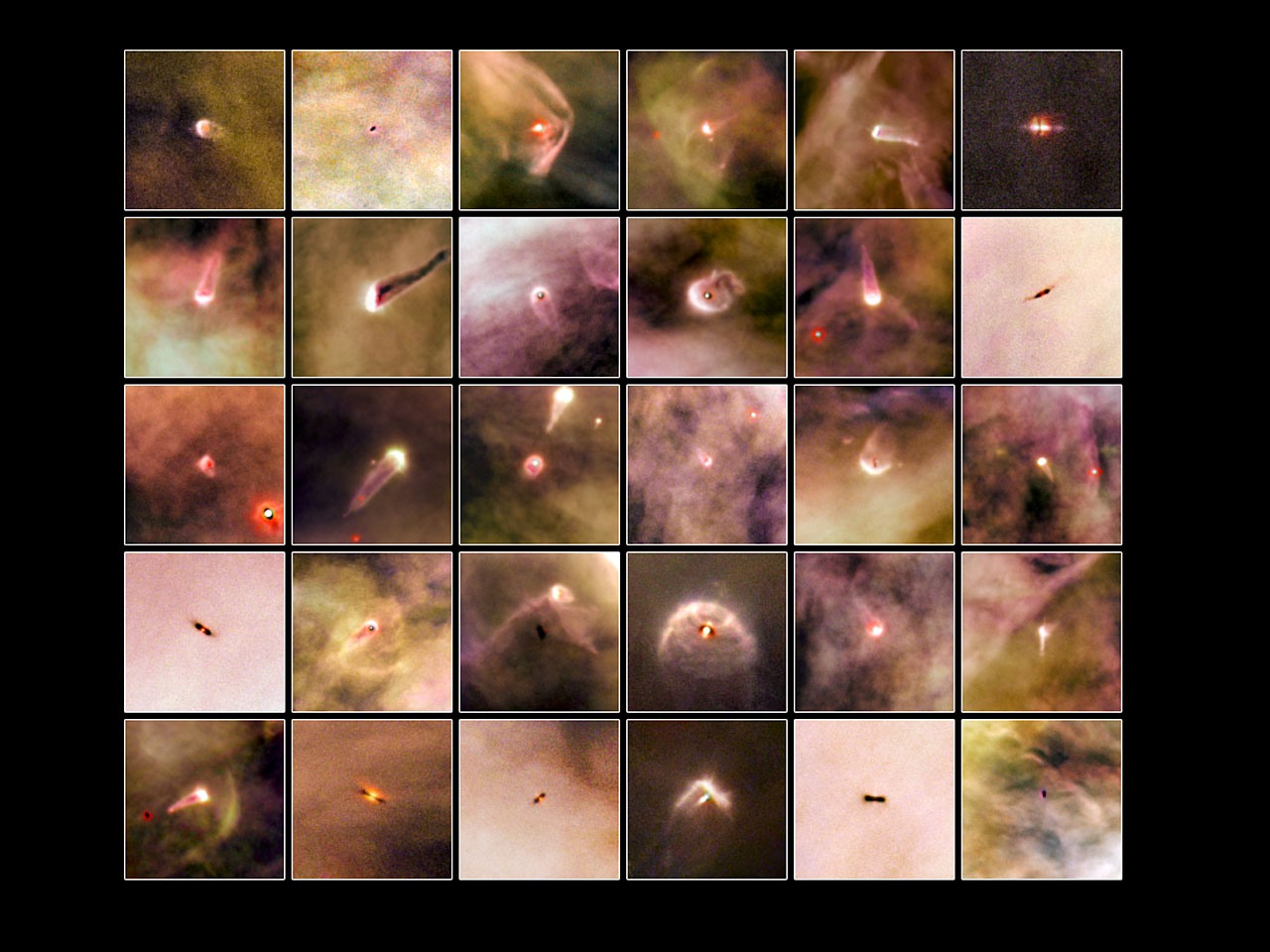}    
    \includegraphics[width=0.4\columnwidth]{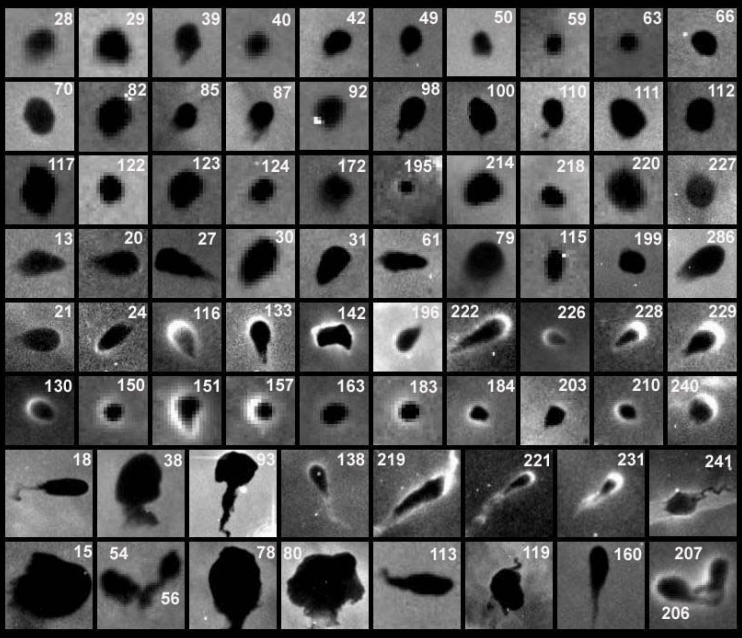}
    \caption{The left hand panel is a gallery of proplyds in the ONC -- evaporating discs around YSO's (Credit: NASA/ESA and L. Ricci). The right hand panel is a gallery of globulettes in Carina from \protect\citep{2014A&A...565A.107G}. Globulettes have radii from hundreds to thousands of au and can  also take on a cometary morphology when externally irradiated, though in many cases do not contain any YSOs, which by our definition would mean that they are not proplyds (note that \protect\cite{2014A&A...565A.107G} never referred to them as such, rather using the term globulette).  }
    \label{fig:globulettes}
\end{figure}

The term ``proplyd'' was originally used to describe any disc directly imaged in Orion with HST in the mid-90s \citep[e.g.][]{O'dell93,  Johnstone98} as a portmanteau of ``protoplanetary disc''. Since then, use of the term has adapted to only refer to cometary objects resulting from the external photoevaporation of compact systems. However this use of the term is ambiguous, since a cometary morphology can result from both externally irradiated protoplanetary discs \textit{and} externally irradiated globules which may be host no embedded star/disc, such as many of the compact globulettes as illustrated in Figure \ref{fig:globulettes} \citep{2007AJ....133.1795G, 2014A&A...565A.107G}. We therefore propose to define a proplyd as follows
\begin{mdframed}
\textbf{Proplyd:} \\ \textit{A circumstellar disc with an externally driven photoevaporative wind composed of a photodissociation region and an exterior ionisation front with a cometary morphology. } \\
\end{mdframed}
We have chosen this definition such that it makes no distinction as to whether EUV or FUV photons drive the wind, but specifically define that for an object to be a proplyd then the wind must be launched from a circumstellar disc. In the absence of a disc it is a globule or globulette \citep[also sometimes referred to as an evaporating gas globules, or EGG, e.g.][]{MesaDelgado16}. To further clarify, a globule or globulette with an embedded YSO (identified through a jet for example) with cometary morphology would also not be defined as a proplyd since it is the ambient material being stripped rather than the disc.

\subsection{Where and what kind of proplyds have been detected?}

Proplyds were originally discovered in the ONC and there are now over 150 known in the region
\citep[e.g.][]{O'dell93, Johnstone98, 2005AJ....129..382S, 2008AJ....136.2136R}. The ONC is around {1{-}3}\,Myr old with the primary UV source the O6V star $\theta^1$C. In general, the surface brightness of proplyds if the local EUV exposure is lower then proplyds can be harder to detect. For example, the surface brightness of $\mathrm{H}\alpha$ is \citep{O'dell98}:
\begin{equation}
    \label{eq:SHalpha}
    \langle S(\mathrm{H}\alpha) \rangle  \approx 7 \times 10^{11} \frac{\alpha_{\mathrm{H}\alpha}^{\mathrm{eff}}}{\alpha_B}\frac{\Phi}{10^{49} \, \mathrm{s}^{-1}} \left( \frac{d}{0.1 \, \mathrm{pc}} \right)^{-2} \, \mathrm{s}^{-1} \, \mathrm{cm}^{-2} \, \mathrm{s.r}^{-1}
\end{equation} where $\alpha_{\mathrm{H}\alpha}^\mathrm{eff} = 1.2\times 10^{-13}$~cm$^{-3}$~s$^{-1}$ and $\alpha_B = 2.6\times 10^{-13}$~cm$^{-3}$~s$^{-1}$ at temperature $T=10^4$~K are hydrogen recombination coefficients \citep[e.g.][]{Osterbrock89}. Nonetheless, in recent years there have been important detections of proplyds in lower EUV environments. {Proplyds can be found in the ONC at separations of up to $\sim 1$~pc from $\theta^1$C \citep{Vicente05,Forbrich21,Vargas-Gonzalez21}.} Meanwhile, \cite{Kim16} found proplyds in the vicinity of the B1V star 42 Ori in NGC 1977, demonstrating that B stars can drive external photoevaporative winds. \cite{Haworth21} also presented the discovery of proplyds in NGC 2024. In that region it appears that both an O8V and B star are driving external photoevaporation. The main significance of proplyds there is the $\sim0.2-0.5$\,Myr age of a subset of the region where proplyds have been discovered. This is important since it implies that external photoevaporation can even be in competition with our earliest stage evidence for planet formation \citep{Sheehan18, Segura-Cox20}. 

In these regions proplyds have been detected with host star masses from around solar down to almost planetary masses \citep[$<15\,M_{\textrm{jup}}$][]{2010AJ....139..950R, Kim16, 2016ApJ...833L..16F,  2022MNRAS.512.2594H}. The UV fields incident upon these proplyds ranges from $>10^5 \, G_0$ down to possibly around $100 \, G_0$ \citep{Miotello12}. Mass loss rates are estimated to regularly be greater than $10^{-7}$\,M$_\odot$\,yr$^{-1}$ and sometimes greater than than $10^{-6}$\,M$_\odot$\,yr$^{-1}$ \citep[e.g.][]{Henney98, Henney99,2002ApJ...566..315H, Haworth21}. Examples of binary proplyds have also been discovered \citep{2002ApJ...570..222G}. For sufficiently close separation binaries the disc winds merge to form a so-called interproplyd shell, which was studied by \cite{2002RMxAA..38...71H}. These nearby regions, the ONC, NGC 1977 and NGC 2024 show the clearest evidence for external photoevaporation.

Due to resolution and sensitivity issues, unambiguous evidence for external photoevaporation is more difficult to obtain in more distant star forming regions than the $D \sim400$\,pc of Orion.  \cite{Smith10} identified candidate proplyds in Trumpler 14, at a distance of $D\sim2.8\,$kpc, and  \cite{MesaDelgado16} subsequently detected discs towards those candidates with ALMA. Although many of those candidates are large evaporating globules (in some cases, with embedded discs detected), some are much smaller and so could be bona fide proplyds. 

There are other regions where ``proplyd-like'' objects have been discovered, including Cygnus OB2 \citep{Wright12, 2014ApJ...793...56G},  W5 \citep{2008ApJ...687L..37K}, NGC 3603 \citep{Brandner00}, NGC 2244, IC 1396 and NGC 2264  \citep{2006ApJ...650L..83B}. However our evaluation of those systems so far is that they are all much larger than ONC propyds and likely evaporating globules. Given the high UV environments of those regions and the identification of evaporating globules we \textit{do} expect external photoevaporation to be significant in the region. However, it remains unclear for many of these objects whether the winds are launched from an (embedded) star-disc system. Future higher resolution observations (e.g. extremely large class telescopes should resolve ONC-like proplyds out to Carina) and/or new diagnostics of external photoevaporation that do not require spatially resolving the proplyd (for example line ratios, or tracers that show a wind definitely emanates from a disc) are required in these regions. 

We provide a further discussion of proplyd demographics and particularly the demographics of discs in irradiated star forming regions in Section \ref{sec:surveys}.

\subsection{Estimating the mass loss rates from proplyds}
\label{sec:MdotEstimate}

As discussed above, proplyds have a cometary morphology with a ``cusp'' pointing towards the UV source responsible for driving the wind and an elongated tail on the far side, pointing away from the UV source. The leading hemisphere that is directly irradiated by the UV source is referred to as the cometary cusp. On the far side of the proplyd is a trailing cometary tail.

The extent of the cometary cusp is set by the point beyond  which all of the incident ionising flux is required to keep the gas ionised under ionisation equilibrium. A higher mass loss rate and hence denser flow increases the recombination rate in the wind and would move the ionisation front to larger radii. Conversely, increasing the ionising flux will reduce the ionisation front to smaller radii. As a result, the ionisation front radius $R_{\textsc{if}}$ (i.e. the radius of the cometary cusp)  is related to the ionising flux incident upon the proplyd and the mass loss rate $\dot{M}_\mathrm{ext}$ \citep{Johnstone98}. This is independent of the actual wind driving mechanism, being enforced simply by photoionisation equilibrium downstream of the launching region of the flow. This provides a means to estimate the mass loss rate from the disc:
\begin{equation}
   \left(\frac{\dot{M}_\mathrm{ext}}{10^{-8}\,M_\odot\,\rm{yr}^{-1}}\right) =  \left(\frac{1}{1200}\right)^{3/2}\left(\frac{R_{\textsc{if}}}{\textrm{au}}\right)^{3/2}\left(\frac{d}{\textrm{pc}}\right)^{-1}\left(\frac{\Phi}{10^{45}\,\mathrm{s}^{-1}}\right)^{1/2} 
   \label{equn:Mdot}
\end{equation}
where $\Phi$ is the ionising photons per second emitted by the source at distance $d$ responsible for setting the ionisation front. This has been applied to estimating mass loss rates in NGC 2024 \citep{Haworth21} and NGC 1977 \citep{2022MNRAS.512.2594H}.
Note that this is neglects  extinction between the UV source and the proplyd and could underestimate the true separation between the UV source and proplyd. Both of these effects would reduce the true ionising flux incident upon the proplyd and so equation \ref{equn:Mdot} provides an upper limit on the mass-loss rate. 

Generally, the mass loss rate could alternatively be inferred if one knows the density and velocity through a surface in the wind enclosing the disc. However, the ionisation front is very sharp, making it an ideal surface through which to estimate the mass loss rate.

Other more sophisticated model fits to proplyds have been made, such as \cite{2002ApJ...566..315H}, who use photoionisation, hydrodynamics and radiative transfer calculations to model the proplyd LV 2,  but equation \ref{equn:Mdot} provides a quick estimate that gives mass loss rates comparable to those more complex estimates.

\subsection{Multi-wavelength anatomy of proplyds}
\begin{figure}
    \centering
    \includegraphics[width=\columnwidth]{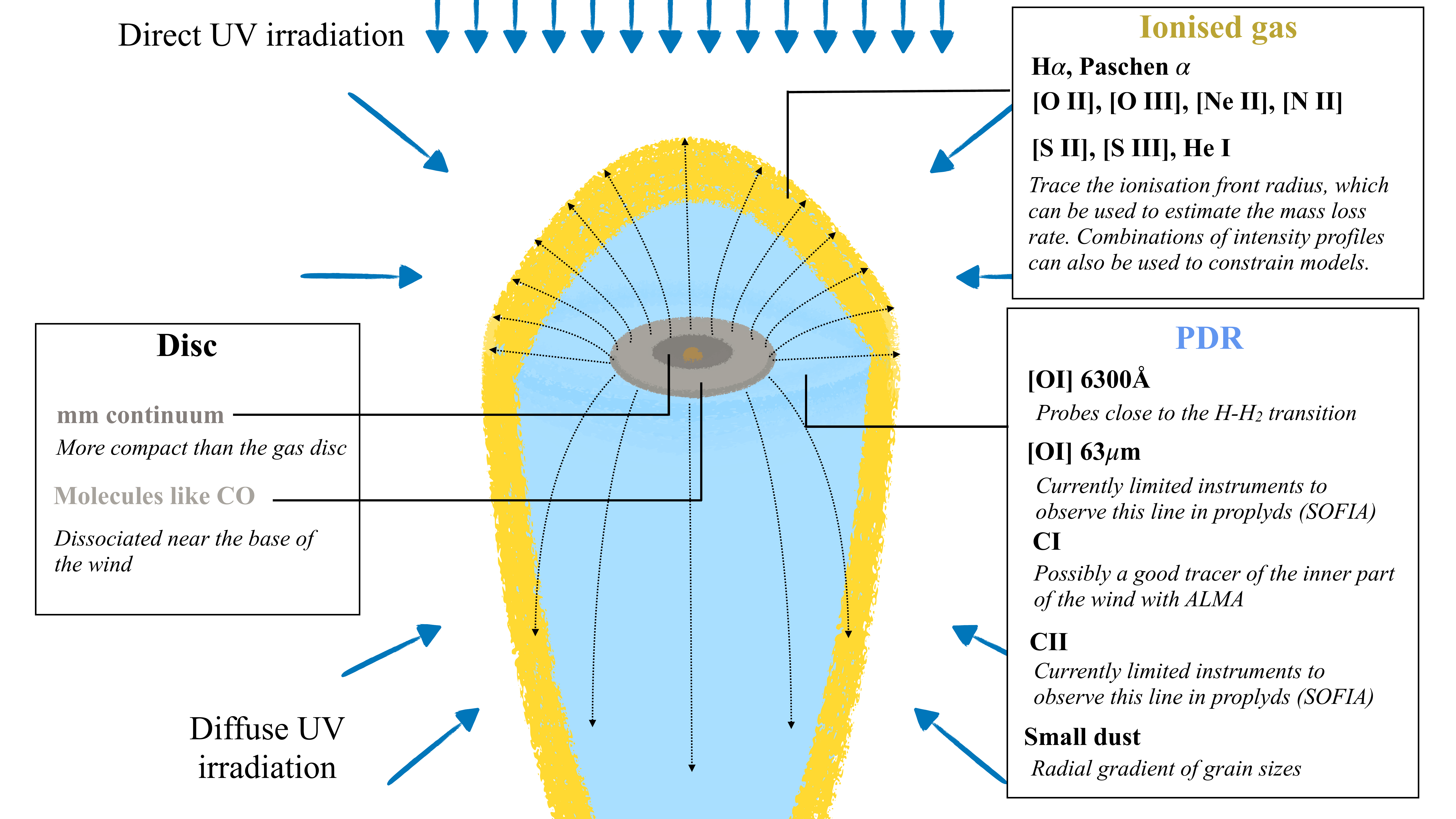}
    \caption{The anatomy of a proplyd, highlighting the observables in different parts of the system. The three main zones are the disc, PDR and ionised gas.  }
    \label{fig:proplydAnatomy}
\end{figure}

Here we provide a brief overview of some key observational tracers of proplyds. We also highlight possible alternative tracers that might prove useful to identify external photoevaporation when proplyds cannot be inferred based on their cometary morphology (i.e. in weaker UV environments and distant massive clusters). A schematic of the overall anatomy and location of various tracers of a proplyd is shown in Figure \ref{fig:proplydAnatomy}. 

\subsubsection{Ionised gas tracers in absorption and emission}
Proplyds are observed in two ways using ionised gas emission lines. Proplyds are typically found in H\,\textsc{ii} regions which are associated with widespread ionised emission lines. Proplyds on the side of the H\,\textsc{ii} region {closest to the observer} can therefore manifest in the absorption of those ionised gas emission lines. In addition, the ionisation front at the rim of the proplyd cusp is an additional source of ionised emission lines that can be directly detected.  Ionised gas tracers in emission probe the region close to or outside of the H\textsc{i}-H\,\textsc{ii} ionisation front \citep[e.g.][]{Henney99}. These ionised gas tracers detected in emission are valuable for estimating mass loss rates using the procedure discussed in \ref{sec:MdotEstimate}.  Prominent ionised gas emission lines include H$\,\alpha$, Paschen$\,\alpha$, [O\,\textsc{iii}], [Ne \textsc{ii}] and [N\,\textsc{ii}]. Combinations of these photoionised gas tracers have also been used to constrain models of proplyds. For example by \cite{2002ApJ...566..315H}, who compared simulated images and observations in H$\,\alpha$, [N\,\textsc{ii}] (6583\AA) and [O\,\textsc{iii}] (5007\AA) to model the proplyd LV 2.

\subsubsection{Disc tracers}

\noindent\textbf{CO} \\
Thanks to its brightness, CO is one of the most common line observations towards protoplanetary discs. \cite{Facchini16} pointed out that because the specific angular momentum in the wind scales as $R^{-2}$ rather than the Keplerian $R^{-3/2}$ that non-Keplerian rotation is a signature of external photoevaporation. Although this deviation grows with distance into the wind, it does not get a chance to do so to detectable levels before CO is dissociated \citep{2016MNRAS.463.3616H, Haworth20}. CO is therefore not expected to kinematically provide a good probe of external winds for proplyds (though it may be useful for more extended discs with slow external winds in low UV environments, as we will discuss in \ref{sec:nonProplydWinds}). However, this does not preclude CO line ratios or intensities showing evidence of external heating in spatially resolved observations.    \\

\noindent\textbf{Dust continuum} \\
External photoevaporation influences the dust by i) entraining small dust grains in the wind \citep{Facchini16} and ii) by heating the dust in the disc. Directly observing evidence for grain entrainment would provide a key test of theoretical models both of external photoevaporation and dust-gas dynamics. In addition to the prediction that small grains are entrained, \citet{2021MNRAS.508.2493O} predict a radial gradient in the grain size distribution.

Evidence for such a radial gradient in grain size was inferred in the ONC 114-426 disc by \cite{Miotello12} \citep[see also][]{2001Sci...292.1686T}. This disc is the largest in the central ONC, on the near side of the H\,\textsc{ii} region. Although the UV field incident upon it is expected to be of order $10^2\, G_0$, it is clearly evaporating, with the wind resulting in  an extended diffuse foot-like structure (though no clear cometary proplyd morphology). They mapped the absorption properties of the transluscent outer parts of the disc, finding evidence for a radially decreasing maximum grain size, which would be consistent with theoretical expectations. Revisiting 114-426 to obtain further constraints on grain entrainment would be valuable, as well as searching for the phenomenon in other systems. JWST will offer the capability to similarly study the dust in the outer parts of discs in silhouette in the ONC, comparing JWST Paschen $\alpha$ absorption with HST H$\alpha$ absorption \citep[as part of PID: GTO 1256][]{2017jwst.prop.1256M}.

The dust in discs is also influenced by the radiative heating from the environment. If a proplyd is sufficiently close to a very luminous external source the grain heating can be comparable to, or in some parts of the disc exceed, the heating from the disc's central star. If this is not accounted for when estimating the dust mass in a proplyd (i.e. if one assumes some constant characteristic temperature, typically $T=20\,$K) the mass ends up increasingly overestimated in closer proximity to the luminous external source \citep{Haworth21b} and so may suppress spatial gradients in disc masses at distances within around 0.1\,pc of an O star like $\theta^1$C \citep[e.g.][]{Eisner18, 2021ApJ...923..221O}. 

\subsubsection{Photodissociation region tracers}
Photodissociation region (PDR) tracers are valuable because they trace the inner regions of the flow. In particular, as discussed in \ref{sec:PDRmicrophysics}, it is the FUV/dissociation region that determines the mass loss rate. PDR tracers in the wind are also valuable because they are what we will rely on to identify externally photoevaporating discs that are non proplyds (discussed further in section \ref{sec:nonProplydWinds}). PDR tracers of external photoevaporation have received a lot of attention in recent years and so are explored somewhat more thoroughly than photoionised gas tracers here. \\

\noindent\textbf{[OI] 6300\AA} \\
\cite{Storzer98} modelled the [OI] 6300\AA\, emission from proplyds motivated by \cite{1998AJ....116..293B} observations of the ONC proplyd 1822-413. They found that in the case of external photoevaporation the line is emitted following the photodissociation of OH, with the resulting excited oxygen having a roughly 50\,per cent chance of de-exciting through the emission of a [OI] 6300\AA\, photon. For the density/temperature believed to be representative in the wind this dissociation is expected to dominate over OH reacting out by some other pathway. This model approximately reproduced the flux of 1822-413. Ballabio et al. (in preparation) generalised this, studying how the [OI] 6300\AA\, line strength varies with UV field strength and star disc parameters. They found that the line is a poor kinematic tracer of external winds because the velocity is too low to distinguish it from [OI] emission from internal winds \citep[though spectro-astrometry, e.g.][and future instrumentation may solve this issue by spatially distinguishing the internal and external winds]{2021ApJ...913...43W}. However, the [OI] luminosity increases significantly with UV field strength. The ratio of [OI] 6300\AA\, luminosity to accretion luminosity is therefore expected to be unusually high in strong UV environments. The ratio usually has a spread of about $\pm1$ orders of magnitude in low UV environments \citep{2018A&A...609A..87N} but the models predict the ratio increases by an additional order of magnitude above the usual upper limit in a $10^5$G$_0$ environment. This could make [OI] a valuable tool for identifying external photoevaporation at large distances where proplyds cannot be spatially resolved. 

There are some observational challenges associated with utlising this diagnostic though. One is that the targeted YSO's emission has to be distinguished from emission from the star forming region. For example [OI] emission from a background PDR. Furthermore, estimating the accretion luminosity of proplyds appears to have only been attempted in a handful of cases, with tracers like H\,$\alpha$ also possibly originating from the proplyd wind. \\

\noindent\textbf{C\,I } \\
As we discussed above, the most commonly used disc gas tracer, CO, is dissociated in the wind. This means that it is ineffective for detecting deviations from Keplerian rotation that are expected in external photoevaporative winds. C\,I primarily resides in a layer outside of the CO abundant zone until it is ionised at larger distances in the wind, so could therefore trace the deviation from Keplerian rotation. \citet{Haworth20} proposed that C\,I offers a means of probing the inner wind kinematically. A key utility of this would be its possible use as an identifier of winds in systems where there is no obvious proplyd morphology. \citet{2022MNRAS.512.2594H} used APEX to try and identify the C\,I 1-0 line in NGC 1977 proplyds, which are known evaporating discs in an intermediate ($F_\mathrm{FUV} \sim 3000 \, G_0$) UV environment. However they obtained no detections, which they explain in terms of those proplyd discs being heavily depleted of mass. However an alternative explanation would be if the disc were just depleted in carbon. Distinguishing these requires independent constraints on the mass in those discs. Overall the utility of C\,I remains to be proven and should be tested on higher mass evaporating discs. Based on the expected flux levels \citep[see also][]{2016A&A...588A.108K} it seems unlikely though that C\,I will be suitable for mass surveys searching for the more subtle externally driven winds when there is no ionisation front. Though it could be used for targeted studies of extended systems that are suspected to be intermediate UV environments.  \\

\noindent\textbf{Far-infrared lines: C\,II 158\,$\mu$m and  [OI] 63$\mu$m} \\
The [C\,II] 158\,$\mu$m  and [O\,\textsc{i}] 63\,$\mu$m lines are both bright tracers of the PDR. They have both been observed with \textit{Herschel} \citep{2010A&A...518L...1P} towards a small number of proplyds by \cite{2017A&A...604A..69C}, who compared the line fluxes with uniform density 1D PDR models with the \textsc{Meudon} code \citep{2006ApJS..164..506L} to constrain parameters such as the mean flow density, which is supported by ALMA observations. They suggested that the proplyd PDR self-regulates to maintain the H-H$_2$ transition close to the disc surface and maintain a flow at $\sim1000$\,K in the supercritical regime ($R_{\mathrm{d}} > R_{\mathrm{g}}$). Their models also pointed towards a number of heating contributions being comparably (or more) important than PAH heating. However those calculations also assumed uniform density and ISM-like dust, whereas we now know it is depleted in the wind. Overall this highlights the need for further detail in PDR-dynamical models. 

Clearly these PDR tracers do have enormous utility for understanding the conditions in the inner part of external photoevaporative winds. The main limitation to using these far-infrared lines now is the lack of facilities to observe them, with \textit{Herschel} out of commission and SOFIA \citep[e.g.][]{2012ApJ...749L..17Y} due to end soon. We are unaware of any short term concepts to alleviate this, but in the longer term there are at least two relevant far-infrared probe class mission concepts being prepared, FIRSST and PRIMA, which would address this shortfall. However, these missions would not launch until the 2040s.

\subsection{External photoevaporation of discs without an observed ionisation front}
\label{sec:nonProplydWinds}
Proplyds are most easily identified because of their cometary morphology. However in weaker UV fields it is still possible to drive a significant wind that is essentially all launched only from close to the disc outer edge (where material is most weakly bound. Recent years have seen the discovery of possible external winds from very extended discs in very weak UV environments, {down to $F_\mathrm{FUV}\lesssim 10\,G_0$.} The first example of this was the extremely large disc IM Lup, which has CO emission out to $\sim1000\,$au. IM Lup had previously been demonstrated to have an unusual break in the surface density profile at around 400\,au in submillimetre array (SMA) observations by \cite{2009A&A...501..269P}. \cite{2016ApJ...832..110C} then observed IM Lup in the continuum and CO isotopologues with ALMA, similarly finding that the CO intensity profile could not be simultaneously reproduced at all radii by sophisticated chemical models, inferring a diffuse outer halo of CO. Using \textit{Hipparcos} to map the 3D locations of the main UV sources within 150\,pc of IM Lup and geometrically diluting their UV with various assumptions on the extinction, \cite{2016ApJ...832..110C} had determined that the UV field incident upon IM Lup is only $F_\mathrm{FUV}\sim 4\,G_0$, so not expected to be sufficient to drive an external wind. \cite{2017MNRAS.468L.108H} demonstrated using 1D radiation hydrodynamic models that the CO surface density profile as a disc/halo could be explained by a slow external photoevaporative wind launched from around 400\,au by an extremely low FUV environment, $F_\mathrm{FUV} \sim4 \, G_0$. This is possible because the disc is so extended that the outer parts are only weakly gravitationally bound, so even modest heating can drive a slow molecular wind. However, 2D models are required to give a more robust geometric comparison between simulations and observations to verify an outer wind in IM Lup. 

\begin{figure}
    \centering
    \includegraphics[width=0.8\columnwidth]{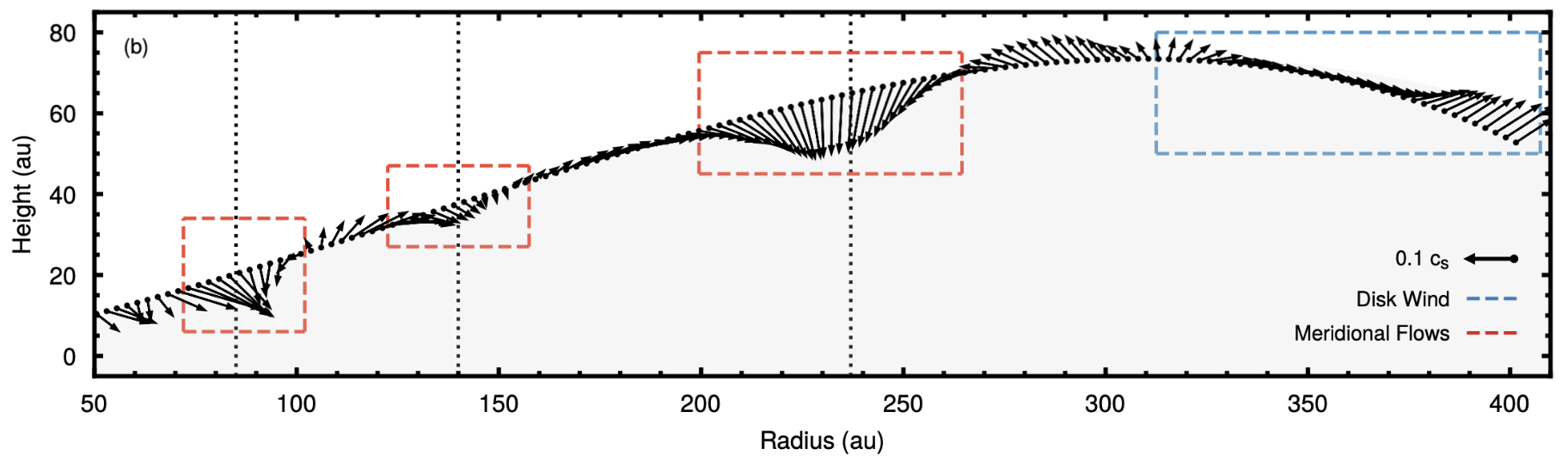}
    \caption{The azimuthally averaged velocity (vectors) at the height of CO emission as a function of radius in HD 163296 by \protect\cite{2021ApJS..257...18T}. In addition to meridional flows, there is detection of a possible outer wind  at $\sim320-400$\,au, highlighted by the blue dashed box on the right.   }
    \label{fig:TeagueHD163296}
\end{figure}

Another candidate external wind in a $F_\mathrm{FUV}<10\,G_0$ UV environment was identified in HD~163296 by \cite{2019Natur.574..378T} and \cite{2021ApJS..257...18T}. They developed a framework in which the 3D CO emitting surface of the disc is traced, which can then be translated into a map of the velocity as a function of radius and height in the disc as illustrated in Figure \ref{fig:TeagueHD163296}. Their main focus was the meridional flows identified at smaller radii in the disc, but they serendipitously discovered evidence for an outer wind launched from $\sim350-400$\,au (see the right hand blue box of Figure \ref{fig:TeagueHD163296}). This is yet to be interpreted with any numerical models or synthetic observations which will be necessary to support the interpretation that it is external photoevaporation that is responsible. 

\cite{2021ApJS..257...18T} also carried out a similar analysis of the disc MCW 480 but found no evidence of an outer wind despite having a similar radial extent and environment. Whether this is a consequence of the face-on orientation ($\sim 33^\circ$ compared to $\sim 45^\circ$ for HD 163296) or because there is no outer wind remains to be determined. Indeed, the HD 163296 kinematic feature that appears to be an outflow may also be due to some other mechanism. Furthermore, a similar approach is yet to be applied to IM Lup to search for kinematic traces of an outer wind. 

Looking ahead, determining whether external irradiation can really launch winds from discs down to FUV fluxes $F_\mathrm{FUV}<10\,G_0$ is important for understanding just how pervasive external photoevaporation is.

\begin{mdframed}
\subsection{Summary and open questions for observational properties of externally photoevaporating discs}
\begin{enumerate}
    \item{External photoevaporation has been directly observed (e.g. proplyds) for almost 30 years. The vast majority of proplyd observations are in the ONC. }
    \item{In recent years direct evidence for external photoevaporation is being identified in more regions, and B stars are now also known to facilitate it. However, the range of environments in which it has been observed is still very limited.}
\end{enumerate}   

\noindent Some of the many open questions are:
\begin{enumerate}    
    \item \textit{What are robust diagnostics and signposts of external winds in weak UV environments?}
    \item \textit{What are diagnostics of external photoevaporation in distant clusters where a cometary proplyd morphology is spatially unresolved?}
    \item \textit{How widespread and significant are external winds from discs in weak UV environments?}
\end{enumerate}
\end{mdframed}

\section{Impact on disc evolution and planet formation}
\label{sec:planet_formation}

In this section we consider how a protoplanetary disc evolves when exposed to strong external UV fields, and the consequences for planet formation. We will only briefly describe some relevant processes for planet formation in isolated star-disc systems. This is a vast topic with several existing review articles on both protoplanetary discs \citep[e.g.][]{Armitage11, Williams11, Andrews20, Lesur22} and the consequences for forming planets \citep[e.g.][]{Kley12, Baruteau14, Zhu21, Drazkowska22} to which we refer the reader. \citet{Adams10} reviewed general influences on planet formation \citep[see also][]{Parker20}, with a focus on the Solar System. Here we focus on a detailed look on how external photoevaporation affects the formation of planetary systems. 




\subsection{Gas evolution}
\label{sec:gas_evolution}
\subsubsection{Governing equations}

The gas dominates over dust by mass in the interstellar medium (ISM) by a factor $\rho_\mathrm{g}/\rho_\mathrm{d}\sim 100$ \citep{Bohlin78, Tricco17}. This ratio is usually assumed to be similar in (young) protoplanetary discs, although CO emission suggests that the gas may be somewhat depleted with respect to the dust \citep{Ansdell16}. Nonetheless, gas is a necessary ingredient in instigating the growth of planetesimals by streaming instablility \citep{Youdin05} and represents the mass budget for assembling the giant planets \citep{Mizuno78, Bodenheimer86}. Thus, how the gas evolves in the protoplanetary disc is one of the first considerations in planet formation physics. 

Despite its importance, the gas evolution of the disc remains uncertain. The observed accretion rates of $\dot{M}_\mathrm{acc} \sim 10^{-10} {-}10^{-7} \, M_\odot$~yr$^{-1}$ onto young stars \citep{Gullbring98, Muzerolle98, Manara12}, depending on the stellar mass \citep{Herczeg08, Manara17}, imply radial angular momentum transport. For several decades, this angular momentum transport has widely been assumed to be driven by an effective viscosity, mediated by turbulence that may originate from the magnetorotational instability \citep{Balbus98}. In the absence of purturbations, the surface density $\Sigma_\mathrm{g}$ of the gaseous disc as a function of cylindrical radius $r$ then follows 
\citep{Lynden-Bell74}:
\begin{equation}
\label{eq:disc_evol}
    {\dot{\Sigma}_\mathrm{g}} = \frac 1 r \partial_r \left[  3 r^{1/2} \partial_r \left( \nu \Sigma_\mathrm{g} r^{1/2} \right)\right] - \dot{\Sigma}_{\rm{int}} -\dot{\Sigma}_{\rm{ext}}.
\end{equation}The loss rates $\dot{\Sigma}_{\rm{int}}$ and $\dot{\Sigma}_{\rm{ext}}$ are the surface density change due to the internally and externally driven winds respectively. The kinematic viscosity $\nu$ is usually parametrised by an $\alpha$ parameter \citep{Shakura73} such that $ \nu(r) = \alpha c_{\rm{s}}^2/\Omega_\mathrm{K}$, for sound speed $c_\mathrm{s}$ and Keplerian frequency $\Omega_\mathrm{K}$. In a disc with a mid-plane temperature $T\propto r^{-1/2} $, this yields $\nu \propto r$ \citep{Hartmann98}. 

In the following discussion, we will assume angular momentum transport is viscous. However, recently several empirical studies of discs have suggested a low $\alpha \sim 10^{-4} {-} 10^{-3}$ \citep{Pinte16, Flaherty17, Trapman20}. This is difficult to reconcile with observed accretion rates. Alternative candidates, particularly magnetohydrodynamic (MHD) winds, have been suggested to drive angular momentum transport \citep[e.g.][]{Bai13}. In this case, angular momentum is not conserved but extracted from the gas disc, with consequences for the disc evolution \citep{Lesur21, Tabone22}. In the following we will assume a standard viscous $\alpha$ disc model, with the caveat that future simulations may offer different predictions by coupling the externally driven photoevaporative wind with MHD mediated angular momentum removal. 

\subsubsection{Implementing wind driven mass loss}

In order to integrate equation~\ref{eq:disc_evol}, we must define the form of $\dot{\Sigma}_{\rm{int}}$ and $\dot{\Sigma}_{\rm{ext}}$. The internal wind may be driven by MHD effects \citep[e.g.][]{Bai13, Lesur14} or thermally due to a combination of EUV \citep[e.g.][]{Hollenbach94,Alexander06}, X-ray \citep[e.g.][]{Ercolano09,Owen10, Owen11} and FUV \citep[e.g.][]{Gorti09} photons, or probably a combination of the two \citep[e.g.][]{Bai16,Bai17, Ballabio20} . We do not focus on the internally driven wind in this review, but note that the driving mechanism influences the radial profile of $\dot{\Sigma}_{\rm{int}}$ \citep[see][for a review]{Ercolano17}. 

Several authors have included the external wind in models of (viscous) disc evolution \citep[e.g.][]{Clarke07,Anderson13,Rosotti17, Sellek20, Concha-Ramirez21, Coleman22}. In general, these studies follow a method similar to that of \citet{Clarke07} in removing mass from the outer edge, because winds are driven far more efficiently where the disc is weakly gravitationally bound to the host star. We discuss the theoretical mass loss rates in Section~\ref{sec:theory}; in brief, the analytic expressions by \citet{Johnstone98} are applied to compute the mass loss rate in the EUV driven wind, while early studies applied the expressions by \citet{Adams04} for an FUV driven wind. The latter has now been improved upon using more detailed models by \citet{Haworth18b}, such that one can interpolate over the {FRIED} grid to find an instantaneous mass loss rate. For typical EUV fluxes, once the disc is (rapidly) sheared down to a smaller size, any severe externally driven mass loss is expected to be driven by FUV rather than EUV photons (see Section~\ref{sec:EUVvFUV}). For this reason, the EUV mass loss is often neglected in studies of disc evolution. 

Since the mass loss rate is sensitive to the outer radius $R_\mathrm{out}$, care must be taken when implementing a numerical scheme that the value of $R_\mathrm{out}$ is sensibly chosen. In practice, a sharp outer radius quickly develops for a disc with initial mass loss rate higher than the rate of viscous mass flux (accretion). For a viscous disc with $\nu\propto r$, the surface density $\Sigma \propto r^{-1}$ in the steady state, which is the same profile as adopted for the numerical models in the {FRIED} grid. One can then interpolate using the total disc mass and outer radius. The latter is evolved each time-step by considering the rate of wind-driven depletion versus viscous re-expansion \citep[e.g.][]{Clarke07, Winter18b}. However, the physically correct way to define the outer radius is to find the optically thin-thick transition, since the flow in the optically thin region is set by the wind launched from inside this radius. Mass loss scales linearly with surface density in the optically thin limit \citep{Facchini16}, such that one can define $R_\mathrm{out}$ to be the value of $r$ that gives the maximal mass loss rate for the corresponding $\Sigma(r)$ in the disc evolution model \citep[see discussion by][]{Sellek20}. Under the assumption of a viscous disc with $\nu \propto r$ both approaches yield similar outcomes, but the approach of \citet{Sellek20} should be adopted generally. For example, this prescription would be particularly important in the case of a disc model incorporating angular momentum removal via MHD winds \citep{Tabone22}.

\subsubsection{Viscous disc evolution with external depletion of gas}

\label{sec:visc_evol_extdep}
\begin{figure}
    \centering
    \includegraphics[width=\textwidth]{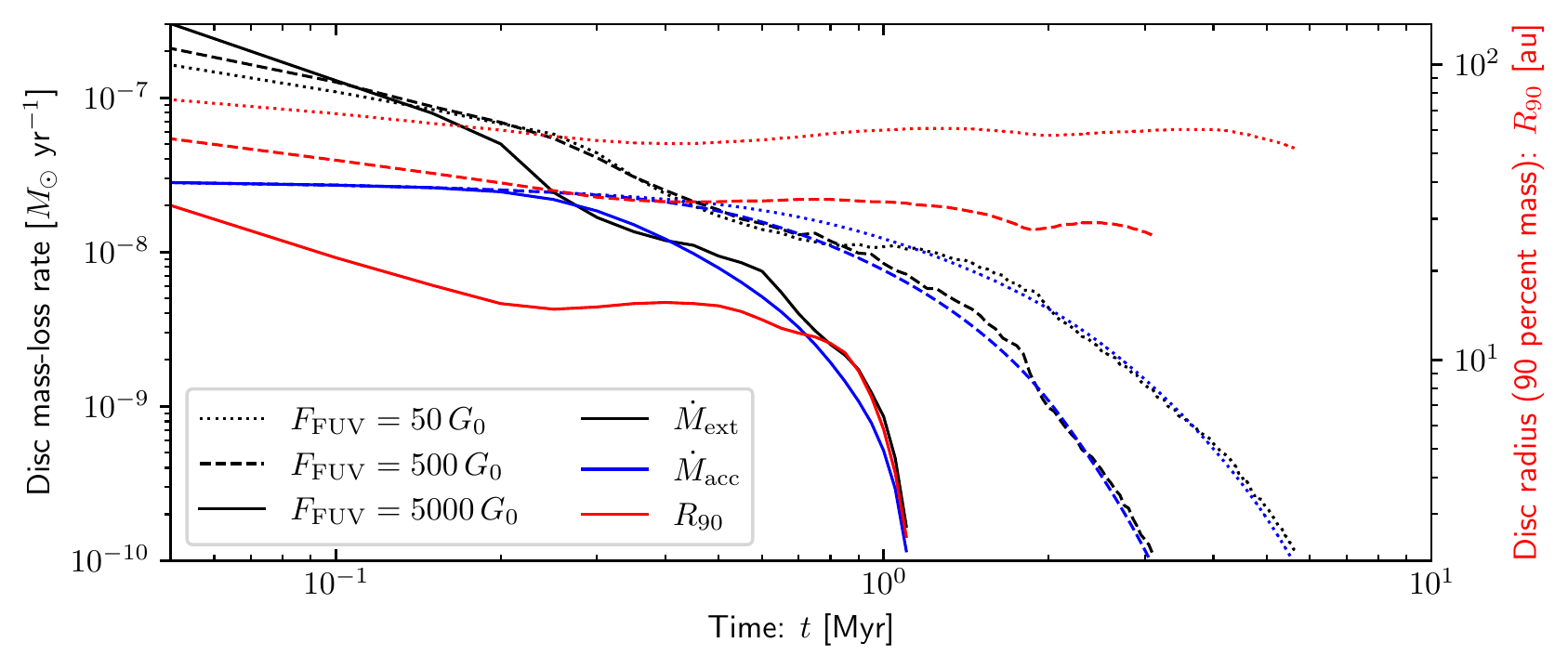}
    \caption{Evolution of the externally driven mass loss rates $\dot{M}_\mathrm{ext}$ (black, from the {FRIED} grid), accretion rates $
    \dot{M}_\mathrm{acc}$ (blue) and radii containing 90~percent of the disc mass (red) for a disc evolving under equation~\ref{eq:disc_evol} with a viscous $\alpha=3\times10^{-3}$, no internal photoevaporation and varying the external FUV flux $F_\mathrm{FUV}= 50 \,G_0$ (dotted), $500\,G_0$ (dashed) and $5000 \, G_0$ (solid). We stop integrating when the mass loss rate  or inner surface density reaches the floor in the {FRIED} grid ($\dot{M}_\mathrm{ext} = 10^{-10}\,M_\odot$~yr$^{-1}$ and $\Sigma_\mathrm{g}(1\,\rm{au}) = 0.1$~g~cm$^{-2}$ respectively).  }
    \label{fig:R_Mdot_evol}
\end{figure}

In Figure~\ref{fig:R_Mdot_evol} we show examples for the evolution of the disc radius that contains 90~percent of the mass, $R_{90}$, and corresponding externally driven mass loss rates \citep[from][]{Haworth18b} due to the numerical integration of equation~\ref{eq:disc_evol} \citep[see also Figures 3 and 4 of][for example]{Sellek20}. To illustrate the variation in mass loss and radius, we choose an initial scale radius $R_\mathrm{s} = 100$~au and mass $M_\mathrm{disc}=0.1\,M_\odot$, truncated outside of $200$~au, and with a viscous $\alpha=3\times 10^{-3}$. For simplicity, we ignore internal winds ($\dot{\Sigma}_\mathrm{int} = 0$ everywhere), in order to highlight the main consequences of the externally driven photoevaporative wind in isolation.

From such simple models, we gain some insights into how we expect disc evolution to be affected by externally driven mass loss. In the first instance, the efficiency of these winds at large $r$ leads to extremely rapid shrinking of the disc. In a viscous disc evolution model, this shrinking continues until the outwards mass flux due to angular momentum transport balances the mass loss due to the wind such that the accretion rate $\dot{M}_\mathrm{acc} \sim \dot{M}_\mathrm{ext}$ \citep{Winter20a, Hasegawa22}. This offers a potential discriminant for disc evolution models: while  $\dot{M}_\mathrm{acc} \sim \dot{M}_\mathrm{ext}$ for a viscously evolving disc in an irradiated environment, this need not be the case if angular momentum is removed from the disc via MHD winds or similar. In either case, the initial rapid mass loss rates of $\dot{M}_\mathrm{ext} \sim 10^{-7} \, M_\odot$~yr$^{-1}$ are only sustained for relatively short time-scales of a few $10^5$~yr. Because the mass loss rate is related to the spatial extent of proplyds (equation~\ref{equn:Mdot}), this implies easily resolvable proplyds should be short-lived and therefore rare.  

This rapid shrinking of the disc has consequences for the apparent viscous depletion time-scale of the disc, leading \citet{Rosotti17} to expound the usefulness of the dimensionless accretion parameter:
\begin{equation}
    \eta \equiv \frac{\tau_\mathrm{age} \dot{M}_\mathrm{acc}}{M_\mathrm{disc}},
\end{equation}where $\tau_\mathrm{age}$, $M_\mathrm{disc}$ are the age and mass of the disc respectively, while $\dot{M}_\mathrm{acc}$ is the stellar accretion rate. For disc evolution driven by viscosity, and where the disc can reach
a quasi steady-state, we expect $\eta\approx 1$. Indeed $\eta\approx 1$ for discs in low mass local SFRs when using the dust mass $M_\mathrm{dust}$ (or, more precisely, sub-mm flux) as a proxy for the total disc mass $M_\mathrm{disc} = 100\cdot M_\mathrm{dust}$. While numerous processes can interrupt accretion and yield $\eta<1$, only an outside-in depletion process can yield $\eta >1$. 
\citet{Rosotti17} showed that this applies to a number of discs in the ONC, hinting that external photoevaporation has sculpted this population. 

With the inclusion of internal disc winds, a number of disc evolution scenarios become possible for an externally irradiated disc. The internal wind drives mass loss outside of some launching radius $R_\mathrm{launch} \approx 0.2 R_\mathrm{g}$, which is inside of the gravitational radius due to hydrodynamic effects \citep{Liffman03}. Once (viscous) mass flux through the disc becomes sufficiently small ($\dot{M}_\mathrm{acc} 
\lesssim \dot{M}_\mathrm{int}$, the mass loss in the internal wind), the internal wind depletes material at $r\approx R_\mathrm{launch}$ faster than it is replenished, leading to gap opening. Subsequently, the inner disc is rapidly drained and inside-out disc depletion proceeds \citep{Skrutskie90, Clarke01}. \citet{Coleman22} discussed how the balance of internal and external photoevaporation can alter this evolutionary picture. In the limit of a vigorous externally driven wind, the outer disc may be depleted down to $R_\mathrm{launch}$. In this case, the internal wind no longer drives inside-out depletion, and the disc is dispersed outside-in. In the intermediate {FUV case ($F_\mathrm{FUV}=500\, G_0$)}, external disc depletion erodes the outer disc without reaching $R_\mathrm{launch}$. In models for disc material where outer disc material is eventually transported inwards, then outer disc depletion should still shorten the lifetime to some degree. In Figure~\ref{fig:R_Mdot_evol} we see that only the disc exposed to $F_\mathrm{FUV}=5000\,G_0$ {(above which we consider `high FUV environments' by our definitions in Section~\ref{sec:flux_units})} is sheared down to $R_\mathrm{out}< 20$~au before the inner surface density becomes small (lower than the {FRIED} grid range). Thus, if angular momentum transport is inefficient beyond this radius, then sustained exposure to {high FUV fluxes} $F_\mathrm{FUV}\gtrsim 5000\, G_0$ should be required to shorten the disc lifetime. In this case, we may expect inside-out depletion for discs exposed to more moderate $F_\mathrm{FUV}$, although the external depletion still reduces their overall mass.


\subsection{Solid evolution}
We now consider how the evolution of solids within the gas disc may be influenced by externally driven winds. The growth of ISM-like dust grains to larger aggregates and eventually planets is the result of numerous inter-related physical processes, covering a huge range of scales. We do not review these processes in detail here, but refer the reader to the recent reviews by \citet[][with a focus on dust-gas dynamics]{Lesur22} and \citet[][with a focus on planet formation]{Drazkowska22}. Due to the complexity of the topic, and the fact that the primary empirical evidence comes from local, low mass SFRs without OB stars, most studies to date have focused on dust growth in isolated protoplanetary discs. We here consider the results of the few investigations focused on dust evolution specifically in irradiated discs. 

\subsubsection{Dust evolution}

\begin{figure}
     \centering
     \begin{subfigure}[b]{0.45\textwidth}
         \centering
         \includegraphics[width=\textwidth]{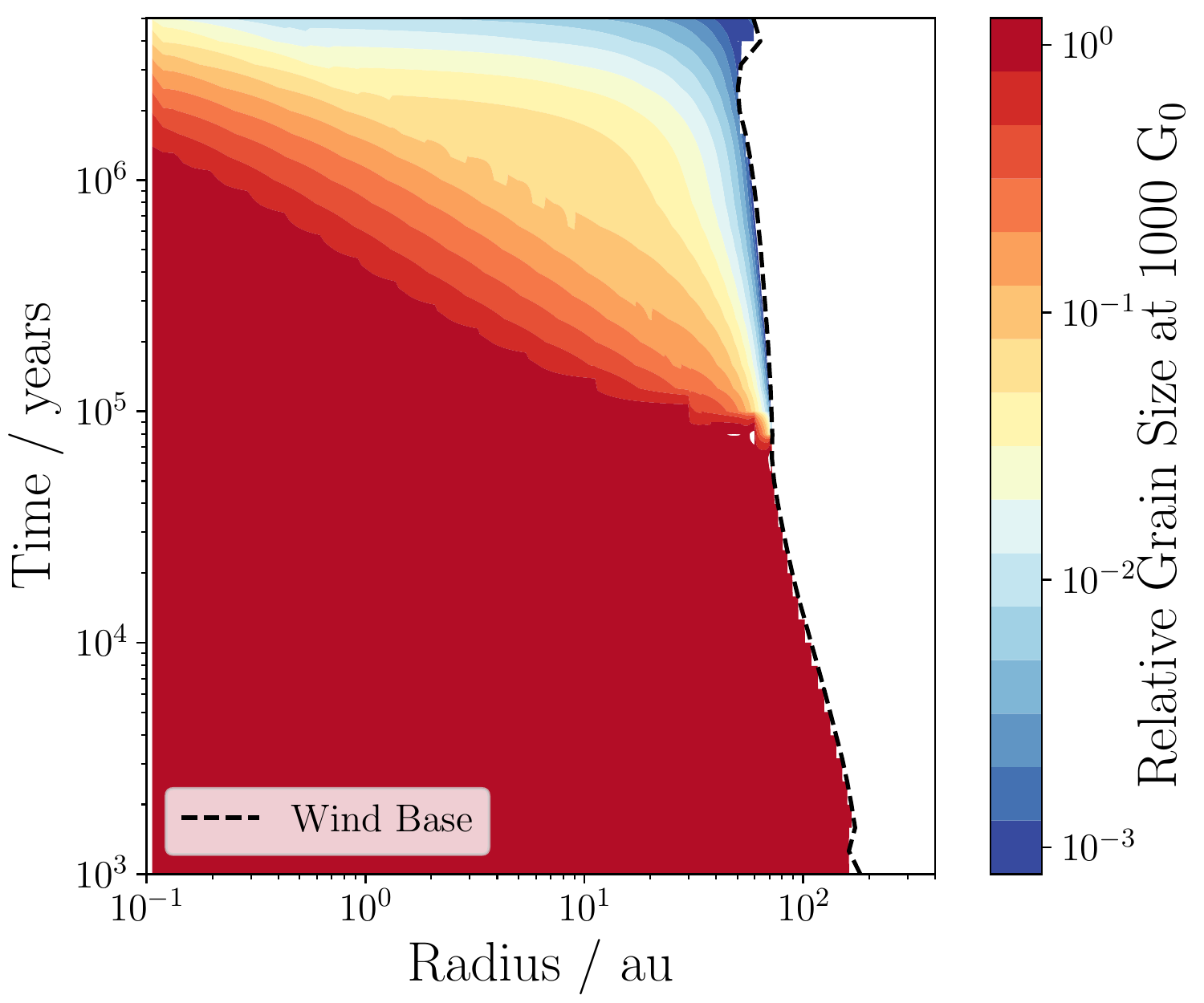}
         \caption{Maximum grain size.}
         \label{fig:gsize}
     \end{subfigure}
     \hfill
     \begin{subfigure}[b]{0.45\textwidth}
         \centering
         \includegraphics[width=\textwidth]{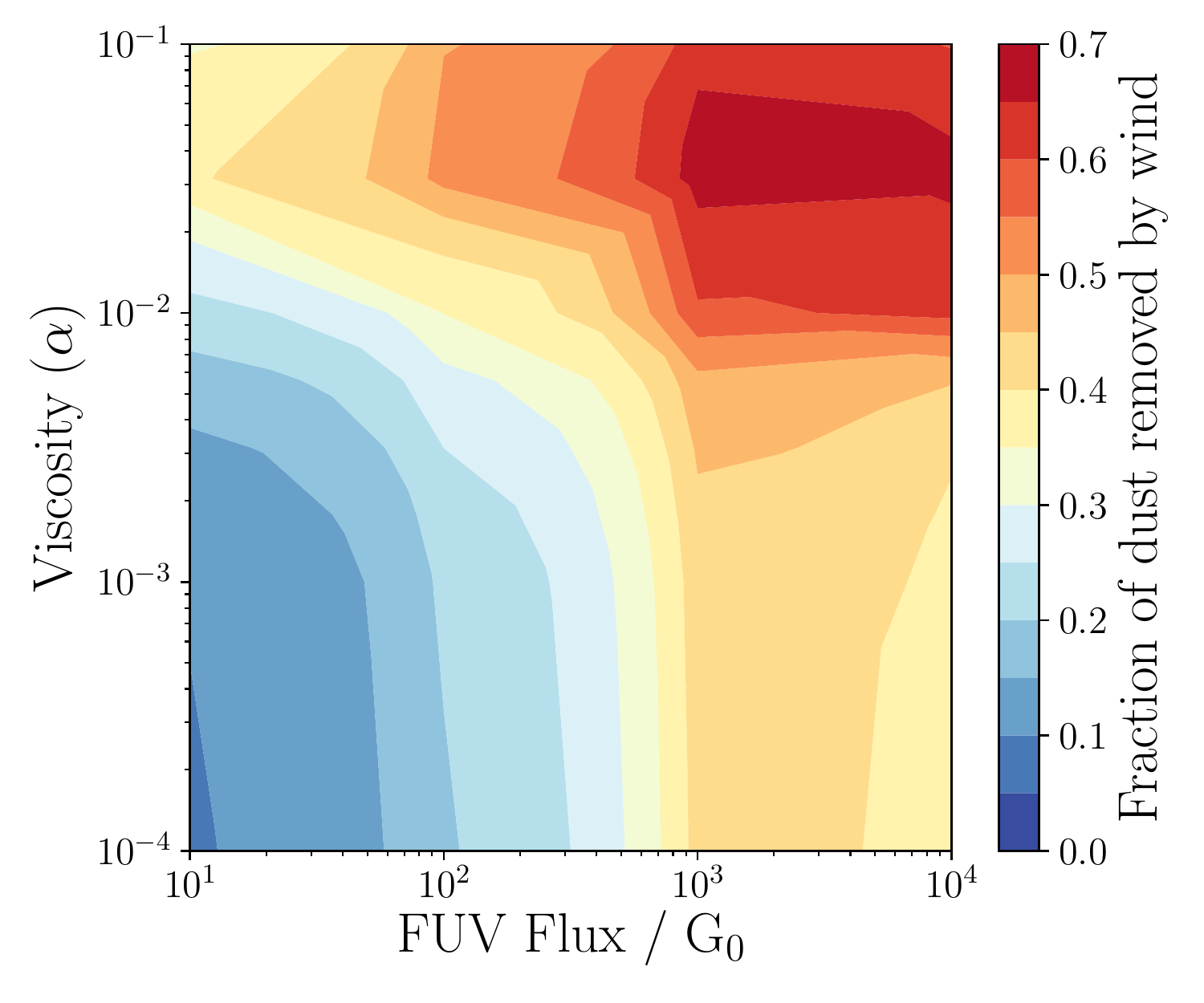}
         \caption{Dust entrainment in an externally driven  wind}
         \label{fig:fd_wind}
     \end{subfigure}
     
     \caption{The effect of external FUV irradiation on the dust content of a protoplanetary disc, adapted from \citet{Sellek20}. Figure~\ref{fig:gsize} shows the maximum grain size, relative to an isolated disc ($F_\mathrm{FUV}=0\, G_0$) as a function of time and radius in an irradiated disc with viscous $\alpha=10^{-3}$, exposed to FUV flux $F_\mathrm{FUV}=1000\, G_0$. Figure~\ref{fig:fd_wind} shows the total fraction of dust removed by the wind over the lifetime of a disc as a function of the viscous $\alpha$ and $F_\mathrm{FUV}$. }
\end{figure}

\citet{Sellek20} investigated the drift and depletion of dust in a viscously evolving protoplanetary disc. Dust is subject to radial drift, wherein dust moves towards pressure maxima  \citep[i.e. inwards in the absence of local pressure traps --][]{Weidenschilling77}, as well as grain growth dependent on the local sticking, bouncing and fragmentation properties \citep[see][and references therein]{Birnstiel12}. The drift velocity is determined by the Stokes number, which in this context is the ratio of the stopping time of the dust grain to the largest eddy timescale $\sim \Omega_\mathrm{K}^{-1}$. Near the midplane and in the Epstein regime, this can be approximated:
\begin{equation}
    \rm{St} \approx \frac {\pi} {2} \frac{ \rho_\mathrm{s} a_{\mathrm{s}}}{\Sigma_\mathrm{g}},
\end{equation}where $a_\mathrm{s}$ and $\rho_\mathrm{s}$ are the grain size and density respectively. The draining of large dust grains can be understood in terms of the balance of viscous mass flux and radial drift \citep{Birnstiel12}, such that the equilibrium value of $\rm{St}$ is:
\begin{equation}
    \rm{St}_{\mathrm{eq}} = 3\alpha \left| \frac{\frac{3}{2} + \frac{\mathrm{d}\ln \Sigma_\mathrm{g}}{\mathrm{d}\ln r} }{\frac{7}{4} - \frac{\mathrm{d}\ln \Sigma_\mathrm{g}}{\mathrm{d}\ln r} } \right|
\end{equation} for the standard $\alpha$ disc model with $\nu \propto r$. For $\rm{St}>\rm{St}_\mathrm{eq}$, dust drifts inwards regardless of the viscosity. At the outer edge of the disc $|{\mathrm{d}\ln \Sigma_\mathrm{g}}/{\mathrm{d}\ln r}| $ becomes large, so that $\rm{St}_{\mathrm{eq}} \rightarrow 3\alpha$. Hence dust in the outer disc that grows above $\rm{St}>3\alpha$ will migrate rapidly inwards. External evaporation acts to increase $|{\mathrm{d}\ln \Sigma_\mathrm{g}}/{\mathrm{d}\ln r}|$ in the outer disc, clearing this region of large dust grains. Figure~\ref{fig:gsize}, adapted from \citet{Sellek20}, shows how the external wind can rapidly evacuate the outer disc of large grains, dependent on the value of $\alpha$. Given that this occurs on short time-scales compared to the disc lifetime, it has consequences for planet formation and observational properties of discs, possibly explaining why the sub-mm flux-radius relationship seen in low mass SFRs 
\citep{Tripathi17} does not hold in the ONC \citep{Eisner18}. 

The clearing of large grains from the outer disc also has consequences for the quantity of solid material that can be lost in the wind. 
\citet{Sellek20} use the entrainment constraints given by equation \ref{equn:entrainedSize} to estimate an entrainment fraction $f_\mathrm{ent}$ for a given grain size distribution. The fraction of dust removed in their global externally irradiated disc model is shown in Figure~\ref{fig:fd_wind}. When viscosity is large ($\alpha \gtrsim 10^{-2}$), this fraction can exceed $50$~percent for $F_\mathrm{FUV}\gtrsim 10^3\, G_0$. This trend of higher dust depletion with higher $\alpha$ is both due to the faster replenishment of disc material in the outer regions where the wind is launched, and because inward drift of large grains is less efficient (higher $\rm{St}_\mathrm{eq}$). However, in general depletion of dust is not efficient, with typically less than half of the dust mass being lost. However, in the models of \citet{Sellek20} this does not result in an enhancement of the dust-to-gas ratio due to the enhanced loss of the large dust grains via rapid inwards drift.

A big caveat of the above discussion is that it does not consider the role of local pressure traps in halting the radial drift \citep[e.g.][]{Pinilla12, Rosotti20}. This local accumulation of solids can also lead to a mutual aerodynamic interaction that leads to a local unstable density growth that can seed planet formation \citep{Youdin05}. If sufficient quantities of dust remain when the gas component is depleted by external photoevaporation, then this may serve to initiate planetesimal formation in the outer disc \citep{Throop05}, {perhaps leading to low mass or rocky planets rather than gas giants}. Future studies may consider how the efficiency of dust trapping in the outer disc affect this picture. 
 
\subsubsection{Planet formation and evolution}

As discussed in the introduction, the influence of external photoevaporation is still rarely included in models for planet formation, despite its apparent prevalence. However, \citet{Ndugu18} studied how increases in the disc temperature due to external irradiation alter the formation process. By increasing the disc scale height (thus decreasing midplane density), this heating reduces the efficiency of pebble accretion and giant planet formation in the outer disc. In this framework, giant planets that do form at high temperatures more frequently orbit with short periods (hot/warm Jupiters) because their planet cores need to form early or in the inner disc. Temperature also has an influence after the formation of a low mass planet core. Low mass planets that have not opened up a gap in the gas disc undergo type I migration, which is due to torques associated with a number of local resonances \citep{Paardekooper11}. These torques are sensitive to thermal diffusion, such that they also depend on local temperature and associated opacity. This may lead to complex migration behaviour for the low mass planets \citep{Coleman14}, which would also be influenced by increasing the disc temperature due to external irradiation \citep{Haworth21b}.  

\begin{figure}
    \centering
    \includegraphics[width=0.95\textwidth]{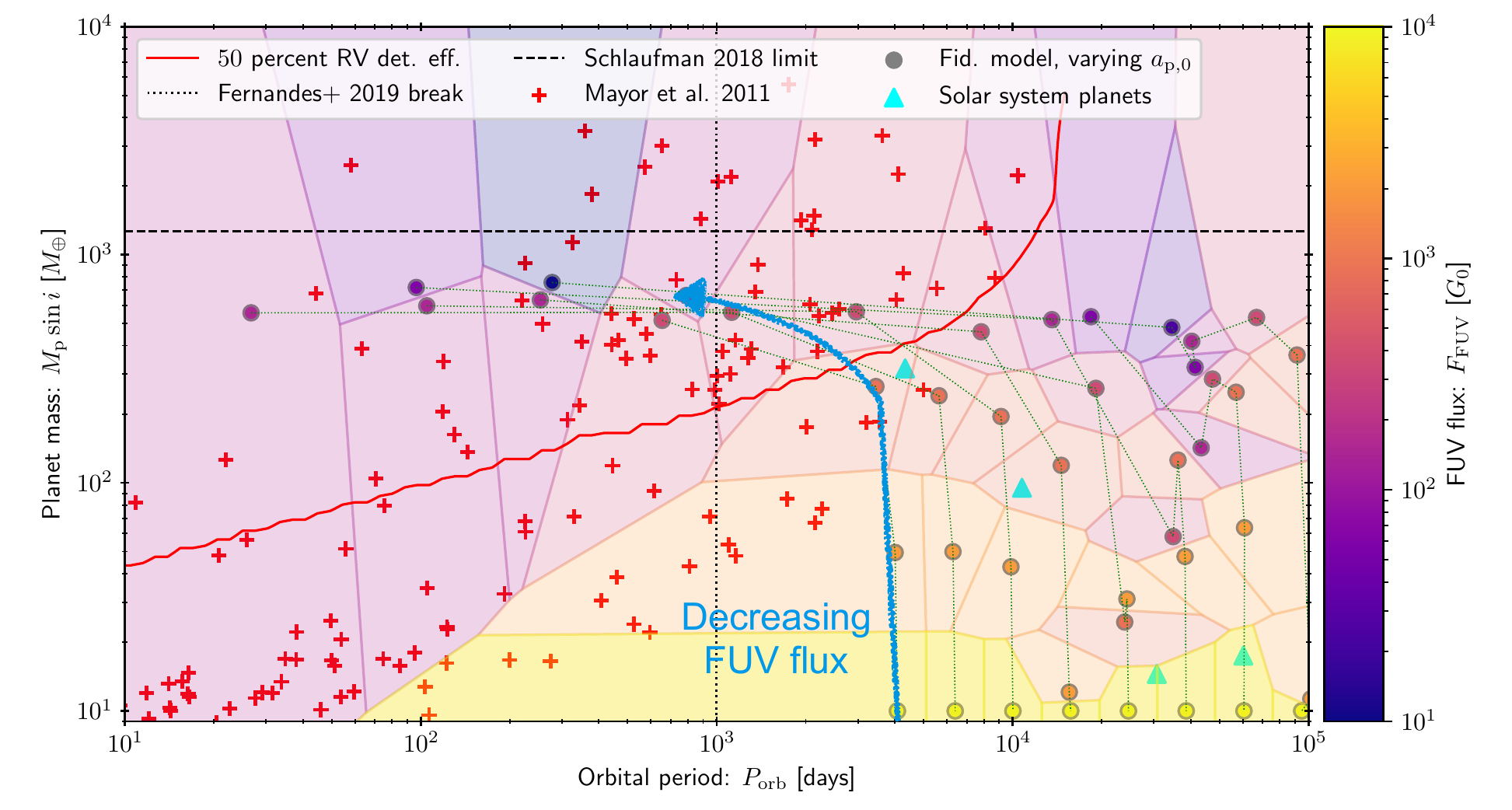}
    \caption{The final masses $M_\mathrm{p}$ and orbital periods $P_\mathrm{orb}$ of giant planets evolving under a growth and evolution model for planetary cores of mass $10\, M_\oplus$ and formation time of $\tau_\mathrm{form} = 1$~Myr with disc viscosity $\alpha=3\cdot 10^{-3}$ \citep[adapted from][]{Winter22}. The circular points, connected by faint dotted green lines for fixed starting semi-major axis $a_{\mathrm{p},0}$, represent the final location of the planet in $P_\mathrm{orb}-M_\mathrm{p}$ space. Points are coloured by the external FUV flux experienced in that model, with the same colour is shown in a Voronoi cell to emphasise trends. Red crosses show the locations of HARPS radial velocity (RV) planet discoveries presented by \citet{Mayor11}. The 50~percent detection efficiency of the HARPS survey is shown by the red contour. We show orbital period at longer than which \citet{Fernandes19} infers a dearth of planets (dotted black line) and the limit above which planets may no longer form by core accretion \citep[black dashed line --][]{Schlaufman18}. We also show the planets in the Solar Sytem (green triangles). }
    \label{fig:aG_exp}
\end{figure}

Perhaps more directly, where there is sufficient mass loss from the disc due to an externally driven wind, this also reduces the time and mass budget available for (giant) planet formation and migration. Internal winds have long been suggested to play an important role in stopping inward planet migration \citep{Matsuyama03}. \citet{Armitage02} and \citet{Veras04} investigated a how type II migration of giant planets (within a gap) can be severely altered by the loss of the disc material in a photoevaporative wind, even leading to outwards migration if the outer disc material is removed. Since then, a number of authors have investigated how giant planet migration can be halted by internally driven disc winds \citep[e.g.][]{Alexander09,Jennings18, Monsch19}. Planet population synthesis models have now started to implement prescriptions for mass loss due to an external wind \citep[such as in the Bern synthesis models --][]{Emsenhuber21}. However, this is presently based on a single (typical) estimated $\dot{\Sigma}_\mathrm{ext}$ that is constant throughout time and with radius $r$, rather than the more physical outside-in, radially dependent depletion models discussed in Section~\ref{sec:gas_evolution}. Recently, \citet{Winter22} looked at how growth and migration is altered by external FUV flux exposure, and we show the outcomes of some of these models in Figure~\ref{fig:aG_exp}. The outside-in mass loss prescription leads to more dramatic consequences than those of \citet{Emsenhuber21}, curtailing growth and migration early. As a result, {low FUV fluxes $F_\mathrm{FUV}\lesssim 100\, G_0$} produce planets that are massive, $M_\mathrm{p}\gg 100\,M_\oplus$ and on short orbital periods ($P_\mathrm{orb}\ll 10^4$~days), similar to those that are most frequently discovered by radial velocity (RV) surveys \citep[e.g.][]{Mayor11}.  The typical FUV fluxes in the solar neighbourhood are $\langle F_\mathrm{FUV}\rangle \sim 10^3 {-}10^4\, G_0$ (see Section~\ref{sec:demographics_SFRs}), which yield typical planet masses and orbital periods closer to those of the massive planets in the Solar System. Such relatively low mass planets fall close to, or below, typical RV detection limits, and the anti-correlation between planet mass and semi-major axis may therefore contribute to the inferred dearth of detected planets with periods $P_\mathrm{orb}\gtrsim 10^3$~days \citep{Fernandes19}. Testing the role of external photoevapation for populations of planets further requires coupling these prescriptions with population synthesis efforts.

While we can try to identify the role of external photoevaporation via correlations in disc populations, more direct ways to connect environment with the present day stellar (and exoplanet) population would be useful. For an example of how this might work in practice, \citet{Roquette21} have highlighted that premature disc dispersal may leave an impact on the rotation period distribution of stars. The rotation of a star is decelerated due to `disc-locking' during the early pre-main-sequence phase accelerates rotation. Thus, by shortening the inner disc lifetime (see Section~\ref{sec:inner_disc}), external photoevaporation may leave an imprint on the stellar population up to several $100$~Myr after formation. This may explain, for example, the increased number of fast rotating stars in the ONC compared to Taurus \citep{Clarke00}. In future, models and rotation rates may be used to interpret the disc dispersal time-scales for main sequence stars, which may complement investigations into planet populations in stellar clusters \citep[e.g.][]{Gilliland00, Brucalassi16, Mann17}.

\label{sec:giant_planets}

\subsection{Disc surveys of local star forming regions}
\label{sec:surveys}

In this section so far, we focus on the theoretical influence of external photoevaporation on planet formation. However, the most important evidence for or against the influence of external photoevaporation on forming planets must be found statistically in the surveys of local protoplanetary disc populations. Such surveys and more detailed observations of individual discs offer the most direct way to probe the physics of planet formation. Here we report the evidence in the literature for the role of external photoevaporation in sculpting disc populations. {We first consider the observational approaches and the challenges in inferring disc properties in Section~\ref{sec:obs_methods_challenges}. We then consider the evidence for variations in (inner) disc survival fractions (Section~\ref{sec:inner_disc}) and outer disc properties (Section~\ref{sec:outer_disc}) with FUV flux exposure.} 

\subsubsection{Methods and challenges}
\label{sec:obs_methods_challenges}
\noindent\textbf{Inner disc lifetimes}

Photometric censuses of young stars in varying SFRs can yield insights into the survival time of discs in regions of similar ages. Young stars exhibit luminous X-ray emission due to the magnetically confined coronal plasma \citep{Pallavicini81}, X-ray surveys with telescopes such as \textit{Chandra} offer the basis for constructing a catalogue of young members of a SFR. These can be coupled with photometric surveys to infer the existence or absence of a NIR excess. Comparison of the fractions of discs with varying either within the same SFR, or between different regions with similar ages, allows one to identify regions where discs have shorter lifetimes than the $\sim 3$~Myr typical of low mass SFRs.  

While this principle appears simple, in fact several steps required in achieving this comparison bear with them numerous pitfalls. One issue is reliable membership criteria, which were historically photometric or spectroscopic \citep[e.g.][]{Blaauw56, deZeeuw99}, improved recently through proper motions (and to a lesser extent, parallax measurements) from \textit{Gaia} DR2 \citep{Arenou18}. 

Uncertainty and heterogeneity in age determination for young SFRs also represent a significant challenge, particularly in comparing across different SFRs \citep[e.g.][]{Bell13}. {\citet{Michel21} point out that considering only low-mass, nearby regions, the isolated disc life-time might be a factor $2{-}3$ longer than the canonical 3 Myr obtained by aggregating across many star forming regions with a ride range of properties. Further, since disc life-times are shorter around high mass stars \citep{Ribas15}, care must be taken when comparing across samples of discs that may have different sensitivity limits.} 
 
Binning young stars by incident FUV flux also carries complications. The three dimensional geometry (projection effects) and dynamical mixing in stellar aggregates may also hide correlations between historic FUV exposure and present day disc properties \citep[e.g.][]{Parker21}. Even empirically quantifying the luminosity and spectral type of neighbour OB stars can be challenging. Massive stars are often found in multiple systems \citep[e.g.][]{Sana09}, which can lead to mistaken characterisation \citep[e.g.][]{MaizApellaniz07}, while these massive stars are also expected to go through transitions in the Hertzsprung–Russell diagram in combination with rapid mass loss \citep[see][for a review]{Vink22}. Statistically, any studies attempting to measure correlations in individual regions must choose binning procedures for apparent FUV flux with care. For example, the number of stars per bin must be sufficient such that uncertainties are not prohibitively large and binning should be performed by projected UV flux rather than separation alone (i.e. controlling for the luminosity of the closest O star). 

Finally, studies of NIR excess probe the presence of inner disc material, which represents the part of the disc expected to be least affected by external photoevaporation (see discussion at the end of Section~\ref{sec:visc_evol_extdep}). Therefore, inner disc survival fractions should be interpreted as the most conservative metric by which to measure the role of external disc depletion. \\

\noindent\textbf{Outer disc properties}

{The outer regions of the protoplanetary disc less weakly bound to the host star than the inner disc, and therefore much easily unbound by externally driven photoevaporative winds. Probing disc mass and radius also offers much more information than simply the presence or absence of circumstellar material. Outer disc properties are frequently inferred via probing the dust content, then assuming a canonical dust-to-gas ratio of $10^{-2}$ to infer a total mass, although this may be significantly higher in some cases \citep{Ansdell16}. In surveys of protoplanetary discs, the dust mass is usually inferred by making the assumption that the disc is optically thin such that:  }
\begin{equation}
\label{eq:Mdust}
    M_\mathrm{dust} \approx \frac{S_{\nu, \mathrm{dust}}D^2}{\kappa_{\nu, \mathrm{dust}} B_{\nu}(T_\mathrm{dust})},
\end{equation}{where $D$ is the distance to the source, $F_{\nu, \mathrm{dust}}$ is the flux from the cool dust at frequency $\nu$, $T_\mathrm{dust}=20$~K is
the dust temperature, $B_{\nu}$ is the Planck function and $\kappa_{\nu, \mathrm{dust}}\approx 2$~cm$^{2}$~g$^{-1}$ is the assumed opacity \citep{Beckwith90}. Even for studies of discs in low mass star forming regions, the estimate from equation~\ref{eq:Mdust} comes with several assumptions, including the fixed dust temperature and opacity. In the context of discs in irradiated environments, this can be further complicated by the heating of the dust by neighbouring stars \citep{Haworth21b}. Even further, in some regions background free-free emission may contribute to the continuum flux, and must therefore be subtracted by extrapolating from cm-wavelength observations \citep[see][and references therein]{Eisner18}. However, since free-free emission may be variable, this is another source of uncertainty.}

{Beyond dust masses, estimates of the outer disc extents can be made using spatially resolved continuum observations by either fitting a surface density profile or defining an effective radius enclosing a fraction of the total disc emission \citep[see discussion by][]{Tripathi17, Tazzari21}. This comes with the significant caveat that the size-dependent inward drift of dust grains mean that the inferred radii may not trace the physical disc radius, with long integration times required to trace the small grains that remain well-coupled to the gas \citep{Rosotti19}. } 

{Probing the gas content of discs cannot be achieved by directly measuring the hydrogen. Molecular hydrogen is a symmetric rotator, thus has no electric dipole moment. Quadrupole transitions are not excited at the typical temperature for the bulk of the gas in protoplanetary discs \citep[e.g.][]{Thi01}. Instead CO isotopologues, and sometimes HD \citep{Trapman17} are commonly used to infer disc masses \citep{Williams14}, which requires some assumption for the carbon and oxygen abundances. Outer disc radii might be inferred from spatially resolved moment 0 maps using a similar approach as discussed for resolved dust observations \citep[e.g.][]{Ansdell18}, or by assuming Keplerian rotation and fitting a model to the gas kinematics \citep[e.g.][]{Czekala15}. All of these methods rely on some assumptions about disc abundances and chemistry, which should be applied with caution in the case of irradiated discs. In particular the heating, winds/sub-keplerian rotation and photodissociation of molecules close to the outer edge of the disc will all influence the observed molecular line emission \citep{Haworth20}.} 
\\

\noindent\textbf{Proplyd definitions }

{Proplyds are discussed in detail in Section~\ref{sec:obs}, and we do not revisit them in detail here. As discussed in that section,  we distinguish between photoevporating globules and proplyds, the latter of which are typically smaller than a few $100$~au and are synonymous with an ionised star-disc system. For this reason, we here consider the examples in Trumpler 14 \citep{Smith10}, Pismis 24 \citep{Fang12}, NGC 3603 \citep{Brandner00} and Cygnus OB2 \citep{Wright12} as candidate proplyds only. These range in size from $\sim 1000$~au to several $10^4$~au in size, and are generally undetected in X-rays. We therefore consider these large photoevaporating objects as candidate proplyds or globules, rather than confirmed proplyds. This does not necessarily mean that true proplyds do not exist in these regions, as these are challenging to resolve at such large distances.}

\begin{table}[]
    \centering
  \rowcolors{2}{gray!25}{white}
    \begin{tabular}{c| c c c c c c c c c}
        Name & $D$  & $\log M_{\rm{tot}}$ & $\log \rho_0$   & $\log L_\mathrm{FUV}$  & $T_\mathrm{age}$ & Evidence & References\\
          &  [kpc] &  [$M_\odot$] & [$M_\odot$~pc$^{-3}$]   & [erg~s$^{-1}$]  &  [Myr] &  & \\
        \hline
        NGC 1977  & $0.4$ & $\sim 1.9 $ & -  &  $37.3$ & $1-3$ & P &  \small{1, 2, 28} \\
        NGC 2024   & $0.4$& $\sim 2.1$ & -  & $\sim 38.1$ & $0.5{-}1$ & P/DD & \small{3, 4, 5, 6, 7, 28} \\
          $\sigma$ Orionis  & $0.4$& 2.2  & 2.7  & $38.3$  & $3{-}5$ & P/DD & \small{8, 9, 10, 11, 12, 28, 39} \\
         ONC  & $0.4$ & $3.6$ & $4.1$ &  $38.7$ & $1{-}3$ & P/DD &  \small{13, 14, 15, 16, 28} \\
        Pismis 24  & $1.7$  & $\sim 2.8$ & $\sim3.8$  & $39.8$   & $1{-}1.5$ & IDL & \small{17, 18, 37} \\
        Quintuplet & $8.0$ & $4.0$ & $2.7$ & $39.8$  & $3-5$ & IDL &  \small{19, 20, 34}\\
        Trumpler 14  & $2.8$ & $3.6$ & $4.3$  & $39.9$ & $0.5{-}3$ & IDL & \small{21, 22, 23, 38, 40} \\
        Cygnus OB2  & $1.6$  & $4.2$ & $1.3$ & $40.0$   & $1{-}3$ & IDL & \small{24, 25, 26, 27, 41} \\
        NGC 3603 & 6.7 &  $4.1$ & $4.8$  & $40.3$  & $0.5{-}1$ & IDL & \small{28, 29, 30, 41}\\
        Arches & $8.0$  & $4.7$ & $5.3$ & $40.5$  & $2{-}3$ & IDL & 31, 32, 33, 34 \\
        Westerlund 1 & 3.8 & $4.8$ & $5.0$  & $40.6$ & $3-4$ & - & 35, 36, 41 
    \end{tabular}
    \caption{{Properties of some notable SFRs that show evidence of external depletion, except Westerlund 1 which we include since it is the subject of an upcoming JWST campaign \citep{JWST_Guarcello}. This compilation is not complete, nor necessarily unbiased. We include only SFRs that exhibit strong evidence of external photoevaporation of the discs, as well as well-studied properties in terms of mass and age.} Columns from left to right are: name of the SFR, heliocentric distance, total stellar mass, central density, total FUV luminosity, age, type of evidence for external disc depletion and references. FUV luminosity is calculated using the luminosity of the most massive members at an age of $1$~Myr. The evidence for external dispersal in each region is listed as proplyds/winds (P), dust/outer disc depletion (DD) or shortened inner disc lifetime (IDL). \\ \small{References -- 1: \citet{Peterson08} 2: \citet{Kim16}, 3: \citet{Skinner03}, 4: \citet{Bik03}, 5: \citet{Levine06}, 6: \citet{vanTerwisga20}, 7: \citet{Haworth21}, 8: \citet{Caballero08}, 9: \citet{Oliveira02} 10: \citet{Oliveira06}, 11:  \citet{Mauco16}, 12: \citet{Ansdell17}, 13: \citet{Hillenbrand98b}, 14: \citet{Palla99} \citet{Beccari17}, 15: \citet{Eisner18}, 16: \citet{vanTerwisga19}, 17:  \citep{Fang12}, 18: \citet{JWST_RamirezTannus}, 19: \citet{Figer99a}, 20: \citet{Stolte15}, 21:  \citet{Sana10}, 22: \citet{Preibisch11}, 23: \citet{Preibisch11b},  24: \citet{Hanson03}, 26: \citet{Wright15}, 27: \citet{Guarcello16}, 28: \citet{Grosschedl18}, 29: \citet{Stolte04}, 30: \citet{Harayama08}, 31: \citet{Najarro04}, 32: \citet{Espinoza09}, 33: \citet{Harfst10}, 34: \citet{Stolte15}, 35: \citet{Mengel07}, 36: \citet{JWST_Guarcello}, 37: \citet{Getman14a}, 38: \citet{Smith10}}, 39: \citet{Schaefer16}, 40: \citet{Cantat-Gaudin20}, 41: \citet{Rate20}}
    \label{tab:SFR_props}
\end{table}

\subsubsection{Disc survival fractions}
\label{sec:inner_disc}

Despite the numerous difficulties, many studies have reported correlations between disc lifetimes and local FUV flux. One of the earliest was \citet{Stolte04}, who found an increase from $20\pm 3$~percent in the central $0.6$~pc of NGC 3603, increasing to $30\pm 10$~percent at separations of $\sim 1$~pc from the centre. Later, \citet{Balog07} presented a \textit{Spitzer} survey of NGC 2244 \citep[of age $\sim {2}$~Myr -- ][]{Hensberge00, Park02} found a marginal drop off in the disc hosting stars for separations $d< 0.5$~pc from an O star ($ 27\pm 11$~percent) versus those at greater separations ($45\pm 6$~percent). \citet{Guarcello07} obtained similar results for NGC 6611 (of age $\sim 1$~Myr), finding $31.1\pm 4$~percent survival in their lowest bolometric flux bin, versus $16\pm 3$~percent in their highest \citep[see also][]{Guarcello09}. Similar evidence of shortened inner disc lifetimes have been reported in Arches \citep{Stolte10, Stolte15}, Trumpler 14, 15 and 16 \citep{Preibisch11},  NGC 6357 \citep[or Pismis 24 --][]{Fang12} and the comparatively low density OB association Cygnus OB2 \citep[][]{Guarcello16}. {We highlight that correlations between location and inner disc fraction are not found ubiquitously -- for example, \citet{Roccatagliata11} do not recover evidence of depleted discs in the central $0.6$~pc of IC 1795. We return to discuss some of these cases in greater detail below.}

Interestingly, the study by \citet{Fang12} of Pismis 24 revealed not only a correlation of disc fraction with FUV flux, but also a stellar mass dependent effect wherein  disc fractions are found to be lower for higher mass stars. While this is similar to the case of non-irradiated discs \citep{Ribas15}, it is the opposite of what might be expected from the dependence of $\dot{M}_\mathrm{ext}$ on the stellar mass; lower mass stars are more vulnerable to externally driven winds due to weaker gravitational binding of disc material \citep[e.g.][]{Haworth18b}. This finding, if generally true for irradiated disc populations, would therefore put constraints on how other processes in discs scale with stellar host mass. For example, more rapid (viscous) angular momentum transport for discs around higher mass stars would sustain higher mass loss rates for longer, due to the mass flux balance in the outer disc (see discussion in Section~\ref{sec:gas_evolution}). However, although accretion rates correlate with stellar mass \citep[][]{Herczeg08,Manara17}, this does not necessarily imply faster viscous evolution for discs with high mass host stars \citep{Somigliana22}; thus the physical reason for the \citet{Fang12} findings remain unclear. Whether discs around high mass stars are generally depleted faster than those around low mass stars may be further confirmed by the upcoming JWST GTO program by \citet{JWST_Guarcello}, who aim to map out disc fractions and properties in Westerlund 1 for stars down to brown dwarf masses. This large dataset should allow to control both for stellar mass and location in the starburst cluster.

All of the regions mentioned above, in which evidence of inner disc dispersal has been inferred, share the property that they are sufficiently massive to host many O stars. A notable example for which shortened disc lifetimes are not observed is in the ONC. Despite rapid mass loss rate for the central proplyds, \citet{Hillenbrand98} inferred a high disc fraction of $\sim 80$~percent. This may be due to {a} large spread in ages \citep[typically estimated at $\sim 1{-}3$~Myr -- e.g.][]{Hillenbrand97, Palla99, Beccari17}, such that the resultant dynamical evolution \citep{Kroupa18, Winter19b} leads to the central concentration of young stars \citep{Hillenbrand97, Getman14, Beccari17}, as discussed in Section~\ref{sec:stellar_dynamics}. {However, as always, interpreting the luminosity spread as a physical spread in ages is not so simple. Luminosity spreads may result from disc orientation, accretion, multiplicity and extinction, while the latter two effects may contribute to systematic gradients in the cluster.} High survival rates may also reflect the inefficiency of inner disc clearing of photoevaporative winds in intermediate $F_\mathrm{FUV}$ environments, which would hint at inefficient angular momentum transport at large radii (due to dead zones, for example).

{Some other studies have also obtained null results when trying to find evidence of spatial gradients in the fraction of stars with NIR excess within SFRs.  Studies searching for spatial correlations in NIR excess fraction are useful because they are not subject to the same large uncertainties in the stellar ages. However, for studies of this kind the question of membership criteria to the region is of utmost importance, since given a constant number of foreground/background contaminants then the outer regions should be more affected, presumably suppressing the apparent disc fraction. One example of a search for these correlations is that of \citet{Richert15}, who found an absence of any correlation in the MYStIX catalogue \citep{Povich13} of NIR sources across several O star hosting SFRs. The methodology of \citet{Richert15} differed from other studies in that they adopted the metric of the aggregate density of disc-hosting or disc-free stars around O stars, rather than distances of these lower mass stars to their nearest O star. Similarly, \citet{Mendigutia22} used \textit{Gaia} photometry and astrometry to compare relative disc fractions in a sample of SFRs, finding no spatial correlations. This study did not use FUV flux/O star proximity, but binned stars into the inner $2$~pc versus the outer $2-20$~pc for each region. Both studies have not attempted an absolute measure of disc survival fractions, but relative numbers. They highlight that environmental depletion of inner discs is not necessarily ubiquitous in high mass SFRs. Physical considerations such as age and dynamics may also play an important role in  determining whether correlations in disc survival fractions can be uncovered.  }

\begin{figure}
    \vspace{-0.5cm}
    \centering
    \includegraphics[width=0.9\textwidth]{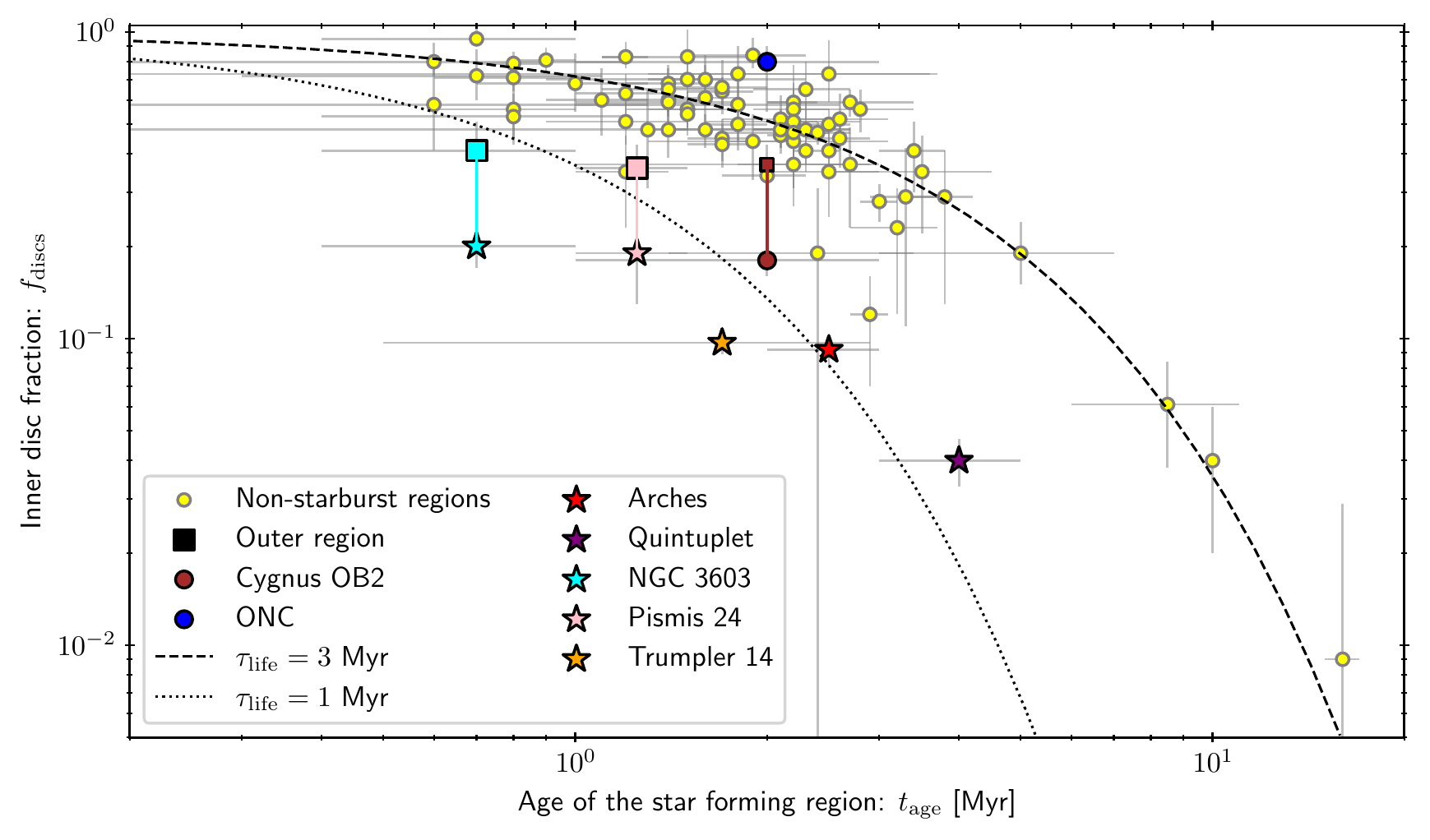}
    \caption{The fraction of surviving inner discs as a function of age of star forming regions. Yellow points show a compilation of various star forming regions using data from \citet{Richert18} and \citet{Mamajek09}. We further show some specific SFRs discussed in the text and listed in Table~\ref{tab:SFR_props}. Massive and dense `starburst' clusters are shown as star symbols. Where a local gradient in the disc fraction has been reported in the literature, we show the outer regions as a square symbol. We show the fraction of discs following $f_\mathrm{discs} = \exp(-t_\mathrm{age}/\tau_\mathrm{life})$ for $\tau_\mathrm{life} = 1$ and $3$~Myr (dotted and dashed lines respectively). {An important caveat for this figure is that disc fractions in distant SFRs are based on samples that may include fewer low mass stars, which typically exhibit longer disc life-times.} }
    \label{fig:lifetimes}
\end{figure}

In Figure~\ref{fig:lifetimes} we show the disc survival fraction versus SFR age for a composite sample, including a number of the massive regions discussed in this section. {This figure is subject to the caveats discussed above, and particularly that observed samples of stars in massive, distant regions may be biased towards higher stellar masses than close by regions. Setting aside the caveats, Figure~\ref{fig:lifetimes} appears to} demonstrate a shortening of disc lifetimes across numerous massive SFRs. Low mass SFRs typically have $\tau_\mathrm{life}\approx 3$~Myr, while the most massive SFRs may have $\tau_\mathrm{life} \lesssim 1$~Myr. This shortening of disc lifetimes has so far only been found in regions with a total FUV luminosity $L_\mathrm{FUV}\gtrsim 10^{39}$~erg~s$^{-2}$ (see Table~\ref{tab:SFR_props}). In the case of many of these massive regions, the shortening of disc lifetimes can also be seen in a local gradient, for which stars that are far from an O star have a greater probability of hosting a disc. {This is more difficult to explain by variations in the stellar masses of surveyed stars, since dynamical mass segregation occurs on much longer time-scales \citep[e.g.][]{Bonnell98b}.} This suggests that external photoevaporation can shorten inner disc disc lifetimes. If this is the case, it implies that if gas giants form in these environments, they must do so early \citep[as suggested by some recent results -- e.g.][]{Segura-Cox20, Tychoniec20}. 



\subsubsection{Outer disc properties}
\label{sec:outer_disc}

In recent times, ALMA has become the principle instrument for surveying outer disc properties in local star forming regions. Nonetheless, prior to this revolution a number of studies had already demonstrated, by degrees, the role of external photoevaporation in sculpting the outer regions of protoplanetary discs. For example, in the ONC, \citet{Henney98} and \citet{Henney99} used HST images and \textit{Keck} spectroscopy to demonstrate the rapid mass loss rates up to $\sim 10^{-6}\, M_\odot$~yr$^{-1}$ of several proplyds in the core of the ONC, and their concentration around the O star, $\theta^1$C. Later, \citet{Mann14} used SMA data to show a statistically significant dearth of discs with high dust masses close to $\theta^1$C \citep[see also][]{Mann10}. This result has since been confirmed via ALMA observations towards the core \citep{Eisner18} and outskirts \citep{vanTerwisga19} of the region. This is a clear demonstration that external photoevaporation depletes the dust content, although it remains unclear if this is via instigating rapid radial drift \citet{Sellek20} or via entrainment in the wind \citep{Miotello12}. \citet{Boyden20} also find the gas component of the disc is more compact than in other SFRs, and correlated with distance to $\theta^1$C, suggestive of truncation by external photoevaporation. 

Dust depletion has also been inferred in $\sigma$ Orionis. Here the dominant UV source is the $\sigma$~Ori multiple system, for which the most massive component has a mass $17\,M_\odot$ \citep{Schaefer16}. From \textit{Herschel} spectral energy distributions of discs in this region, \citet{Mauco16} found  evidence of external depletion in the abundance of compact discs ($R_\mathrm{out}<80$~au) consistent with the models of \citet{Anderson13}. The authors also found evidence of ongoing photoevaporation in one disc exhibiting forbidden [Ne\,\textsc{ii}] line emission \citep[see also][]{Rigliaco09}. Later, \citet{Ansdell17} presented an ALMA survey that uncovered an absence of discs with high dust masses at distances $\lesssim 0.5$~pc from $\sigma$~Ori, consistent with models of dynamical evolution and disc depletion \citep{Winter20b}. 

In NGC 2024, an early survey of the discs with the SMA presented by \citet{Mann15} revealed a more extended tail of high mass discs than those of other SFRs. The authors suggested this is indicative of a young population of discs that had not (yet) been depleted by external photoevaporation. However, \citet{vanTerwisga20} presented an ALMA survey of the region and found evidence of two distinct disc populations. The eastern population is very young ($\sim 0.2{-}0.5$~Myr old) and still embedded close to a dense molecular cloud that may shield them from irradiation. The western population is slightly older, exposed to UV photons from the nearby stars IRS 1 and IRS 2b. The western discs are lower mass than those in the east and those in similar age SFRs, which is probably due to external depletion. This conclusion is supported by the subsequent discovery of proplyd (candidates) in the region \citep{Haworth21}.

We have focused here on regions where evidence of external depletion appears to be present. It is challenging to convincingly demonstrate the converse by counter example, particularly considering the many potential issues discussed above (e.g. accurate stellar aging, projection effects and dynamical histories). As an example, in an ALMA survey of discs in $\lambda$~Orionis, \citet{Ansdell20} reported no correlation between disc dust mass and separation from the OB star $\lambda$~Ori. As discussed by the authors, a number of explanations for this are possible. One explanation is that, given the older age ($\sim 5$~Myr) of the region, the discs may have all reached a similar state of depletion, with spatial correlations washed out by dynamical mixing \citep[e.g.][]{Parker21}. Meanwhile, in a survey of non-irradiated discs, \citet{vanTerwisga22} demonstrated that discs in SFRs of similar ages appear to have similar dust masses, suggesting that any observed depletion in higher mass SFRs is probably the result of an environmental effect. 

\subsubsection{Summary and omissions}

For quick reference, in Table~\ref{tab:SFR_props} we summarise the properties of some local SFRs in which some observational evidence of external photoevaporation has been uncovered, and note some references for further reading. {However, we emphasise that this is not a complete or representative list; we have chosen SFRs that exhibit strong evidence of external photoevaporation and well studied stellar populations. Some notable exclusions include Trumpler 13/IC 1396A (the Elephant Trunk nebula) and IC 1805 (the Heart nebula). Trumpler 13 contains an  O6.5V and O9V central binary \citep{Tokovinin97}, and the young stellar objects exhibit marginal spatial gradients in the disc survival fraction \citep{Getman12}. However, given the abundance of class I discs that indicate ongoing star formation, this may also originate from heterogeneous ages \citep{Silverberg21}. Similarly, in the $2-3.5$~Myr old cluster IC 1805, with mass $M_\mathrm{tot}\approx 2700\, M_\odot$ ($L_\mathrm{FUV} \approx 10^{39.4}$~erg~s$^{-1}$), any gradient in the surviving disc fraction is marginal \citep{Sung17}. A complex dynamical evolution, involving collapse and ejection of members \citep{Lim20}, may confuse signatures of external disc depletion.}

\subsection{Missing aspects of disc evolution and planet formation in irradiated environments}

Notably, we have not discussed chemistry in this section. The chemistry of discs \citep[e.g.][]{Bruderer12, Kamp17} and the planets they produce \citep[e.g.][]{Cridland16, Cridland19, Bitsch20} is a complex function of the stellar metallicity and local disc temperature. For irradiated discs, temperatures increase in the outer regions and disc surface layers with respect to isolated discs, altering the chemistry in these regions, although not necessarily in the disc mid-plane \citep[e.g.][]{Nguyen02, Walsh13}. However, how this alters planet formation and chemistry may dependent on disc evolution, and this remains a matter for future investigation. This issue may also soon be addressed empirically via JWST observations, with the GTO programme by \citet{JWST_RamirezTannus} aiming to probe the chemistry of discs of similar age but varying FUV flux histories in NGC 6357. 

Apart from chemistry, this section has highlighted the many gaps in the understanding of the role of external photoevaporation in shaping the evolution of the disc and the formation of planets. From both theory and observations, we understand that gas disc lifetimes are shortened by external photoevaporation. This shortened lifetime of the gas disc alone should be enough to influence the formation and disc induced migration of giant planets. Meanwhile, how the aggregation of dust produces planets in irradiated discs has only just started to be addressed \citep[e.g.][]{Ndugu18, Sellek20}. When coupled with the apparent prevalence of external photoevaporation in shaping the overall disc population, the role of star formation environment must be considered as a matter of urgency for planet population synthesis efforts \citep[e.g.][]{Alibert05, Ida05, Mordisini09, Emsenhuber21}. Meanwhile, the connection between stellar rotation periods and premature disc dispersal via external photoevaporation may in future offer a window into the birth environments of mature exoplanetary systems up to $\sim 1$~Gyr after their formation \citep{Roquette21}.

\begin{mdframed}
\vspace{-0.8cm}
\subsection{Summary and open questions for irradiated protoplanetary disc evolution}
While many aspects of disc evolution in irradiated evironments remain uncertain, we can be relatively confident in the following conclusions:
\begin{enumerate}
    \item External photoevaporation depletes (dust) masses and truncate the outer disc radii in regions of high FUV flux, at least for  $F_\mathrm{FUV} \gtrsim 10^3\, G_0$. 
    \item Inner disc lifetimes appear to be shorted for discs in regions where the total FUV luminosity $L_\mathrm{FUV} \gtrsim 10^{39}$~erg~s$^{-1}$, while does not appear to be strongly affected for regions with lower $L_\mathrm{FUV}$. 
    \item Dust evolution, low mass planet formation and giant planet formation all have the potential to be strongly influenced by external photoevaporation through changes in the temperature, mass budget, and time available for planet formation. 
\end{enumerate}

\noindent Some of the many open questions in this area include:
\begin{enumerate}
    \item \textit{Do the externally driven mass loss rates in photoevaporating discs balance with accretion rates, as expected from viscous angular momentum transport?}
    \item \textit{If angular momentum transport is extracted via MHD winds rather than viscously transported, how does this influence the efficiency of external photoevaporation? }
    \item \textit{Is the lack of correlation shortened disc lifetime with FUV flux in intermediate mass environments related to a dead-zone or similar suppression of angular momentum transport?  }
    \item \textit{Is the observed dust depletion in near to O stars due to entrainment in the wind or rapid inward drift? }
    \item \textit{In relation to this, how do external winds influence dust trapping and the onset of the streaming instability?}
    \item \textit{Conversely, how does the trapping of dust influence the dust depletion and dust-to-gas ratio in discs? }
    \item {\textit{How does disc chemistry vary between isolated and externally FUV irradiated discs?}}
    \item \textit{How does the local distribution of FUV environments influence the planet population as a whole?}
\end{enumerate}
\end{mdframed}

\section{Demographics of star forming regions}
\label{sec:SFRs}
\label{sec:sf_physics}

In this section, we consider the properties of star forming regions and the discs they host from an observational and theoretical perspective. We are interested in understanding: `what is the role of external photoevaporation of a typical planet-forming disc in the host star's birth environment?' The degree to which  protoplanetary discs are influenced by external irradiation depends on exposure to EUV and FUV photons. Hence the overall prevalence of external photoevaporation for discs (and probably the planets that form within them), depends critically on the physics of star formation and the demographics of star forming regions.

\subsection{OB star formation} 
\label{sec:OBstar}
OB stars have spectral types earlier than B3, and are high mass stars with luminosities $> 10^3\, L_\odot$ and masses $>5 \, M_\odot$. These stars are responsible for shaping a wide range of physical phenomena, from molecular cloud-scale feedback on Myr timescales \citep{Krumholz14, Dale15} to galactic nucleosynthesis over cosmic time \citep{Nomoto13}. The radiation feedback of these stars on their surroundings already plays a significant role during their formation stage, where stars greater than a few $10\,M_\odot$ in mass must overcome the UV radiation pressure problem \citep{Wolfire87}. Forming OB stars may do this through monolithic turbulent collapse \citep[e.g.][]{McKee03}, early core mergers \citep[e.g.][]{Bonnell98}, or competitive accretion \citep[e.g.][]{Bonnell01} -- see \citet{Krumholz15} for a review. 

Two questions regarding OB stars are important with respect to local circumstellar disc evolution:  `when do they form?' and `how many of them are there?'. The former determines when the local discs are first exposed to strong external irradiation. The latter determines the strength and, importantly, the spatial uniformity of the UV field (we discuss this point further in Section~\ref{sec:stellar_dynamics}).  

The question of the timescale for formation of massive stars is empirically tied to the frequency of ultra-compact HII (UCHII) regions. These regions are the small (diameters $\lesssim 0.1$~pc) and dense (HII densities $\gtrsim 10^4 $~cm$^{-3}$) precursors to massive stars. During the main sequence lifetime of the central massive star, the associated HII region will evolve from the embedded ultra compact state to a diffuse nebula. In the context of the surrounding circumstellar discs, the lifetime of a UCHII region represents the length of time for which the surroundings of massive stars are efficiently shielded from UV photons. These lifetimes can be inferred by comparing the the number of main-sequence stars to the number of observed UCHII regions, yielding timescales of a few $10^5$~yrs \citep{Wood89,Mottram11}. The star-less phase prior to excitation of the UCHII region is short \citep[$\sim 10^4$~yr --][]{Motte07, Tige17}, hence the UCHII lifetime represents the effective formation timescale for massive stars in young clusters and associations. This formation timescale is much shorter than the typical lifetime for protoplanetary discs evolving in isolation \citep[$\sim 3$~Myr -- e.g.][]{Haisch01, Ribas15}. Therefore, in regions with many OB stars we expect the local discs to be irradiated throughout this lifetime. However, this need not be the case when the number of OB stars is $\sim 1$, in which case the time at which an OB star forms may be statistically determined by the spread of stellar ages in the SFR. We discuss this point further in Section~\ref{sec:stellar_dynamics}.

The relative contribution of stars of a given mass to the local UV radiation field depends on the relative numbers of stars with this mass, or the initial mass function (IMF), $\mathrm{d} n /\mathrm{d} \log m_*$. We can write the mean total luminosity $\langle L_\mathrm{SFR} \rangle$ of a star forming region (SFR) with $N$ members: 
\begin{equation}
\label{eq:L_mean}
   \langle L_\mathrm{SFR} \rangle = N\langle L \rangle = N \int \frac{\mathrm{d} n }{\mathrm{d} \log m_* } \cdot L(m_*) \,\mathrm{d} \log m_* ,
\end{equation}where $L$ is the luminosity of a single star. Hence, to understand the contribution of massive stars to the UV luminosity, we are interested in the shape of the IMF. The IMF in local SFRs exhibits a peak at stellar masses $m_* \sim 0.1{-}1\, M_\odot$ and a steep power law $\mathrm{d} n /\mathrm{d} \log m_* \propto m_*^{-\Gamma}$ with $\Gamma \approx 1.35$ at higher masses \citep[see][for a review]{Bastian10}. This power-law appears reasonable at least up to $m_* \sim 100\, M_\odot$ \citep{Massey98, Espinoza09}. 

\begin{figure}
    \centering
    \subfloat{\includegraphics[width=0.5\textwidth]{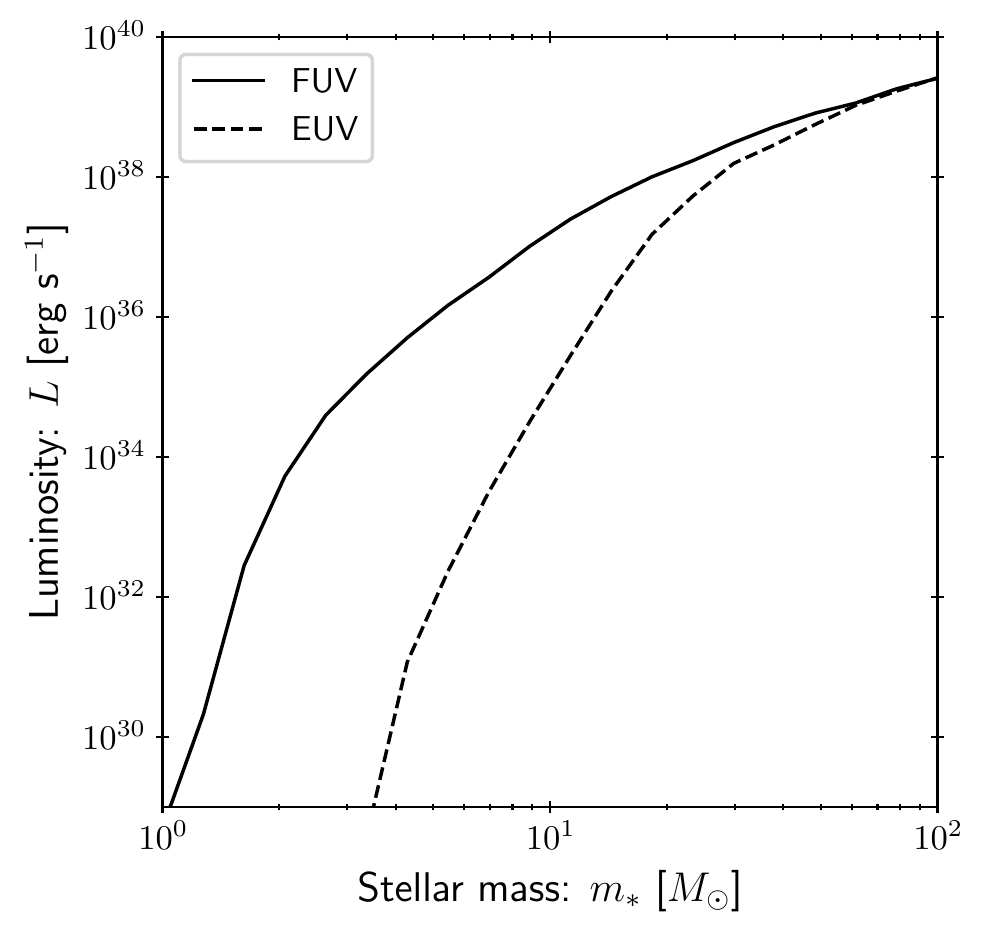}}
    \subfloat{\includegraphics[width=0.5\textwidth]{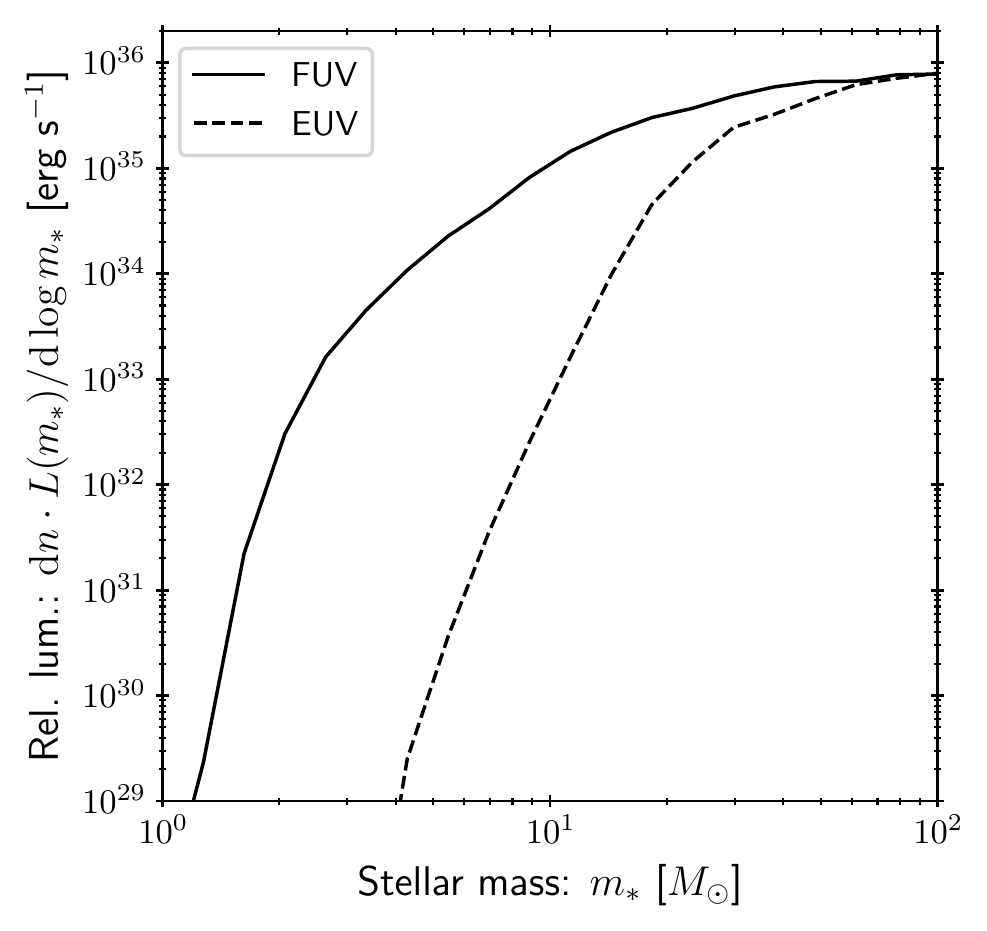}}
    \caption{\textit{Left panel:} stellar FUV and EUV luminosity as a function of stellar mass. The total luminosity is calculated using the effective temperature and total luminosity from the model grids of \citet{Schaller92}, with metallicity $Z=0.02$ and output time closest to $1\,$~Myr. We apply the atmosphere models of \citet{Castelli03} to give the wavelength-dependent luminosity. \textit{Right panel:} The relative contribution of stars of a given mass to the total luminosity averaged across the IMF. Here we use a log-normal \citet{Chabrier03} sub-solar IMF, and power-law with $\Gamma=1.35$ for $m_*>1\, M_\odot$. }
    \label{fig:stellar_lums}
\end{figure}

We combine the mass-dependent luminosity for young stars, as in the left panel of Figure~\ref{fig:stellar_lums}, with the \citet{Chabrier03} IMF to produce the right panel of Figure~\ref{fig:stellar_lums}. Here we multiply the stellar mass dependent luminosity by the IMF, as in the integrand on the RHS of equation~\ref{eq:L_mean}, to yield the average contribution of stars of certain mass to the FUV and EUV luminosity of a SFR. Note that this is only accurate for a very massive SFR, where the IMF is well sampled. However, the figure demonstrates that for low mass SFRs -- which here means where the IMF is not well sampled for stellar masses $m_*\gtrsim 30\, M_\odot$ -- both EUV and FUV luminosities are dominated by individual massive stars, rather than many low mass stars. In the following, we consider this in the context of the properties of local star forming regions. 

\subsection{Demographics of star forming regions}
\label{sec:demographics_SFRs}

We now consider the FUV flux experienced by protoplanetary discs in typical SFRs. We will focus on FUV, since as discussed in Section~\ref{sec:gas_evolution}, the FUV is expected to dominate disc evolution over the lifetime of the disc. The word `typical' in the context of SFRs is in fact dependent on the local properties within a galaxy, and we here refer exclusively to the solar neighbourhood (at distances $\lesssim 2$~kpc from the sun). A number of studies have approached this problem in differing ways \citep{Fatuzzo08,Thompson13,Winter20a, Lee20}, however all such efforts require estimating the statistical distribution ${\mathrm{d} n_\mathrm{SFR}}/{\mathrm{d} N}$ in the number of members $N$ of SFRs, which dictates how many OB stars there are and therefore the local UV luminosity. For example, \citet{Fatuzzo08} used two different approaches to this problem. One was to directly take the distribution from the list of nearby SFRs compiled by \citet{Lada03}. While this is the most direct, it suffers from small number statistics for massive SFRs. The other approach by \citet{Fatuzzo08} is to assume a log-uniform distribution in the number of stars existing in a SFR with a number of members $N$, truncated outside of a minimum $N_\mathrm{min}=40$ and maximum $N_\mathrm{max}=10^5$. Alternatively, one can produce a similar distribution using a smooth \citet{Schechter76} function \citep[see e.g.][]{Gieles06b}:
\begin{equation}
\label{eq:SFRmfunc}
    \frac{\mathrm{d} n_\mathrm{SFR}}{\mathrm{d} N} \propto  N^{-\beta} \exp\left(\frac{-N}{N_\mathrm{max}} \right) \exp\left(-\frac{N_\mathrm{min}}{N} \right),
\end{equation}
where $\beta\approx 2$ is expected from the hierarchical collapse of molecular clouds \citep{Elmegreen96} and is consistent with empirical constraints \citep[e.g.][]{Gieles06b, Fall12, Chandar15}, as well as the log-uniform distribution adopted by \citet[][n.b. that to obtain the fraction of \textbf{stars}, one must multiply the mass function described by equation~\ref{eq:SFRmfunc} by a factor $N$]{Fatuzzo08}. 

In order to interpret equation~\ref{eq:SFRmfunc}, we further need to estimate $N_\mathrm{min}$ and $N_\mathrm{max}$. The upper limit $N_\mathrm{max}$ can be inferred empirically \citep{Gieles06b, Gieles06}, or by appealing to the theoretical limits. As elucidated by \citet{Toomre64}, the origin of the maximum mass of a SFR can be understood in terms of the maximum size of a region that can overcome the galactic shear (Toomre length) and local kinetic pressure (Jeans length). This length scale, the Toomre length, may then be coupled to a local surface density of a flattened disc to obtain a maximum mass for a SFR. However, in the outer regions of the Milky Way, the stellar feedback in massive SFRs can interrupt star formation and further reduce $N_\mathrm{max}$. Adopting the simple model presented by \citet{ReinaCampos17} for this limit with a typical stellar mass $m_*\sim 0.5\,M_\odot$, this yields a maximum $ N_\mathrm{max} \sim 7 \times 10^4$ for all SFRs (not necessarily gravitationally bound). This is broadly consistent with the most massive local young stellar clusters and associations \citep{PortegiesZwart10}. 

Meanwhile, for the minimum number of members in a SFR, \citet{Lamers05} used the sample of \citet{Kharchenko05} to infer a dearth of SFRs with $N<N_\mathrm{min} \sim 280$. Such a lower limit might be understood as the point at which the lower mass end of the hierarchical molecular cloud complex merges due to high star formation efficiency and long formation timescales \citep{Truijillo19}. However, obtaining constraints on $N_\mathrm{min}$ in general is made difficult by the lack of completeness in surveys of low mass stellar aggregates, and it remains empirically unclear how $N_\mathrm{min}$ varies with galactic environment. Note that generally SFR demographics may depend on galactic environment, and therefore also vary over cosmic time \citep[see][for a review]{Adamo20}.

\begin{figure}
    \centering
    \includegraphics[width=\textwidth]{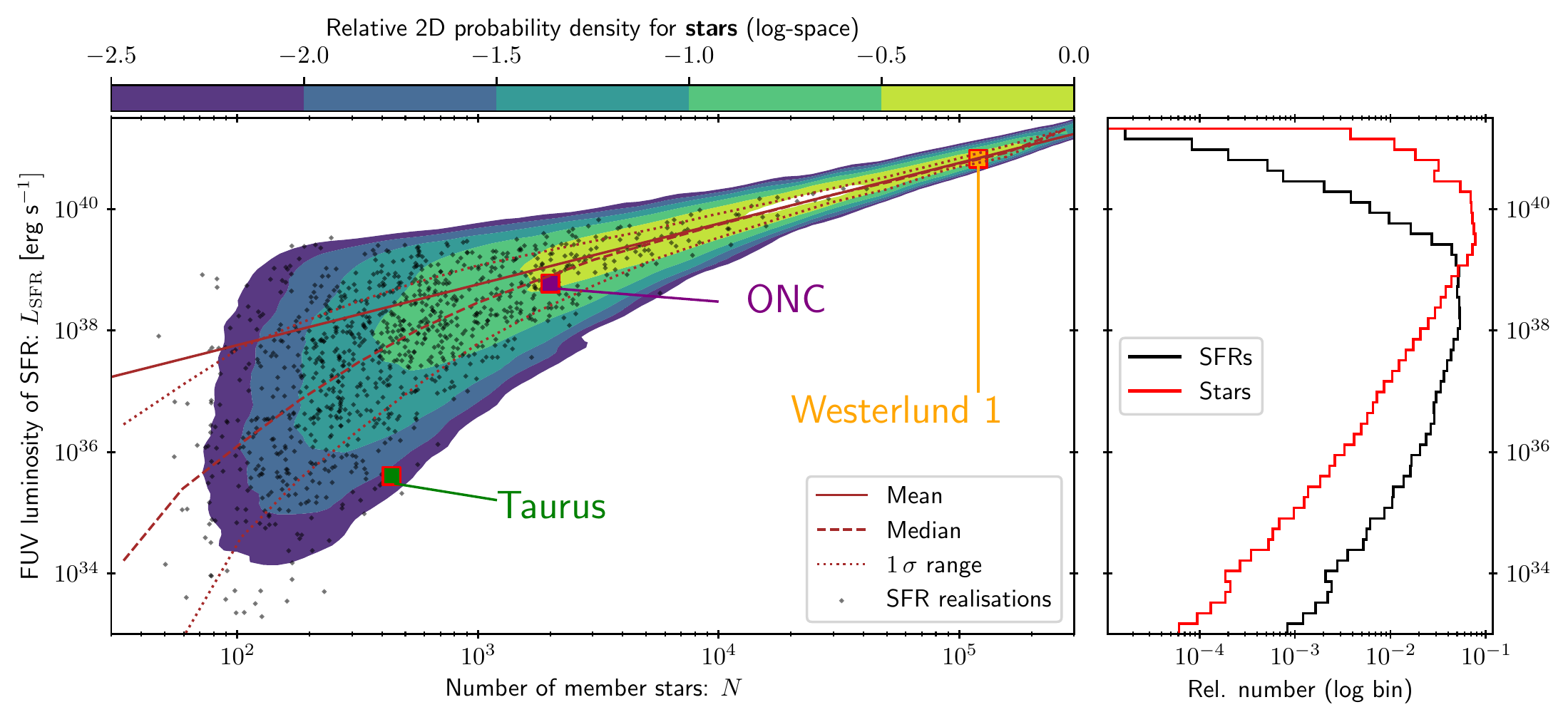}
    \caption{\textit{Left panel:} the FUV luminosities at an age of $1$~Myr of $6\times 10^4$ realisations drawing star forming regions (SFRs), with the number of members drawn from equation~\ref{eq:SFRmfunc} \citep[cf.][Figure 5b]{Fatuzzo08}. We draw individual stellar masses from a \citet{Chabrier03} IMF in the range $0.08{-}100\, M_\odot$. The black points represent a subset of 1000 of the synthetic SFRs. The solid brown line follows the mean luminosity as a function of the number of members (equation~\ref{eq:L_mean}), the dashed brown line is the median, while the dotted lines bound the $14^\mathrm{th}{-}86^{\mathrm{th}}$ percentile range. The colour bar shows an estimate for the relative density of \textit{stars} (i.e. density of SFRs multiplied by the number $N$ of members) in logarithmic space using a Gaussian kernel density estimate with a bandwidth $0.1$~dex. We estimate the numbers of stars and total luminosity of three local SFRs: Taurus (green), the ONC (purple) and Westerlund 1 (orange). \textit{Right panel:} the corresponding distribution in the number of stars (red) and SFRs (black) per logarithmic bin in total FUV luminosity space.    }
    \label{fig:demographics}
\end{figure}

With the above considerations, we can now generate the distribution of UV luminosities experienced by stars in their stellar birth environment. We adopt $N_\mathrm{min}=280$ and $N_\mathrm{max}=7\times10^4$ and couple equation~\ref{eq:SFRmfunc} with the stellar mass dependent luminosity discussed in Section~\ref{sec:OBstar}. The resultant FUV luminosity distribution when drawing $6\times 10^4$ SFRs, each with $N$ members drawn from equation~\ref{eq:SFRmfunc}, is shown in Figure~\ref{fig:demographics}. For context, we estimate the FUV luminosity in Taurus using the census of members by \citet{Luhman18}, assuming the local luminosity is dominated by four B9 and three A0 stars in that sample of 438 members. We also add the ONC, whose total UV luminosity is dominated by the O7-5.5 type star $\theta^1$C, with a mass of $\sim 35 \, M_\odot$ \citep[e.g.][]{Kraus09, Balega14}, and Westerlund 1 which has a total mass of $\sim 6\times 10^4 \, M_\odot$ \citep{Mengel07, Lim13} and thus a well sampled IMF. Regions like Taurus with only a few hundred members are common, and therefore we expect to find such regions nearby. However, each one hosts $\sim 1/1000$ as many stars as a starburst cluster like Westerlund 1. Taurus thus represents an uncommon stellar birth environment in terms of the FUV luminosity. The most common type of birth environment for stars lies somewhere between the ONC and Westerlund 1, with FUV luminosity $\sim 10^{40}$~erg~s$^{-1}$ (at an age of $ 1$~Myr). 

In order to understand the typical flux experienced by a circumstellar disc, we further need to consider the typical distance between it and nearby OB stars -- i.e. the geometry of the SFR. To this end, \citet[][see also \citealt{Adams06}]{Fatuzzo08} and \citet{Lee20} use the \citet{Larson81} relation that giant molecular clouds and young, embedded SFRs have a size-independent surface density \citep[see also][]{Solomon87}; hence the radius $R$ of the SFR scales as $R \propto \sqrt{N}$ \citep{Adams06, Allen07}. However, this relationship does not hold for very young massive clusters  \citep[see][for a review]{PortegiesZwart10}, with several exhibiting comparatively high densities \citep[e.g. Westerlund 1 --][]{Mengel07}, while older clusters follow a shallower mass-radius relationship $R \propto N^\alpha$ with $\alpha \sim 0.2{-}0.3$ \citep{Krumholz19}. Instead of the Larson relation, \citet{Winter20a} attempt to relate SFR demographics to galactic-scale physics by using the lognormal density distribution of turbulent, high Mach number flows combined with a theoretical star formation efficiency. This yields the stellar density distribution in a given galactic environment. In massive local SFRs, the local FUV flux $F_\mathrm{FUV} \approx 1400 \, (\rho_*/10^3 M_\odot\,\rm{pc}^{-3})^{1/2} \, G_0$, for stellar density $\rho_*$, due to their radial density profiles \citep{Winter18b}. Using these two ingredients, it is possible to estimate the distribution of FUV fluxes experienced by stars in the solar neighbourhood from a mass function for SFRs. Despite the differences in the two approaches, both \citet{Fatuzzo08} and \citet{Winter20a} find that, neglecting extinction, young stars in typical stellar birth environments experience external FUV fluxes in the range $\langle F_\mathrm{FUV} \rangle \sim 10^3{-}10^4 \, G_0$. 

While these efforts produce some intuition as to the typical FUV radiation fields, this is not the end of the story for understanding how discs in SFRs are exposed to UV photons. In Sections~\ref{sec:extinction} and~\ref{sec:stellar_dynamics} we will discuss the role of interstellar extinction and the dynamical evolution of stars in SFRs. 

\begin{figure}
    \centering
    \includegraphics[width=\textwidth]{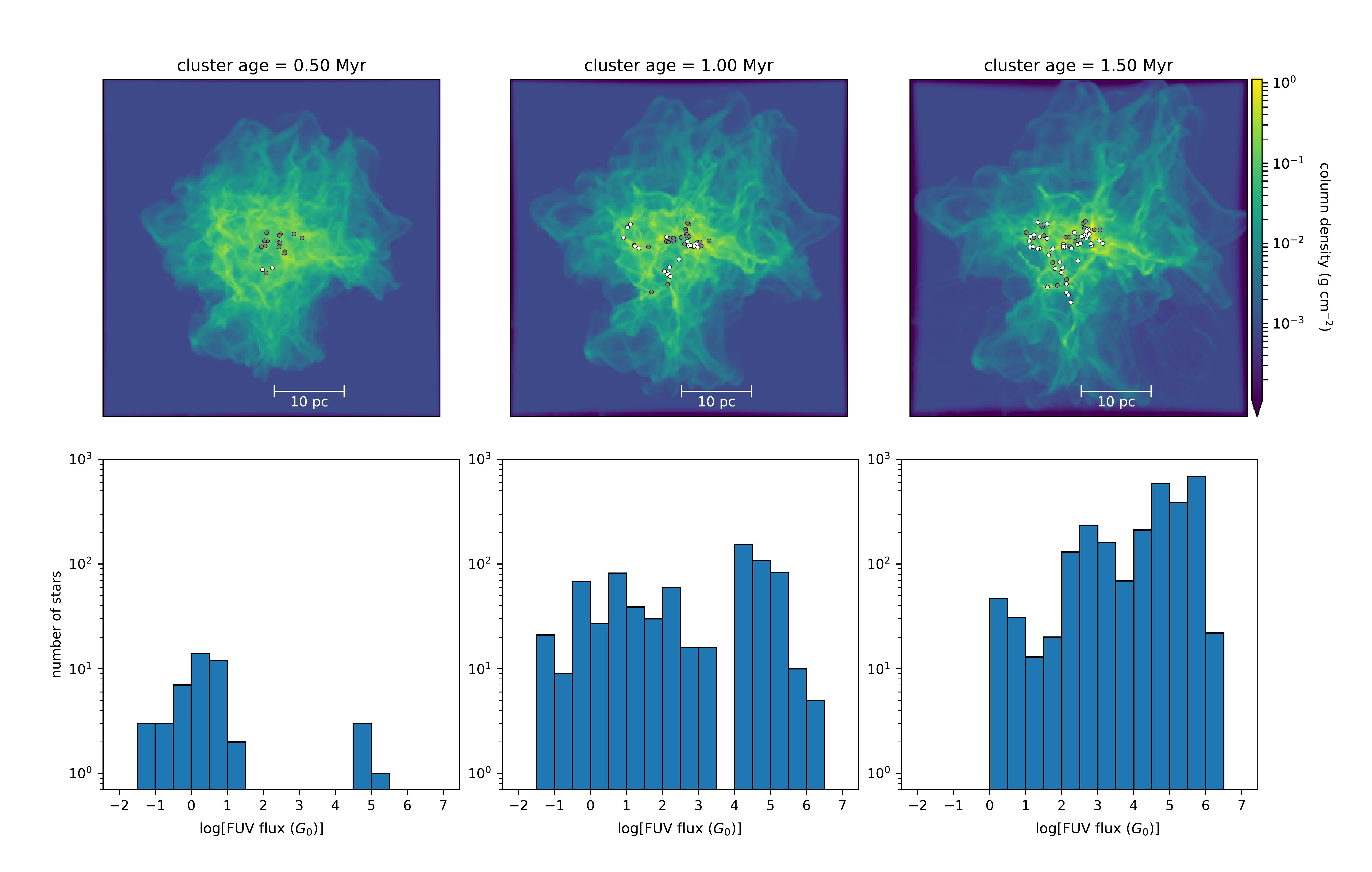}
    \caption{FUV flux experienced by stars formed in the hydrodynamic simulations of \citet{Ali21}, following the collapse of a molecular cloud of mass $10^5 \, M_\odot$ at time $0.5$, $1.0$ and $1.5$~Myr from left to right. \textit{Top row:} the gas column density and location of sink particles as points, with white points representing ionising sources. \textit{Bottom row:} histogram of the number of sources binned by instantaneous FUV flux \citep[see also][]{Qiao22}. Comparatively few stars are born in the early stages of the cluster formation, where attenuation of FUV photons is efficient.  }
    \label{fig:extinction}
\end{figure}
\subsection{Extinction due to the inter-stellar medium}
\label{sec:extinction}

After stars begin to form, residual gas and, more importantly, the dust in the inter-stellar medium (ISM) attenuates FUV photons such that circumstellar discs may be shielded from external photoevaporation. The column density required for one magnitude of extinction at visual magnitudes is $N_\mathrm{H}/A_\mathrm{V} = 1.8\times10^{-21}$~cm$^{-2}$~mag$^{-1}$ \citep{Predehl95}, and this can be multiplied by a factor $A_\mathrm{FUV}/A_\mathrm{V} \approx 2.7$ \citep{Cardelli89} to yield the corresponding FUV extinction. The main problem comes in estimating the column density between OB stars and the cluster members. Both \citet{Fatuzzo08} and \citet{Winter20a} approach this problem by adopting smooth, spherically symmetric density profiles with some assumed star formation efficiency to estimate this column density. However, even during the embedded phase this approach overestimates the role of extinction because a more realistic, clumpy geometry of the gas makes the attenuation inefficient \citep[e.g.][]{Bethell07}. In addition, stellar feedback from massive stars acts to drive gas outflows from the SFR even before local supernovae ignition \citep[e.g.][]{Lyon1980,Matzner02, Jeffreson21}, reducing the quantity of attenuating matter. 

Due to the above concerns, understanding the influence of extinction on shielding the young disc population requires direct simulation of feedback in the molecular cloud. To approach this problem, \citet{Ali19} simulated feedback from a single $34\, M_\odot$ mass star in a molecular cloud of mass $10^4 \, M_\odot$, similar to the conditions in the ONC. The authors find that many discs are efficiently shielded for the first $\sim 0.5$~Myr of evolution, while the discs may experience short drops in UV exposure at later stages. This result is somewhat dependent on cloud metallicity, since lower metallicity increases the efficiency of feedback by lengthening the cooling time \citep{Ali21}. 

If more than one massive star forms in a SFR, this further increases feedback efficiency and geometrically reduces the efficiency of extinction. \citet{Qiao22} investigated the role of feedback in the simulations of a molecular cloud of mass $10^5 \, M_\odot$ by \citet{Ali21}. The resultant FUV flux experienced by the stellar population is shown in Figure~\ref{fig:extinction}. For such a massive region with several O stars, lower mass disc-hosting stars typically experience extremely strong irradiation by the time they reach an age of $\sim 0.5$~Myr. Although extinction may be efficient for the stars that form early, the majority that form later are quickly exposed to $F_\mathrm{FUV} \sim 10^5 \, G_0$. Hence the embedded phase of evolution in massive SFRs can only shield discs early on. {While the FUV flux experienced by discs in real SFRs also depends on the density of the regions, inefficient shielding may partially explain why disc life-times appear to be shortened in SFRs with several O stars (see Table~\ref{tab:SFR_props}).} Any giant planet formation in such a region must therefore initiate early and may be strongly influenced by their environment (see Section~\ref{sec:giant_planets}).

\subsection{The role of stellar dynamics}
\label{sec:stellar_dynamics}

In some instances, the FUV exposure history of stars may strongly depend on the dynamical evolution of the SFR. This is particularly true only one dominant OB star is present. For example, the difference in $F_\mathrm{FUV}$ for a disc at a separation $d = 0.05$~pc from a single massive star (as for some of the brightest proplyds in the ONC) and those at $d=2$~pc (a typical distance for discs in the ONC) is a factor $1600$. Hence the dynamical evolution of the star and the SFR can play a major role in the historic UV exposure of any observed star-disc system. 

Many studies have performed N-body simulations of SFRs to track the exposure of star-disc systems, either with aim of quantifying general trends \citep{Holden11, Nicholson19,Concha-Ramirez19,Parker21} or modelling specific regions \citep{Scally01, Winter19b, Winter20b}. These studies generally adopt the FUV driven mass loss rates given by the {FRIED} grid \citep{Haworth18b} with an external flux computed directly from N-body simulations. In general, these studies find that SFRs with typical stellar densities $\rho_* \gtrsim 100\, M_\odot$ are sufficient to rapidly deplete protoplanetary discs \citep[e.g.][]{Nicholson19, Concha-Ramirez19, Concha-Ramirez21}. Similarly, discs are more rapidly depleted in sub-virial initial stellar populations that undergo cold collapse, as a result of the higher densities and therefore UV flux exposure \citep{Nicholson19}. However, initial substructure has a minimal effect on a external photoevaporation since UV fields are generally dominated by the most massive local stars and not necessarily nearest neighbours \citep{Nicholson19, Parker21}. Thus it is volume-averaged rather than star-averaged density measures in a SFR that act as a proxy for the typical external UV flux. This is not necessarily true for the bolometric flux, which may still be dominated by the closest neighbours in a highly structured SFR \citep{Lee20}. 

For an example of how the dynamics in SFRs may alter disc properties, we need only look at the closest intermediate mass SFR to the sun: the ONC. This region has historically dominated the study of externally photoevaporation protoplanetary discs since the discovery of proplyds \citep[e.g.][]{O'Dell94}. Surprisingly, up to $\sim 80$~percent of stars exhibit a NIR excess \citep{Hillenbrand98}, indicating inner-disc retention. This finding is apparently inconsistent with the $\sim 1{-}3$~Myr age \citep{Hillenbrand97, Palla99} when accounting for reasonable stellar dynamics, mass loss rates and initial disc masses \citep[e.g.][]{Churchwell87, Scally01}. The ONC contains one O star, $\theta^1$C, that dominates the local UV luminosity, tying the FUV history of the local circumstellar discs precariously to the history of this single star. Indeed, for this reason intermediate mass star forming regions may be subject to multiple bursts of star formation due to the periodic ejection of individual massive stars \citep{Kroupa18}. This possibility seems to be supported by the clumped age distribution of stars in the ONC, hinting at multiple populations or phases of star formation \citep{Beccari17, Jerabkova19}. {This possibility bears the usual caveat that luminosity spreads do not necessarily imply age spreads, as discussed in Section~\ref{sec:inner_disc}.} However, \citet{Winter19b} showed how, when considering the gravitational potential of residual gas, such episodic star formation can yield inward migration of young stars and outward migration of older stars. This yields a stellar age gradient as observed in the ONC \citep{Hillenbrand97}. As a result, much younger discs experiencing the strongest UV fluxes such that inner disc survival over their lifetime is feasible. Similar core-halo age gradients have been reported in other star forming regions \citep[e.g.][]{Getman14}, highlighting the importance of interpreting disc properties through the lens of the dynamical processes in SFRs.

Unlike regions with one O star, dynamical evolution may not be such an important consideration for discs in massive SFRs. For example, tracking the UV fluxes experienced by stars in the neighbourhood of numerous OB stars, \citet{Qiao22} find that the vast majority of stars experience $F_\mathrm{FUV} \sim 10^5 \, G_0$ at an age of $1$~Myr. The relatively uniform flux in regions with multiple OB stars may go someway to explaining the absence of correlation between disc mass and location found in some simulations \citep{Parker21}. By contrast, \citet{Winter20b} reproduce a disc mass-separation correlation in the relatively low mass region $\sigma$ Orionis region \citep{Ansdell17}, which is occupied by a single massive star (see Section~\ref{sec:surveys}). In either case, it is clear that the physics of star formation cannot be ignored when interpreting the properties of protoplanetary discs, and probably the resultant planetary systems.

\begin{mdframed}
\vspace{-0.8cm}
\subsection{Summary and open questions for the demographics of star forming regions}
Based on the previous dicussion, the following conclusions represent the current understanding of the demographics of SFRs in the context of external disc irradiation:
\begin{enumerate}
    \item When accounting for a standard IMF, the total FUV and EUV luminosity of a SFR is dominated by stars of mass $\gtrsim 30 \, M_\odot$. 
    \item Although most SFRs are low mass and have few OB stars, the majority of stars form in regions with a total FUV luminosity greater than that of the ONC ($\gtrsim 10^{39}$~erg~s$^{-1}$).
    \item In the absence of interstellar extinction, typical FUV fluxes experienced by star-disc systems in the solar neighbourhood are $\langle F_\mathrm{FUV} \rangle \sim 10^3{-} 10^4 \, G_0$. This is enough to shorten their lifetime with respect to the typical $\sim 3$~Myr for isolated discs.
    \item Extinction due to residual gas in SFRs can shield some young circumstellar discs, with ages $\lesssim 0.5$~Myr. However, at later times extinction is inefficient at shielding discs, and unattenuated FUV photons may reach protoplanetary discs. 
\end{enumerate}
Some of the open questions that remain in quantifying the typical external environments for planet formation:
\begin{enumerate}
    \item \textit{How important are the dynamics of SFRs in determining FUV exposure for populations of discs, and how much does this vary between SFRs? }
    \item \textit{How does FUV exposure end external photoevaporation of typical protoplanetary discs vary with cosmic time?}
    \item \textit{Do particular types of planetary system preferentially form in certain galactic environments? }
\end{enumerate}
\end{mdframed}

\begin{figure}
    \centering
    \includegraphics[width=\textwidth]{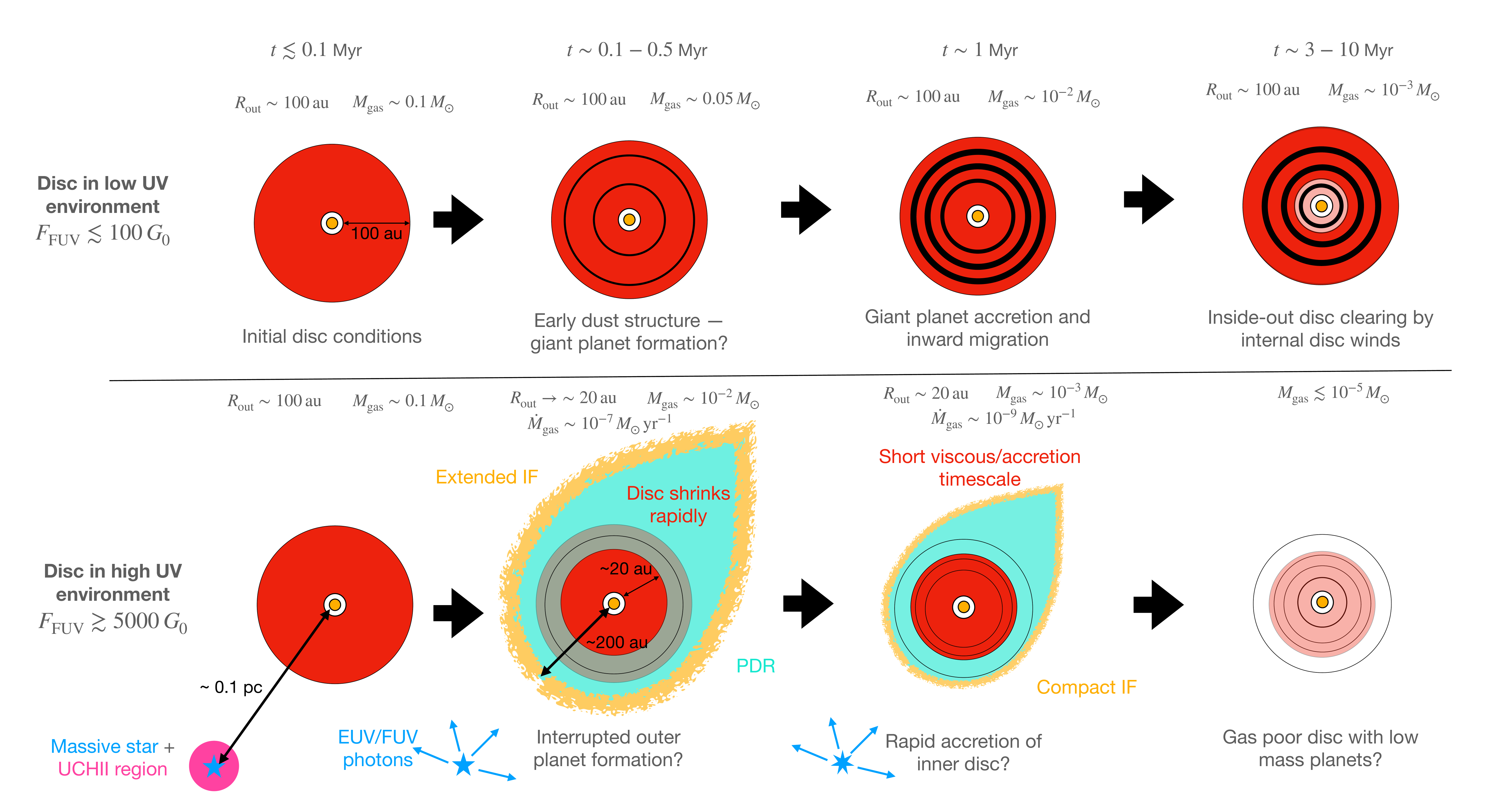}
    \caption{A cartoon for how protoplanetary disc evolution and planet formation proceeds in weakly-irradiated, low mass SFRs (top) and strongly-irradiated, high mass SFRs (bottom). We consider two identical, initially large discs with outer radii of $\sim 100$~au. In the high mass SFR, any neighbouring massive stars may initially be extincted by an UCHII region for $\sim 10^5$~yrs. However, before the star-planet system is $0.5$~Myr old, extinction in the SFR will typically become inefficient and the disc is irradiated by a strong UV field. This produces a bright, extended ionisation front (IF) and rapidly truncates the disc down to a few $10$~au. This may also interrupt the early giant planet formation that proceeds in the outer regions of an isolated disc in a low FUV environment. The irradiated disc is now small, which results in slow mass loss rates and a smaller IF. Given its compact size, the time-scale for viscous draining of the irradiated disc becomes short compared to its isolated counterpart. This can lead to premature clearing of disc material earlier than the typical $\sim 3{-}10$~Myr for which isolated discs survive.    }
    \label{fig:summary_fig}
\end{figure}

\section{Summary and future prospects}

\label{sec:summary}
\label{sec:concluding_summary}

In this review, we have discussed many aspects of the process of external photoevaporation, covering basic physics, observational signatures, consequences for planet formation and prevalence across typical star forming regions. While numerous open questions remain, with the current understanding we can make an educated guess at how protoplanetary discs evolve in different star forming regions. In Figure~\ref{fig:summary_fig} we show a cartoon of how disc evolution proceeds in low and high FUV flux environments. The early period of efficient extinction in SFRs typically lasts less than $0.5$~Myr of a disc's lifetime \citep{Qiao22}. After this, studies of individual externally irradiated discs in proplyd form have demonstrated mass loss rates up to $\sim 10^{-7}{-}10^{-6} \, M_\odot$~yr$^{-1}$ \citep{Weidenschilling77,O'Dell94, Henney99}. This bright and extended proplyd state is short-lived, but the disc is rapidly eroded outside-in during this period. The influence of the external wind on the outer disc can result in dust loss via entrainment in the wind \citep{Miotello12} or due to rapid inward migration of large grains \citep{Sellek20}. It can also interrupt any giant planet formation that occurs on this time-scale in the outer disc \citep{Sheehan18, Segura-Cox20, Tychoniec20}. Given the shorter viscous time-scale of the compact disc, this can lead to a rapid clearing of the disc material similar to expected after gap opening in internally photoevaporating discs \citep{Clarke01}. This is corroborated by the observed shortening of inner disc lifetimes in several massive local SFRs \citep[e.g.][]{Preibisch11, Stolte15,Guarcello16}, however only those in which there are several O stars and a total FUV luminosity $L_\mathrm{FUV} \gtrsim 10^{39}$~erg~s$^{-1}$. In these extreme environments, where high FUV fluxes are sustained over the disc lifetimes, external photoevaporation presumably shuts off accretion onto growing planets and curtails inward migration, possibly leaving behind relatively low mass outer planets \citep{Winter22}. 

Within this framework several questions arise, requiring both theoretical and empirical future study. As an example, models for the chemical evolution of planetary discs and planets in irradiated environments, and their expected observational signatures, are urgently needed. Upcoming JWST investigations of high UV environments may shed some light on disc chemistry from an observational perspective \citep[e.g.][]{JWST_RamirezTannus}. Inferring mass loss rates for moderately photoevaporating discs that do not exhibit bright ionisation fronts is also a crucial test of photodissociation region models. Meanwhile, perhaps the biggest problem for planet population synthesis efforts is determining how the solid content evolves differently in photoevaporating discs; when and where do solid cores form in the discs in high mass versus low mass SFRs? These are just some of the numerous questions that remain open towards the goal of understanding external photoevaporation.

In conclusion, the process of external photoevaporation appears to be an important aspect of planet formation, although the full extent of its influence remains uncertain. Future efforts in both theory and observations are required to determine how it alters the evolution of a protoplanetary disc, and the consequences for the observed (exo)planet population.  

\section*{Acknowledgements}

We thank Cathie Clarke for helpful comments and Arjan Bik for discussions and tables regarding disc lifetimes. We further thank Stefano Facchini, Sabine Richling, Tiia Grenman, Richard Teague, Andrew Sellek, and Ahmad Ali for permission to use the material in Figures 3, 6, 7, 9 and 16 respectively. AJW thanks the Alexander von Humboldt Stiftung for their support in the form of a Humboldt Fellowship. TJH is funded by a Royal Society Dorothy Hodgkin Fellowship.

\section*{Data availability}

All data used in this manuscript are available from previously published articles. Data pertaining to the outcomes of numerical models are available from the authors upon reasonable request. 

%
%
%

%
\bibliographystyle{apalike}
\bibliography{references}
%
%
%

\end{document}